\documentclass[12pt]{article}

\usepackage{graphicx}
\usepackage{scicite}
\usepackage{url}
\usepackage{amssymb}
\usepackage{ref_macros_sci}
\usepackage{caption}

\usepackage{times}

\newenvironment{sciabstract}{%
\begin{quote} \bf}
{\end{quote}}

\newcounter{lastnote}

\title{The Most Energetic Transients:\\Tidal Disruptions of High-Mass Stars} 

\author
{Jason T. Hinkle,$^{1\ast}$ Benjamin J. Shappee,$^{1}$ Katie Auchettl,$^{2,3}$ \\ 
Christopher S. Kochanek,$^{4,5}$ Jack M.~M.~Neustadt,$^{4}$ Abigail Polin,$^{6,7,8}$ \\
Jay Strader,$^{9}$ Thomas W.-S. Holoien,$^{1,6}$ Mark E. Huber,$^{1}$ \\
Michael A. Tucker,$^{4,5}$ Christopher Ashall,$^{1,10}$ Thomas de Jaeger,$^{11}$ \\ Dhvanil D. Desai,$^{1}$ Aaron Do,$^{12}$ Willem B. Hoogendam,$^{1}$ \\
Anna V. Payne$^{13}$
\\
\small{$^{1}$Institute for Astronomy, University of Hawai`i, 2680 Woodlawn Drive, Honolulu, HI 96822, USA}\\
\small{$^{2}$School of Physics, The University of Melbourne, Parkville, VIC 3010, Australia}\\
\small{$^{3}$Department of Astronomy and Astrophysics, University of California, Santa Cruz, CA 95064, USA}\\
\small{$^{4}$Department of Astronomy, The Ohio State University, 140 West 18th Avenue, Columbus, OH 43210, USA}\\
\small{$^{5}$Center for Cosmology and Astroparticle Physics, The Ohio State University,}\\
\small{191 W.~Woodruff Avenue, Columbus, OH 43210, USA}\\
\small{$^{6}$The Observatories of the Carnegie Institution for Science, 813 Santa Barbara St., Pasadena, CA 91101, USA}\\
\small{$^{7}$TAPIR, Walter Burke Institute for Theoretical Physics,}\\
\small{350-17, Caltech, Pasadena, CA 91125, USA}\\
\small{$^{8}$Department of Physics and Astronomy, Purdue University,}\\
\small{525 Northwestern Avenue, West Lafayette, IN 47907, USA}\\
\small{$^{9}$Center for Data Intensive and Time Domain Astronomy, Department of Physics and Astronomy,}\\
\small{Michigan State University, East Lansing, MI 48824, USA}\\
\small{$^{10}$Department of Physics, Virginia Tech, Blacksburg, VA 24061, USA}\\
\small{$^{11}$CNRS/IN2P3 (Sorbonne Université, Université Paris Cité),}\\
\small{Laboratoire de Physique Nucléaire et de Hautes Énergies, 75005 Paris, France}\\
\small{$^{12}$Institute of Astronomy and Kavli Institute for Cosmology, Madingley Road, Cambridge, CB3 0HA, UK}\\
\small{$^{13}$Space Telescope Science Institute, 3700 San Martin Drive, Baltimore, MD 21218, USA}\\
\\
\small{$^\ast$To whom correspondence should be addressed; E-mail: jhinkle6@hawaii.edu.}
}

\date{}

\begin{document} 


\baselineskip24pt


\maketitle 


\begin{sciabstract}
We present the class of extreme nuclear transients (ENTs), including the most energetic single transient yet discovered, Gaia18cdj. Each ENT is coincident with its host-galaxy nucleus and exhibits a smooth ($<$$10$\% excess variability), luminous ($2\times$$10^{45}$ to $7\times$$10^{45}$ erg s$^{-1}$), and long-lived ($>$$150$ days) flare. ENTs are extremely rare ($\geq$$1$$\times$$10^{-3}$ Gpc$^{-3}$ yr$^{-1}$) compared to any other known class of transients. They are at least twice as energetic ($0.5\times10^{53}$ to $2.5\times10^{53}$ erg) as any other known transient, ruling out supernova origins. Instead, the high peak luminosities, long flare timescales, and immense radiated energies of the ENTs are most consistent with the tidal disruption of high-mass ($\gtrsim$$3$ M$_{\odot}$) stars by massive ($\gtrsim$$10^8$ M$_{\odot}$) supermassive black holes (SMBHs). ENTs will be visible to high redshifts ($z\sim4$ to $6$) in upcoming surveys, providing an avenue to study the high-mass end of the SMBH mass distribution, complementing recent studies of actively accreting SMBHs at high redshifts with the James Webb Space Telescope.
\end{sciabstract}

\section*{Introduction}

Accretion onto supermassive black holes (SMBHs) powers many of the most luminous events in the universe. At a redshift of $z \approx 1$, roughly 10\% of SMBHs are actively accreting mass (e.g., \cite{zou24}) and are observed as active galactic nuclei (AGNs). AGN light curves commonly show stochastic variability at a broad range of timescales from minutes to years (e.g., \cite{macleod12}), with some AGNs showing long-term photometric trends often accompanied by dramatic changes in their spectra (e.g.,\cite{shappee14, denney14}). AGNs can also, albeit rarely, exhibit large, coherent flares \cite{graham17}, although the physical mechanisms for powering them are unclear.

With the recent growth of optical transient surveys, several classes of flares coincident with the nuclei of their host galaxies have been detected. These include tidal disruption events (TDEs; \cite{holoien16a, gezari21}), rapid turn-on AGNs \cite{wyrzykowski17, trakhtenbrot19a}, and ambiguous nuclear transients (ANTs; \cite{neustadt20, frederick21, hinkle22a}). Accretion-powered transients share several key observational properties, including bright ultraviolet (UV) emission \cite{holoien16a, hinkle20a, vanvelzen20b, vanvelzen21, gezari21}, strong emission lines \cite{trakhtenbrot19a, vanvelzen21, frederick21}, and often X-ray emission \cite{holoien16a, auchettl17, neustadt20, frederick21}. The smooth flares of nuclear transients on several-month timescales \cite{frederick21, vanvelzen21} are distinct from the stochastic variability typical of AGNs.

A TDE results from the disruption of a star as it passes too close to a SMBH (e.g., \cite{rees88, evans89}). Most observed TDEs appear consistent with the disruption of a main-sequence star with a mass of $\sim0.5 \mbox{--} 2$ M$_{\odot}$ \cite{mockler22}, although there is appreciable scatter in these estimates \cite{ryu20_mass}. Nevertheless, characteristics like enhanced N/C ratios \cite{kochanek16a, mockler22} suggest a population of TDEs resulting from more massive stars. The host galaxies of TDEs typically do not host a strong AGN, although this is likely driven in part by selection effects \cite{gomez23, stein24}. Recently, an increasing number of TDE candidates have been discovered for which their host galaxies exhibit signs of weak AGN activity (e.g., \cite{onori22, hoogendam24}).

\section*{Results}

Owing to the establishment of long-baseline all-sky surveys \cite{shappee14, tonry18, hodgkin13}, we are now sensitive to rare and unexpected classes of transients. One such survey, Gaia Alerts, uses the Gaia spacecraft to monitor the transient sky at approximately monthly cadence with a per-transit 5$\sigma$ depth of 21 mag. From the Gaia Alerts \cite{hodgkin13, hodgkin21} transient stream we selected a sample of flares with three primary characteristics: (1) large amplitudes of $\geq$ 1 mag, (2) smooth light curves with $<$10\% excess variability about the flare evolution, and (3) a long timescale of $\geq 1$ year. Gaia is ideal for such a search as it has observed the full sky since late 2014 and, as a space-based mission, it typically has shorter seasonal breaks than ground-based surveys. Our search yielded two transients, Gaia16aaw (AT2016dbs) and Gaia18cdj (AT2018fbb). We combine these events with the recently published object ZTF20abrbeie (AT2021lwx; \cite{subrayan23, wiseman23}) as a sample of events we will refer to as extreme nuclear transients (ENTs).

The observed properties of the ENTs are reminiscent of extreme versions of ANTs, which are transients occurring in an AGN host galaxy. The light curves of the ENTs, shown in Figure \ref{fig:opt_ir_lcs}, each exhibit a long ($\geq$ 100 day) rise to a high peak luminosity. The ENTs decline slowly after peak, taking more than 150 days to fade to half of their peak luminosity. The ENTs detected prior to the flare show tentative signs of pre-flare variability, suggesting weak AGN activity within their host galaxies. After the UV/optical emission peaks, the ENTs show an IR excess, indicative of transient heating of circumnuclear dust and re-emission at longer wavelengths. Much like the ENT hosts, the host galaxies of AGNs typically have large amounts of nuclear dust.

The ENTs Gaia16aaw and Gaia18cdj are located within $0.68 \pm 0.80$ kpc and $0.25 \pm 0.60$ kpc of their host-galaxy centers, respectively, confirming that they are nuclear transients. Their nuclear origin, long timescales, and high peak luminosities are immediately suggestive of a transient resulting from accretion onto a SMBH. We cannot measure the host offset of AT2021lwx as it has no detected host galaxy prior to the flare, at levels of $>-21$ absolute mag in the rest-frame blue bands. Nevertheless, the similar timescales and peak luminosities of AT2021lwx indicate that it is also powered by accretion, consistent with previous studies.

The ENTs are located at a relatively high redshift of $z \approx 1$, as measured from optical and near-IR follow-up spectra. Using the stellar population synthesis and AGN model of the Code Investigating GALaxy Emission (CIGALE; \cite{burgarella05}), we find that the host galaxies of Gaia16aaw and Gaia18cdj each have a stellar mass of $\approx 9 \times 10^{10}$ M$_\odot$ and star formation rates (SFRs) of $\approx 75 - 110$ M$_\odot$ yr$^{-1}$. While undetected, the luminosity limits for the host galaxy of AT2021lwx combined with a conservative mass-to-light ratio of 3 yields a mass upper-limit of $M \leq 1 \times 10^{11}$ M$_\odot$. The upper limit on the [O II] emission means that the SFR for AT2021lwx is $<4$ M$_\odot$ yr$^{-1}$. 

From typical galaxy-SMBH scaling relations, with a scatter of $\sim$0.4 dex \cite{mcconnell13}, the stellar masses imply SMBH masses of $10^{8.4}$ M$_\odot$ for Gaia16aaw and Gaia18cdj and a 3$\sigma$ upper limit on the mass of $< 10^{8.5}$ M$_\odot$ for AT2021lwx, which are more massive SMBHs than those in the majority of known nuclear transient hosts. The detected ENT hosts are more massive and display higher SFRs than the host galaxies of local nuclear transients such as TDEs and ANTs. Furthermore, the detected ENT host-galaxy masses are within the top few percent of stellar masses at $z = 1$, when the universe was half its current age. Thus, the inferred SMBH masses are similarly extreme at this redshift. In contrast, the prodigious SFRs of the hosts of Gaia16aaw and Gaia18cdj are only moderately high, in the $\sim$70th percentile of specific star formation rate at $z = 1$, when the star formation density in the universe was a factor of 6 higher than today \cite{madau14}.

The rest-frame spectra of the ENTs, shown in Figure \ref{fig:comparison_spec}, exhibit blue spectra with broad lines from the Balmer series of hydrogen and singly-ionized magnesium (Mg II). These are similar to the comparison spectra of the luminous nuclear transients AT2019brs \cite{frederick21}, ASASSN-17jz \cite{holoien21}, PS1-10adi \cite{kankare17}, ASASSN-18jd \cite{neustadt20}, PS16dtm \cite{blanchard17}, and AT2019dsg \cite{vanvelzen21} and broadly similar to the energetic SLSN-II SN2018lzi \cite{pessi24} all shown in Fig.\ \ref{fig:comparison_spec}. ASASSN-15lh \cite{dong16, leloudas16} also shows a blue continuum but does not have strong emission lines. The spectrum of the energetic SLSN-I SN2020qlb \cite{west23} is distinct from the ENT spectra. The persistent blue continua and broad lines are inconsistent with known classes of supernovae, but fully consistent with SMBH accretion. The Mg II and H$\alpha$ emission of the ENTs is broad ($\sim5000  \mbox{--}10000$ km s$^{-1}$) and luminous ($(0.6 \mbox{--} 5) \times 10^{43}$ erg s$^{-1}$), very similar to AGNs.  While Mg II emission has not been seen for TDEs, the H$\alpha$ emission is consistent with TDEs if the lines are equally as overluminous as the broadband emission.

The ENT spectral energy distributions (SEDs) are well-fit by a blackbody model, which is consistent with the super-Eddington accretion expected from TDEs and some supernovae (SNe). In contrast, AGNs with broad lines typically exhibit power-law-like SEDs as a result of viewing the accretion disk directly. The resulting bolometric light curves are shown in Figure \ref{fig:comparison_phot}, along with the comparison objects from Fig.\ \ref{fig:comparison_spec}. The ENTs are extremely luminous, with peak luminosities of $(2 \mbox{--} 7) \times 10^{45}$ erg s$^{-1}$. This is $\sim$$1000$ times more luminous than typical core-collapse SNe, $\sim$$100$ times the average Type Ia SN peak luminosity, $\sim$$30$ times more luminous than the median Type I superluminous supernovae (SLSN-I; \cite{chen23a}), and $\sim$$70$ times more luminous than the average SLSNe-II \cite{pessi24}. The most energetic SLSN-I (SN2020qlb) and SLSN-II (SN2018lzi), shown in Figure \ref{fig:comparison_phot}, are $<20$\% as luminous as the ENTs and an order of magnitude less energetic. Only ASASSN-15lh, suggested to either be SLSN-I \cite{dong16}) or TDE \cite{leloudas16}, rivals these high peak luminosities.

The blackbody properties of the ENTs are also consistent with some form of accretion onto a SMBH rather than an exotic class of SN. The temperatures of the ENTs are hot at $\sim 1.5 \times 10^4$ K and show little or very slow evolution during the flare. This is inconsistent with SLSNe, which quickly cool as the ejecta expands. However, this behavior is similar to TDEs and ANTs, with a remarkable agreement between the ANT and ENT blackbody temperatures. The large effective radii of the ENTs are consistent with SLSNe and some ANTs, although the decreasing blackbody radii in time are more typical of TDEs and ANTs.

The light curve decay timescale of the ENTs is also far longer than most transients, which is consistent with TDEs occurring on massive SMBHs. The rest-frame durations for the flares to fade by half are $(171 \pm 15)$ days for Gaia16aaw, $(155 \pm 10)$ days for Gaia18cdj, and $(205 \pm 20)$ days for AT2021lwx. Figure \ref{fig:transients} shows the position of these ENTs in the parameter space of peak absolute magnitude and characteristic timescale. The ENTs stand out in this parameter space for being very luminous and long-lived. ASASSN-15lh, which has a similar peak luminosity to the ENTs, decays more quickly, with a timescale of $(57 \pm 10)$ days. The position of the ENTs in the upper right corner of this space indicates a high total radiated energy.

The radiated energies of the ENTs are $(5.2 \pm 0.2) \times 10^{52}$ erg for Gaia16aaw, $(2.5 \pm 0.5) \times 10^{53}$ erg for Gaia18cdj, and $(2.2 \pm 0.1) \times 10^{53}$ erg for AT2021lwx. These extreme radiated energies correspond to high accreted masses of $0.3 \mbox{--} 1.4$ M$_{\odot}$ for a typical 10\% accretion efficiency, far greater than typical TDEs and ANTs. The ENT flares are at least twice as energetic as the next most energetic known flares, PS1-10adi \cite{kankare17} and ASASSN-15lh \cite{dong16, leloudas16}, and up to an order of magnitude higher in the cases of Gaia18cdj and AT2021lwx. While the estimates for the accreted mass are extreme compared to local transients, they are fully consistent with the tidal disruption of a high-mass ($\gtrsim$3 M$_{\odot}$) star.

The dust properties of ENTs are very similar to other SMBH accretion-powered transients. To study the environment directly surrounding the SMBHs, we used NEOWISE data to probe the emission from hot dust as it reprocesses the intense UV/optical emission from the flare. By fitting the WISE IR SEDs as blackbodies, we find peak luminosities of $\sim (0.3 \mbox{--} 3) \times 10^{45}$ erg s$^{-1}$, temperatures of $\sim 1500 \mbox{--} 3000$ K and radii of $\sim 0.05 \mbox{--} 0.15$ pc, all consistent with hot dust in nuclear environments \cite{ma13, vanvelzen16b}. From the ratio of the peak IR luminosity to the peak UV/optical luminosity we estimate dust covering fractions of $\sim 0.2 \mbox{--} 0.4$, which we confirm with models of the optical and IR light curves. These covering fractions are consistent with AGN dust covering fractions at similar SMBH masses as well as dust-obscured TDE candidates \cite{masterson24}. This indicates the presence of dense gas and dust near the SMBH, which likely supports the existence of AGN activity, whether weak or in the past, in each of the ENT hosts.

Accretion-powered transients often show X-ray emission, as do two of the three ENTs in our sample. Gaia16aaw and AT2021lwx both show X-rays at levels of $(0.3  \mbox{--} 1) \times 10^{45}$ erg s$^{-1}$ in the rest-frame 0.3--10 keV band throughout the flare, similar to luminous AGNs \cite{ricci17}. Gaia18cdj is undetected in the X-rays at $<4 \times 10^{44}$ erg s$^{-1}$. While significantly more luminous, the X-ray to UV/optical ratios of the ENTs are broadly similar to TDEs and ANTs as well as within the typical range of AGNs, again supporting an accretion-based origin for these events.

As the ENTs are likely powered by accretion, it is important to consider the presence of previous AGN activity. Through a combination of the CIGALE fits, WISE IR colors, narrow emission lines, and X-ray emission we find evidence of a strong AGN in Gaia16aaw. Gaia18cdj likely hosts a weak AGN based on an [O III] line luminosity similar to Seyferts, but the CIGALE fits and MIR colors rule out a strong quasar that dominates the observed emission. The rest-frame UV luminosity and [O III] line luminosity limits derived from the pre-flare properties of AT2021lwx rule out a strong quasar but remain consistent with a weaker AGN. The fact that some ENT host galaxies do not host a strong AGN suggests that prior strong AGN activity is not a requirement to power an ENT.

The rate of luminous, long-lived, accretion-powered events like these ENTs can also be used to understand their potential physical origins. The high redshift and low observed number of ENTs is suggestive of a low intrinsic rate. From the 3 detected ENTs, their peak absolute magnitudes, and the survey parameters for Gaia and ZTF, we estimate a lower limit on the rate of $\gtrsim 1 \times 10^{-3}$ Gpc$^{-3}$ yr$^{-1}$. Further examination of selection effects and biases in searching for these extreme transients is needed to refine estimates of the intrinsic rate of ENTs. Nevertheless, our estimated rate implies that ENTs are roughly ten thousand times less common than SLSNe and TDEs at $z \approx 1$ \cite{prajs17, kochanek16b}.

\section*{Discussion}

With the observed properties of the ENT flares and their estimated rates, we consider potential physical models and plausible origins. First, we examine strong gravitational lensing, which can magnify transient events (e.g., \cite{kelly15}). Constraints on the lack of a foreground lens galaxy from photometry and spectra, even when considering the effects of magnification bias \cite{turner80}, and the high required magnifications ($>$$10$ for normal SNe) make this a remote possibility. Typical radioactively-powered SNe are ruled out based on the unphysically high ($>$$300$ M$_{\odot}$) $^{56}$Ni masses that they would require. There are several luminous classes of SNe \cite{gal-yam12}, including those powered by magnetar spin-down and interactions with circumstellar material (CSM). A magnetar-powered event is ruled out as the most energetic ENT would require a neutron star spinning at breakup to be 5.5 M$_{\odot}$, even assuming 100\% efficiency. This mass is well above the Tolman–Oppenheimer–Volkoff upper limit on the mass of a neutron star (e.g., \cite{tolman39}). CSM interactions are also ruled out as they predominantly produce narrow emission lines, which are not seen for all of the ENTs, and the required CSM masses are of order $1000$ M$_{\odot}$. Thus, no stellar transient can be responsible for the ENTs.

As there is evidence for AGN activity, albeit typically weak, in most of the ENT host galaxies, an AGN origin for the flares must be considered. From studies on quasar variability, a flare resulting from an extreme stochastic variability event is ruled out \cite{macleod12} given estimated rates of smooth AGN flares an order of magnitude below the ENT rate. Another class of transients requiring interaction between an AGN disk wind and the broad line region clouds \cite{moriya17} is unlikely given the high required masses and the sub-Eddington pre-flare accretion rates. Finally, we find it unlikely that instabilities within an AGN disk cause the ENTs due to the similar accreted masses and timescales of the events despite a large range in pre-flare Eddington ratios of the AGNs within the ENT host galaxies, although they cannot be ruled out entirely.

The most plausible physical scenario for these ENTs is the tidal disruption of a high-mass star and the subsequent return of material onto the SMBH. The high masses of the SMBHs naturally provide long-duration flares consistent with the ENT timescales, as the flare timescale scales as $M_{BH}^{1/2}$. Thus, for a TDE occurring on the $\approx 10^{8.4}$ M$_\odot$ SMBHs of the ENT hosts, we would expect a timescale at least 5 times longer than typical TDEs occurring on $\lesssim 10^{7}$ M$_\odot$ SMBHs. As roughly half of the disrupted stellar mass in a TDE leaves the system, the total radiated energies provide a lower limit on the stellar masses of $\gtrsim3$ M$_{\odot}$. The timescales and luminosities for the disruption of $\sim3 \mbox{--} 10$ M$_{\odot}$ stars match the ENT observables well \cite{stone13, kochanek16b}. Scaling from the known local TDE rate and assuming that the TDE rate is proportional to the number of stars, using an IMF with high-mass slope of $\alpha = -2.35$, and their stellar lifetimes we estimate the rate of $3 \mbox{--}10$ M$_{\odot}$ TDEs at $z \simeq 1$ to be $\approx 1.5 \times 10^{-2}$ Gpc$^{-3}$ yr$^{-1}$, consistent with the estimated ENT rate. As the ENT rate is formally a lower limit, we note that top-heavy stellar initial mass functions \cite{lu13} can increase the expected rates. We additionally find several plausible explanations for the difference in observed rates and the theoretical estimate above. Each ENT exhibits a large dust echo, which implies that a large fraction might be obscured and therefore missing from optical surveys. Additionally, as Gaia did not trigger on AT2021lwx, these surveys are incomplete. Both of these effects will result in a higher intrinsic ENT rate than our estimate. Finally, the theoretical rate estimate is weighted towards the lowest mass stars. The theoretically expected rates would decrease if the ENTs are powered by stars more massive than our assumed $3$ M$_{\odot}$ lower bound.

Given the natural explanation of smooth flares \cite{mockler19, price24}, the compatible timescale and luminosities, and several multi-wavelength similarities we propose that the ENTs are the product of the tidal disruption of high-mass stars. The presence of AGNs in the ENT host galaxies is likely related to the order of magnitude increase in the AGN fraction at $z = 1$ as compared to local galaxies \cite{zou24}. Additionally, recent work suggests that AGN disks may increase TDE rates substantially \cite{y_wang24, kaur25}. Both effects make it more likely that TDEs at high redshifts will occur in AGN host galaxies.

These events represent the upper bound of accretion-powered transients to date. For analogs at higher redshift, these ENTs will be an unparalleled window into transient accretion in the early universe given their extreme luminosities. High redshift ENTs will simultaneously probe both the high-mass end of the stellar IMF and SMBH mass distribution in the early universe. Similar events will be visible to the Legacy Survey of Space and Time (LSST; \cite{ivezic08}) on the Vera Rubin Observatory out to a redshift of $z \sim 2 - 3$, although the rates may drop given that the SMBH number density declines by a factor of $\sim5 \mbox{ -- } 30$ as compared to the local density at these redshifts \cite{hopkins07}. Future IR monitoring of the sky from surveys such as the Roman Space Telescope \cite{spergel15} will capture the redshifted rest-frame UV light from these events out to even higher redshifts of $z \sim 4 - 6$. With already 3 well-studied examples from comparatively shallower surveys, ENTs are poised as an ideal beacon to guide our way towards a more complete understanding of the extremes of transient events in the universe.

\section*{Materials \& Methods}

\subsection*{Observational Data}

\paragraph*{Sample Selection} \label{sec:sample}

We select our initial sample of extreme nuclear transients (ENTs) from the Gaia Alerts \cite{hodgkin13, hodgkin21} transient stream. Our criteria for inclusion were designed to select smooth, luminous, and long-lived events. We had five selection criteria: (1) a flare of $\geq 1$ mag over the pre-flare baseline, (2) an observed timescale of that flare of $\geq 1$ year, (3) a flare that was smooth as defined by having a monotonic flare profile and no strong short-term variability ($<$10\% excess variability; below the stochastic variability of AGNs \cite{macleod12}) during the flare, (4) a flare with a peak luminosity of $>10^{45}$ erg s$^{-1}$, and (5) a source without radio and/or gamma-ray detections that would suggest a jetted AGN. This selection resulted in two sources: Gaia16aaw and Gaia18cdj.

We additionally find that the known source ZTF20abrbeie (AT2021lwx) \cite{subrayan23, wiseman23} meets the above criteria to be considered an ENT, although it was not triggered on by the Gaia Alerts team. We suspect that this may be due to a bright nearby star 16.0" away which is $\approx5.2$ mag brighter than ZTF20abrbeie at peak. Since ZTF20abrbeie passes our selection criteria, we include it in our sample of ENTs.

\paragraph*{Archival and Transient Photometry} \label{sec:phot}

We obtained archival photometry of the ENT host galaxies in the $V$-band from the Catalina Real-Time Transient Survey (CRTS; \cite{drake09}), the $W1$ and $W2$ bands of the Wide-field Infrared Survey Explorer (WISE; \cite{wright10}), and the $griz$ bands of the Dark Energy Survey \cite{DES05}. We also obtained survey photometry of the ENTs from the Asteroid Terrestrial-impact Last Alert System  (ATLAS; \cite{tonry18}) in the $c$ and $o$ bands. Each of the ENTs was observed by the Neil Gehrels Swift Gamma-ray Burst Mission (\textit{Swift}; \cite{gehrels04}), from which we measured observer-frame UV/optical photometry. Finally, we acquired light curves from the discovering survey for each source. For Gaia16aaw and Gaia18cdj, this was $G$-band photometry from Gaia and for AT2021lwx this was ZTF photometry in the $g$ and $r$ bands. Details on the reductions of the photometry are given in the ``Archival and Transient Photometry'' section of the Supplementary Materials.

\paragraph*{Offset from Host-Galaxy Center} \label{sec:host_offset}

To constrain the offset of the ENTs from the center of their host galaxies, we aligned the images and then measured the offset between the positions before and during the transient. We elected to measure the relative position before and during transient emission rather than absolute positions to avoid uncertainty in the distortion terms that going to a full world coordinate solution would induce. We first used a modified version of the ISIS image subtraction package \cite{alard98, alard00} interpolation function which matches sources identified by \textsc{sextractor} \cite{bertin96, siverd12} to align the DES images at the location of Gaia16aaw and Gaia18cdj. We can not measure the offset for AT2021lwx as the host galaxy was not detected in any archival images. For the Gaia sources, we first retrieved the DES images from the NOIRlab image servers. Because the image cutouts are not aligned, we first did a rough alignment of all retrieved cutouts including all DES filters. We then removed image cutouts with too small an area to have a sufficient number of alignment stars, images where the source was near the edge, and images where the interpolation did not converge. Despite these cuts, we were left with a sufficient number of images to measure the centroids. We then trimmed the images to the intersecting area and re-interpolated the images. This left 20 images for Gaia16aaw and 8 images for Gaia18cdj.

Next, we used photutils \cite{bradley23} to determine the centroid of the host galaxy before the transient and then again when each transient was near peak.  For Gaia16aaw, the DES images that pass our cuts are from Dec 2013, Jan 2014, Jan 2016, Feb 2016, Nov 2016, Dec 2016, Dec 2017, and Jan 2018. The available Gaia light curve begins in Oct 2014, and it appears that the host+transient emission may already be brighter than the host alone because it was $\sim 0.3$ mag brighter in 2014 than in 2023, indicating that the source was already on the rise. Thus we use the 8 DES images that passed our cuts from Dec 2013 and Jan 2014 for our host galaxy image without the transient flux. The light curve of Gaia16aaw peaks on 2016 March 28 and fades by only $\sim 0.3$ mag by Dec 2016. We use the 9 DES images from 2016 to centroid the transient as these are dominated by the transient emission. We average the centroids from the host galaxy and transient epochs and use the standard deviation of these measurements as an estimate of the statistical error. Finally, we use 10 stars of comparable brightness to Gaia16aaw and centroid those stars in all our interpolated DES images. We then take the median of the standard deviation of each star's positions as an estimate of our systematic uncertainty. This yielded an offset from the host galaxy's center of $0.083" \pm 0.065"_{\textrm{stat}} \pm 0.072"_{\textrm{sys}}$ which corresponds to a physical offset of $0.68 \pm 0.80$ kpc at the distance of Gaia16aaw.

We then repeat this process for Gaia18cdj. Gaia18cdj has DES epochs from Oct 2013, Jan 2014, Oct 2016, Nov 2016, Nov 2018, and Dec 2018. The light curve of Gaia18cdj peaks on 2018 October 24. There is no detectable additional transient flux from Nov 2016 and earlier, so we use the 5 DES images that pass our cuts during these times for the host galaxy and the 3 DES images in 2018 to centroid the transient position. We then follow a similar procedure to Gaia16aaw including 10 comparison stars. This yielded an offset from the host galaxy's center of $0.032" \pm 0.048"_{\textrm{stat}} \pm 0.058"_{\textrm{sys}}$ which corresponds to a physical distance offset of $0.25 \pm 0.60$ kpc at the distance of Gaia18cdj. 

Thus the positions of both Gaia16aaw and Gaia18cdj are both consistent with the center of their host galaxies. We verified this result with full ISIS image subtraction using the DES $g$-band images before the transient to construct a transient-free reference image and then subtract it from images including the transient emission.  We centroided on the reference image and the subtracted image and find that they agree to a similar precision. 

\paragraph*{Follow-up Spectroscopy} \label{sec:spec}

After identification of the two Gaia sources as ENTs, we obtained follow-up spectra. For Gaia16aaw, we obtained an optical spectrum on MJD 59210.1 (854 rest-frame days after peak) using the Inamori-Magellan Areal Camera and Spectrograph (IMACS; \cite{dressler11}) on the 6.5-m Magellan Baade telescope. For Gaia18cdj, we obtained an optical spectrum on MJD 59168.2 (388 rest-frame days after peak) using the Goodman High Throughput Spectrograph \cite{clemens04} on the Southern Astrophysical Research telescope. These spectra were reduced and calibrated with standard IRAF \cite{tody86, tody93} procedures such as bias subtraction, flat-fielding, 1-D spectroscopic extraction, and wavelength calibration. For AT2021lwx, we obtained data for the first observer-frame optical spectra presented in \cite{subrayan23} from the Keck Observatory Archive and reduced it using PypeIt \cite{prochaska20}. The flux calibrations were initially performed using standard star spectra and then scaled to match concurrent Gaia $G$ photometry for Gaia16aaw and Gaia18cdj and ZTF $r$-band photometry for AT2021lwx. These spectra are shown in figure \ref{fig:spectra}. We additionally obtained the follow-up optical spectra from \cite{subrayan23} and \cite{wiseman23} for AT2021lwx, and again used ZTF $r$-band photometry to refine the flux calibration. Finally, to examine the late-time behavior of the narrow lines seen for AT2021lwx, we obtained a spectrum with the Keck Cosmic Web Imager (KCWI) \cite{morrissey18}. This spectrum was reduced with the Keck-supported KCWI data reduction pipeline.

From these spectra, we can estimate the redshifts of the sources. For AT2021lwx, we confirm the redshift of $z = 0.995$ \cite{subrayan23, wiseman23}. For Gaia18cdj there is a clear Mg II absorption doublet that places the source at $z = 0.93747$. The fact that there is Mg II emission at the same redshift as the absorption doublet confirms that this absorption feature is not a foreground absorber. The spectrum of Gaia16aaw does not exhibit any strong narrow emission or absorption features, and the redder portions of the spectrum are noisy due to night sky lines and the faint magnitude of the source at the time the spectrum was taken ($G = 20.7$ AB mag). However, there is a strong broad feature that we interpret as Mg II, placing the source at $z = 1.03$. Our identification of this feature as Mg II is supported by the broad Fe feature just blueward of the Mg II feature. Such a feature is seen in the spectra of both Gaia18cdj and Gaia16aaw. We also evaluated the possibility that this broad line was a Balmer or He feature. For each possible identification, other expected strong lines are not seen, further supporting the Mg II interpretation.

In addition to our optical spectra, we obtained a near-infrared (NIR) spectrum of Gaia18cdj with the Folded port InfraRed Echellette (FIRE; \cite{simcoe08}) in Prism Mode. We reduced this spectrum using PypeIt \cite{prochaska20}, wavelength calibrating our data using an arc lamp, flux calibrating the extracted spectrum with a nearby A0V star, and doing telluric calibration with the \textsc{poly} model within PypeIt. For AT2021lwx, we obtained the NIR spectrum presented in \cite{wiseman23}.

\subsection*{Host-Galaxy Properties}

\paragraph*{Stellar Mass and Star Formation Rate} \label{sec:mass_and_sfr}

The host galaxies of Gaia16aaw and Gaia18cdj are readily detected in archival surveys (see Fig. \ref{fig:opt_ir_lcs}), so we model their properties. We used the Code Investigating GALaxy Emission (CIGALE; \cite{burgarella05, boquien19, yang22}) because it allows for simultaneous contributions from AGN activity and stellar emission. We fit the optical ($griz$) data from co-adding the pre-flare epochs of DES imaging, near-infrared ($JK_s$) data from the VISTA Hemisphere Survey \cite{mcmahon13}, and mid-infrared ($W1W2$) data from the AllWISE catalog \cite{wright10}.

We used a delayed star formation history, the Bruzual \& Charlot \cite{bruzual03} stellar population models, a Salpeter initial mass function \cite{salpeter55}, a CCM \cite{cardelli89} extinction law with R$_V$ = 3.1, and the SKIRTOR AGN model \cite{stalevski12, stalevski16}. From our CIGALE fits, we find that the host galaxy of Gaia16aaw has a stellar mass of $M_* = (9.5 \pm 4.7)$ $\times 10^{10}$ M$_{\odot}$, an age of $(3.4 \pm 1.2)$ Gyr, and a star formation rate of SFR $= (110 \pm 42)$ M$_{\odot}$ yr$^{-1}$. For the host of Gaia18cdj the values are $M_* = (9.4 \pm 1.8)$ $\times 10^{10}$ M$_{\odot}$, an age of $(2.0 \pm 0.8)$ Gyr, and a SFR $= (74 \pm 23)$ M$_{\odot}$ yr$^{-1}$. We used synthetic photometry computed from the best-fitting host-galaxy models to subtract the host-galaxy fluxes from photometry as needed to isolate the transient flux.

The host galaxy of AT2021lwx is undetected even in deep pre-outburst optical and IR imaging, precluding any detailed analysis of its properties. However, previous works have estimated upper limits on the stellar mass and star formation rate (SFR). From the Pan-STARRS upper limits and assuming a mass-to-light ratio of 2, \cite{wiseman23} compute a mass limit of $<7 \times 10^{10}$ M$_{\odot}$. Since there is considerable scatter in the mass-to-light ratio for the bluer rest-frame bands being probed by the Pan-STARRS data, up to a $M/L \sim 3$ \cite{bell01}, we will instead adopt a slightly more permissive limit of $M_{*} <1.1 \times 10^{11}$ M$_{\odot}$ here. From limits on the [O II] luminosity in their follow-up spectra, \cite{wiseman23} calculate an upper limit on the SFR of $<3.7$ M$_{\odot}$ yr$^{-1}$, which we confirm using a deep follow-up spectrum shown in fig.\ \ref{fig:spectra}.

The stellar masses and SFRs from CIGALE for the hosts of Gaia16aaw and Gaia18cdj and the estimated limits for AT2021lwx are shown in Figure \ref{fig:host_props}. Along with these three ENTs, we show host properties for a sample of tidal disruption events \cite{hinkle21b} in blue and a sample of ambiguous nuclear transients \cite{hinkle21b, hinkle22a} in gold. We additionally show two broader galaxy samples in gray, one local sample from SDSS \cite{eisenstein11} and the MPA-JHU catalog \cite{brinchmann04} and a sample of galaxies at $z \approx 1$ from Cosmic Evolution Survey 2020 data release (COSMOS; \cite{weaver22}). We have also indicated the SFRs expected for the star-forming main sequence at $z = 1$ \cite{popesso23} with the dashed line. In addition to the individual source host properties, we show kernel density estimates (KDE) computed using \textsc{scipy.stats.gaussian\_kde} and Scott's Rule to model the underlying distributions.

The stellar masses and SFRs of the ENT hosts are generally higher than any of the hosts in comparison samples. As compared to the TDE sample, the stellar masses are more than an order of magnitude larger than the typical TDE host and the SFRs are several orders of magnitude higher. When compared to the ANTs, the difference is less stark in mass, although the SFRs are still much higher. Finally, as compared to the local galaxy population, the host masses are consistent with the massive end of the blue sequence, but with SFRs nearly two orders of magnitude higher than the expected SFR for this mass. However, this is expected as these galaxies reside at $z \approx 1$. When compared to the location of the star-forming main sequence at this redshift as probed by the COSMOS data, the difference is less dramatic. In particular, when compared to the expected SFR \cite{popesso23} given the star-forming main sequence at the redshift and stellar mass of Gaia16aaw and Gaia18cdj, they lie 0.45 and 0.36 dex above the relation respectively. As compared to the typical dispersion about the star-forming main sequence of $0.2 - 0.5$ dex \cite{guo13, popesso23}, these hosts appear high, but consistent with, a typical massive star-forming galaxy at $z \approx 1$. The masses of the hosts for Gaia16aaw and Gaia18cdj are quite high for $z = 1$, far more massive than the typical galaxy at that redshift. 

Using the scaling relationship between stellar mass and SMBH mass \cite{mcconnell13, mendel14}, with a typical scatter of $\sim 0.4$ dex, we can estimate the central black hole masses. For the hosts of Gaia16aaw and Gaia18cdj, which have similar stellar masses, we find a $M_{BH} \sim 10^{8.4}$ M$_{\odot}$. Taking the limit on stellar mass for AT2021lwx, we find $M_{BH} < 10^{8.5}$ M$_{\odot}$, which is consistent with previous estimates \cite{subrayan23, wiseman23}. Despite being an upper limit, we can obtain a conservative lower bound by assuming a peak luminosity that is capped at $10\times$ the Eddington limit and calculating the corresponding SMBH mass. This yields a mass of $>$$10^{6.7}$ M$_{\odot}$. Thus, the SMBH mass for AT2021lwx may be consistent with normal TDEs, although only if the peak luminosity is substantially super-Eddington.

\paragraph*{Presence of AGN Activity} \label{sec:agn_activity}

Our CIGALE fits also allow us to examine the contribution of AGN activity to the pre-flare emission of the Gaia sources. For Gaia16aaw, the fits prefer an AGN luminosity of $4.6 \times 10^{45}$ erg s$^{-1}$ as compared to a stellar output of $3.3 \times 10^{45}$ erg s$^{-1}$. This suggests that Gaia16aaw hosts a relatively strong AGN, comparable to the combined stellar luminosity. For Gaia18cdj the AGN luminosity is $2.4 \times 10^{45}$ erg s$^{-1}$ and the stellar luminosity is $2.6 \times 10^{45}$ erg s$^{-1}$, consistent with a slightly weaker AGN relative to the stellar output. The fractional errors on the AGN luminosity estimates are much higher ($\approx 50-100$\%) than the $\approx 20-25$\% fractional uncertainties for the stellar luminosity. 

We can also assess the presence of AGN activity through the WISE $W1 - W2$ color and associated selection criteria \cite{assef10, assef13}. Gaia16aaw has a color of $W1 - W2 = 1.03 \pm 0.05$ Vega mag. This is redder than the $\sim 0.8$ mag threshold for local galaxies and the $0.33$ mag color expected for star-forming galaxies at this redshift. However, the position of AGNs within this color space changes with redshift. For the AGN template of \cite{assef10} at z = 1.03, the $W1 - W2$ color is 1.41 Vega mag, suggesting that while the host of Gaia16aaw likely hosts an AGN, it may not be dominating the MIR emission.

The $W1 - W2$ color of Gaia18cdj is $0.49 \pm 0.10$ Vega mag, bluer than both the local threshold and the expected AGN color of $1.39$ mag at the redshift of Gaia18cdj. It is redder than the $0.26$ mag color for a typical star-forming galaxy at this redshift, supporting the likelihood that the host of Gaia18cdj hosts a relatively weak AGN. From the observed NIR (rest-frame optical) spectrum of Gaia18cdj, we measured the flux of the narrow [O III] $\lambda5007$ line by fitting it as a Gaussian and estimating errors through Monte Carlo resampling. We find a flux of $(1.8 \pm 0.3) \times 10^{-17}$ erg s$^{-1}$ cm$^{-2}$, with a full-width at half-maximum (FWHM) of $800 \pm 150$ km s$^{-1}$, and an equivalent width (EW) of $(4.2 \pm 0.8)$ \AA. At the distance of Gaia18cdj, this flux corresponds to a luminosity of $8.1 \times 10^{40}$ erg s$^{-1}$. The luminosity and EW of the [O III] line are modest compared to quasars but consistent with Seyferts (e.g., \cite{zakamska03, hao05, bongiorno10}). Additionally, the high line width supports an AGN origin, although the line is only marginally resolved in the R $\sim 500$ FIRE spectrum.

While the host of AT2021lwx is undetected and therefore we cannot use CIGALE or typical color selections to constrain pre-flare AGN activity, the archival photometry and follow-up spectra can be used to place constraints. AT2021lwx is undetected in Pan-STARRS with $g > 23.3$ mag \cite{wiseman23}. If we assume that the corresponding rest-frame UV emission ($\lambda_{rest} \approx 2400$\AA) is solely tracing light from an AGN, this places a limit on the AGN luminosity of $\lambda L_{\lambda} < 10^{44}$ erg s$^{-1}$, ruling out a quasar. Additionally, the spectra in \cite{subrayan23} and \cite{wiseman23} show no signs of strong narrow emission lines from AGN activity. We computed an upper limit on the [O III] $\lambda5007$ luminosity following the procedure of \cite{leonard01}. We assumed a line width of 1000 km s$^{-1}$ and obtained 3$\sigma$ flux limits as

\begin{equation}
F(3\sigma) = 3C_{\lambda}\Delta I \sqrt{W_{line}\Delta X}
\end{equation}

\noindent where $C_{\lambda}$ is the continuum flux at wavelength $\lambda$, $\Delta I$ is the RMS scatter about the normalized continuum, $W_{line}$ is the width of the line profile, and $\Delta X$ is the pixel scale of the spectrum. This yields a 3$\sigma$ upper-limit of $<3.4 \times 10^{41}$ erg s$^{-1}$ which also rules out a luminous quasar but is consistent with a Seyfert (e.g., \cite{zakamska03, hao05, bongiorno10}).

Finally, we can use archival X-ray data to probe pre-flare AGN activity in these galaxies. From ROSAT All-Sky Survey (RASS) data we calculate count rates for the host galaxies of our ENTs and convert them to rest-frame $0.3-10$ keV fluxes assuming a photon index of $\Gamma = 2$ and the Galactic foreground column density \cite{HI4PI16}. Gaia16aaw is detected in ROSAT, with a rest-frame $0.3-10$ keV luminosity of $(2.3 \pm 0.6) \times 10^{45}$ erg s$^{-1}$, significantly higher than the $\sim 5 \times 10^{41}$ erg s$^{-1}$ expected from the high SFR \cite{gilfanov04, riccio23}, thus confirming the presence of an AGN. Gaia18cdj is undetected, with a 3$\sigma$ upper-limit of $<2.6 \times 10^{45}$ erg s$^{-1}$. Gaia18cdj is also undetected in the later \textit{Swift} XRT coverage, providing a deeper limit of $<3.6 \times 10^{44}$ erg s$^{-1}$. This further rules out a strong quasar, but this limit is consistent with a weaker AGN. For AT2021lwx, the limit from RASS (using $\Gamma = 0.6$) is $<1.1 \times 10^{46}$ erg s$^{-1}$, which does not constrain the pre-flare AGN activity.

\subsection*{Context Within the Transient Landscape} \label{sec:context}

With our optical and IR light curves of Gaia16aaw, Gaia18cdj, and AT2021lwx, we can compare these ENTs to other classes of transient events. In particular, the parameter space of peak absolute magnitude and the characteristic timescale of a transient often separates distinct source classes as shown in Figure \ref{fig:transients}. 

To place the ENTs on this diagram, we first computed the peak magnitudes by fitting a smooth spline to the best-sampled optical light curve. For Gaia16aaw and Gaia18cdj this was the Gaia $G$ light curve and for AT2021lwx this was the ZTF $r$ light curve. These filters have similar effective wavelengths and the sources lie at comparable redshifts, making this a reasonable comparison. Our uncertainties were estimated by taking the standard deviation of the peak magnitudes from 10,000 Monte Carlo iterations. Gaia16aaw peaked at an apparent magnitude of m$_G = 19.37 \pm 0.01$ on MJD $= 57476 \pm 13$. Gaia18cdj peaked at an apparent magnitude of m$_G = 18.21 \pm 0.01$ on MJD $= 58416 \pm 9$. AT2021lwx peaked at an apparent magnitude of m$_r = 18.02 \pm 0.03$ on MJD $= 59342 \pm 15$.

The corresponding peak absolute magnitudes, accounting for foreground extinction and applying a flat K-correction of $-2.5\textrm{log}_{10}(1+z)$, are $M_G = -24.11 \pm 0.01$, $M_G = -25.06 \pm 0.01$, and $M_r = -25.65 \pm 0.03$ for Gaia16aaw, Gaia18cdj, and AT2021lwx respectively. The uncertainties are only statistical errors from the Monte Carlo procedure. We also show the source ASASSN-15lh as a comparison object given its status as a similarly overluminous transient \cite{dong16, leloudas16}. From the same procedure, applied to the ASAS-SN $V$-band light curve, we estimate a peak absolute magnitude of $M_V = -23.18 \pm 0.06$.

Next, we calculated the characteristic timescale of the ENTs. We again fit a smooth spline to the optical light curves and define the characteristic timescale as the rest-frame time difference between the time of peak and when the source had faded to half of the peak flux. We estimated the uncertainty through Monte Carlo resampling of the flare and added this in quadrature with the uncertainty on the peak time to compute the total uncertainty. The characteristic timescales are $(171 \pm 15)$ days, $(155 \pm 10)$ days, and $(205 \pm 20)$ days for Gaia16aaw, Gaia18cdj, and AT2021lwx, respectively. For reference, the next most luminous transient, ASASSN-15lh, has a characteristic timescale estimated in the same manner of $(57 \pm 10)$ days.

Figure \ref{fig:transients} compares the ENTs to a number of different types of transients. In blue we show various classes of supernovae, ranging from the fast and faint Calcium-Rich transients (e.g., \cite{kasliwal12}) to the SLSNe (\cite{gal-yam19}). The green boxes show transients related to stellar mergers and/or mass transfer (e.g., \cite{berger09, pastorello19}). In red shades, we show transients powered by accretion onto SMBHs, including tidal disruption events \cite{gezari21}. We have also estimated the absolute magnitude and characteristic timescale range for the growing class of ambiguous nuclear transients \cite{frederick21, holoien21, hinkle22a, hinkle22_dust, hinkle23b}. While their connection to SMBH accretion is likely, we have elected to present them in a light shade of red to indicate the uncertainty regarding their physical origin. It is clear that the ENTs studied here lie at higher peak luminosities than any other source class and are among the longest-lived. The only other transients that rival the long characteristic timescales are the core-collapse supernovae, which are $\sim 7$ mag fainter, and some ANTs, which are still several magnitudes fainter at peak.

The ranges shown for the various classes of transients are broadly based on \cite{kasliwal12}. Given the large increase in discoveries since those works, we have augmented the regions for certain classes, such as TDEs and SLSNe, to reflect the full range of properties better. Nevertheless, it is clear that these ENTs reside in a region currently unoccupied by other transient classes.

As our sample of ENTs clusters in this parameter space rather tightly, we draw a red box in Figure \ref{fig:transients} as a proposed selection method for similar events. To cleanly separate the ENTs from other classes of luminous transients, we propose a threshold in peak absolute magnitude of $M \leq -24$. While a separation in terms of the characteristic timescales for these events is less clear, we propose a threshold of $\tau \geq 125$ days, which is 20\% lower than for our shortest timescale ENT. Any transient meeting these criteria would emit an energy comparable to our ENTs and far more than any normal class of supernova or accretion-powered transient.

\paragraph*{Other Groups of Luminous Flares}

While the ENTs are the most extreme transients yet discovered, there are other groups of luminous nuclear flares with similar properties to the ENTs. One such group is the AGN flares discovered by CRTS \cite{graham17}. Many of the CRTS AGN flares are luminous and long-lived, but they typically are not as smooth, with many showing short-term variability (i.e. $\lesssim 100$ rest-frame days) or long-timescale (i.e. $\gtrsim 500$ rest-frame days) re-brightening episodes. Following the same procedure as for the ENTs, we quantified this by computing the excess variability of the spectroscopically-confirmed CRTS flares.

Of the monotonic AGN flares in the CRTS sample, the median excess variability is 8\%, higher than any of the ENTs, with only 5 sources having excess variability equal to or less than the ENTs. Nevertheless, 14 of these flares pass our 10\% excess variability cut, indicating a population of truly smooth flares within this sample. To further examine the nature of these smooth  AGN flares, we compared their absolute magnitudes and energies \cite{graham17} to the ENTs in Figure \ref{fig:luminous_flares_comp}. Despite several of these events having smooth flare profiles and luminous peak absolute magnitudes, the energies are considerably lower than the ENTs. Most are at least an order of magnitude less energetic and even the most energetic flare is a factor of 2 lower in energy. Thus, these AGN flares do not belong to the same population as the ENTs.

Recently, \cite{wiseman25} searched ZTF for luminous, long-duration ANTs, finding a sample of 10 events with high peak luminosities and long timescales. This includes AT2021lwx, considered an ENT in this work. Two events, ZTF19aamrjar and ZTF20aaqtncr, which initially showed smooth, singular flares have non-monotonic behavior in recent ZTF photometry. Nevertheless, the light curves, SEDs, and spectra of these events are reminiscent of ENTs. We compare this sample of ANTs to the ENTs and CRTS AGN flares in Fig. \ref{fig:luminous_flares_comp}. The range of peak absolute magnitudes between the ANTs and CRTS AGN flares is similar, but the ANTs are longer-lived and have higher emitted energies. Many of the ANTs are less luminous and energetic than the ENTs, but they appear to lie along a sequence, suggesting similar physical origins. Based on their properties and rates, \cite{wiseman25} found that these ANTs are consistent with the tidal disruption of intermediate-mass stars. If the ENTs are physically related, this may suggest that ENTs result from the disruption of higher-mass stars.

Of the ANTs in \cite{wiseman25}, only ZTF19aamrjar has a peak absolute magnitude meeting our proposed ENT threshold. It additionally has a long timescale, making it more energetic than the ENTs studied here. Thus, of the sample of $\sim 25$ other luminous nuclear transients from CRTS and ZTF, only one additional source is a potential ENT candidate, affirming the rarity of the ENTs. However, as previously noted, ZTF19aamrjar has recently exhibited a large re-brightening, with the second peak within 0.5 mag of the first peak. This behavior is not seen for any of the ENTs despite similar rest-frame baselines, underscoring the fact that ENTs are unique as luminous, energetic, and monotonic flares.

\subsection*{Properties of the Flares}

\paragraph*{Spectral Energy Distribution} \label{sec:SED}

Using the well-sampled multi-wavelength light curves from DES ($griz$; for Gaia16aaw and Gaia18cdj) and ZTF ($gr$; for AT2021lwx), we can study the time evolution of the spectral energy distributions (SEDs) of these ENTs. As each of these ENTs is at a redshift of $z \approx 1$, the observed $griz$ bands correspond to the rest-frame near-UV and blue, with wavelengths of $\sim 2400$\AA, $\sim 3210$\AA, $\sim 3910$\AA, and $\sim 4580$\AA \ respectively. In figure \ref{fig:color_plots}, we show the light curves from these surveys along with the corresponding color evolution. As the cadence in each band is not identical, we have interpolated the time series and calculated colors at the times corresponding to the bluer filter of the pair. In this figure, we have not removed any host-galaxy flux, so that we can directly compare the source color both prior to and during the flare.

For Gaia16aaw, the source is moderately red in the first DES epoch and becomes significantly bluer by peak emission. After peak, Gaia16aaw becomes redder again, approaching the colors of the earliest DES epoch. In the case of Gaia18cdj, the pre-flare emission from the host is red, both in comparison to the first epoch of Gaia16aaw and to the near-peak DES epoch of Gaia18cdj. The redder color in quiescence is yet again suggestive of a weaker AGN component than Gaia16aaw. Near peak, the color of Gaia18cdj is similar to the other two sources, if not slightly bluer. For AT2021lwx, the ZTF photometry spans the whole flare, showing a largely monotonic change from blue to red colors as the flare progresses. The $g -r$ color of AT2021lwx at peak is redder than either of the other two sources near their peak.

We can go beyond just the colors of the sources and estimate their properties through blackbody fits to their time-evolving SEDs. Using Markov Chain Monte Carlo (MCMC) methods and a forward-modeling approach, we fit the available epochs of host-subtracted and foreground extinction-corrected multi-band photometry as a blackbody to obtain the bolometric luminosity, temperature, and effective radius for the ENTs. To keep our fits relatively unconstrained, we adopted wide flat temperature priors of 1000 K $\leq T \leq$ 50000 K and flat radius priors of $10^{13}$ cm $\leq R \leq$ $10^{17}$ cm. For Gaia16aaw, we fit three epochs of DES data, for Gaia18cdj we fit two epochs of DES data and one epoch of \textit{Swift} photometry, and for AT2021lwx we fit six epochs of \textit{Swift} data. For the DES data, we added a 2\% uncertainty in quadrature with the photometric uncertainties before fitting, to avoid underestimating the true uncertainties on the fitted blackbody parameters. These fits are given in table \ref{tab:BB_fits}. We find that a blackbody model adequately describes the observed emission, with a median reduced $\chi^2$ of 2.3, 0.8, and 0.7 for Gaia16aaw, Gaia18cdj, and AT2021lwx respectively. The photometry and best-fitting blackbody models are shown in figure \ref{fig:BB_fits}.

In Figure \ref{fig:BB_comp} we show the results for the ENT luminosity, radius, and temperature compared to a sample of well-observed TDEs \cite{hinkle21b}, SLSNe-I from \cite{hinkle23}, and ANTs \cite{hinkle21b, hinkle22_dust}. In terms of luminosity, the ENTs are clearly more luminous and longer-lived, as expected given their position in Fig.\ \ref{fig:transients}. Their decline slopes are flatter than all of the TDEs and SLSNe, but similar to some of the ANTs. Two transients, ASASSN-15lh (either a SLSN or TDE; \cite{dong16, leloudas16, godoy-rivera17}) and AT2019brs (an ANT; \cite{frederick21}) have similar peak luminosities, although the initial decline rate for ASASSN-15lh is much steeper and thus the integrated energy is lower.

The effective radii of our ENTs are significantly larger than the TDEs, but consistent with the SLSNe and ANT comparison samples. In terms of the radius evolution, the ENTs appear to show a slow decline in radius post-peak. This is consistent with the TDE and ANT evolution, but not with SLSNe whose effective radii increase with time. The effective temperatures of the ENTs lie between those of the SLSNe and TDE samples and are strikingly similar to the ANTs. The temperature of Gaia16aaw slowly cools after peak, whereas Gaia18cdj slightly increases during the flare. Such behavior is more consistent with either TDEs or ANTs than the cool temperatures of SLSNe that drop dramatically after peak as the ejecta expand and cool (e.g., \cite{nicholl15}).

For each of our sources, we also created a bolometric light curve by scaling the host-subtracted Gaia $G$ (for Gaia16aaw and Gaia18cdj) and ZTF $r$ (for AT2021lwx) light curves to match the bolometric luminosity estimated from the blackbody fits. For epochs between blackbody fits we linearly interpolated the bolometric correction. For data outside of this range we used a flat bolometric correction corresponding to either the first or last fitted blackbody epoch. This is similar to what we have done for previous events \cite{hinkle20a, hinkle21b, hinkle22a} and gives a peak luminosity for AT2021lwx consistent with previous estimates \cite{subrayan23, wiseman23}. 

Figure \ref{fig:comparison_phot} compares the ENT bolometric light curves to luminous nuclear transients of various classes. As in Fig.\ \ref{fig:BB_comp} the ENTs are generally the most luminous transients, although ASASSN-15lh has a higher bolometric luminosity than Gaia16aaw at peak. It is clear that the ENTs decay more slowly than the majority of other transients. 

Finally, we used the bolometric light curves to quantify the smoothness of the ENT flares. We used a Savitzky–Golay filter \cite{savitzky64} with a window of 100 rest-frame days and a cubic polynomial to compute the long-term trend. We then normalized the bolometric light curves by this long-term trend and computed the RMS scatter as the characteristic fractional variability of the flare about the overall flare profile. This was 12\% for Gaia16aaw, 9\% for Gaia18cdj, and 3\% for AT2021lwx. After subtracting the median fractional uncertainty of the bolometric light curve in quadrature (e.g., \cite{vaughan03}), Gaia16aaw has an excess variability of 6\% and both Gaia18cdj and AT2021lwx are consistent with noise. These confirm that the ENT flares are smooth compared to typical quasars (e.g., \cite{macleod12}).

\paragraph*{Radiated Energy} \label{sec:energies}

From the bolometric light curves, we can calculate a lower limit on the radiated energy through a trapezoidal integral in time. Using \textsc{numpy.trapz}, we estimate radiated energies of $>4.9 \times 10^{52}$ erg for Gaia16aaw, $>1.5 \times 10^{53}$ erg for Gaia18cdj, and $>2.1 \times 10^{53}$ erg for AT2021lwx. This estimate for AT2021lwx is consistent with earlier estimates \cite{subrayan23, wiseman23}, especially when considering the longer temporal baseline in this study. These lower limits are higher than those for any other known optical transients.

We also estimated the total radiated energy by fitting a Gaussian to the early-time rise and an exponential decline to the late-time decline. This allowed us to smoothly extrapolate to times without observational constraints using rise and decay slopes motivated by the existing data. We integrated the Gaussian fit before the first epoch of data and the exponential decline after the last epoch, adding these energies to the energy computed by directly integrating the bolometric light curve. Given the high S/N data for these ENTs, the statistical uncertainties on the emitted energy from the bolometric light curve are small. Therefore, we conservatively estimate the uncertainty on the energies as half of the difference between the directly-integrated energy and the total energy from adding the fits to unobserved portions of the transient evolution. The total radiated energy for Gaia16aaw was $(5.2 \pm 0.2) \times 10^{52}$ erg, for Gaia18cdj it was $(2.5 \pm 0.5) \times 10^{53}$ erg, and for AT2021lwx the energy was $(2.2 \pm 0.1) \times 10^{53}$ erg. Unsurprisingly, the estimated total energies are not much higher than the directly-integrated energies since each of the ENTs has been observed late into their evolution. Assuming these flares are powered by accretion with an efficiency of 10\%, the energies correspond to accreted masses of $\approx 0.3 \mbox{--} 1.4$ M$_{\odot}$.

These energies are significantly higher than other known transients. Following the same procedure, we estimate a total emitted energy of $(2.4 \pm 0.1) \times 10^{52}$ erg for ASASSN-15lh, less than half that of Gaia16aaw, already the least energetic of the three flares studied here. The energetic transient PS1-10adi \cite{kankare17}, with a total energy of $(2.3 \pm 0.5) \times 10^{52}$ erg also lies below these flares. These flares are roughly an order of magnitude more energetic than typical well-observed SLSNe (e.g., \cite{hinkle23}), TDEs (e.g., \cite{mockler22}), and ANTs (e.g., \cite{holoien21}) and at least several times more energetic than the most energetic examples of each class. While an extreme SN origin can plausibly explain sources like ASASSN-15lh and PS1-10adi, such a scenario is ruled out for the ENTs.

\paragraph*{Dust Covering Fraction} \label{sec:dust}

Each of the ENTs in our sample has good coverage of a flare in WISE data, allowing us to estimate the properties of dust in the nuclei of the host galaxies. At the redshift of these sources, the WISE $W1$ and $W2$ bandpasses observed during NEOWISE roughly correspond to the rest-frame $H$ and $K$ bandpasses. These data probe hot dust emission and can constrain the dust covering fraction for these host galaxies.

We follow a similar procedure as \cite{hinkle22_dust}. Briefly, we subtract the pre-flare emission from the $W1$ and $W2$ light curves and fit the host-subtracted and extinction-corrected WISE photometry as a blackbody. To keep our fits relatively unconstrained, we adopted flat temperature priors of 100 K $\leq T \leq$ 5000 K and flat radius priors of $10^{15}$ cm $\leq R \leq$ $10^{19}$ cm. Given a fit to only two data points per epoch, there is a degeneracy between temperature and radius. Additionally, with the relatively large photometric errors, we find average fractional uncertainties of $\approx$20\% in temperature, $\approx$50\% in radius, and $\approx$20\% in luminosity per epoch. Regardless, these luminosity constraints are sufficient to estimate the dust covering fraction. From the luminosity evolution in the IR and UV/optical, we fit for the peak luminosity in each band. The dust covering fraction (f$_c$) is estimated as the ratio of the peak IR luminosity to the peak UV/optical luminosity (e.g., \cite{jiang21b}).

Our blackbody fits to the NEOWISE data yield dust temperatures of $\sim 1500 \mbox{--} 3000$ K and effective radii of $\sim 0.05 \mbox{--} 0.15$ pc with the temperatures decreasing and the radii increasing in time, consistent with previous studies of hot dust in galactic nuclei \cite{vanvelzen16b, jiang21b, hinkle22_dust}. From the luminosity ratios, we find only a lower limit of $f_c \geq 0.2$ for Gaia16aaw, as the peak of the IR luminosity is not observed. For Gaia18cdj we find $f_c = 0.22 \pm 0.07$ and for AT2021lwx we find $f_c = 0.42 \pm 0.12$, where the uncertainties are estimated from Monte Carlo resampling. Further discussion of the covering fraction estimates can be found in the ``Dust Covering Fraction'' section of the Supplementary Materials.

These covering fractions are shown in Figure \ref{fig:dust_covering}, where they are compared to previously estimated covering fractions for TDEs \cite{jiang21b, masterson24} and ANTs \cite{hinkle22_dust}. The ENT dust covering fractions are much higher than for optically-selected TDEs, but very similar to both ANTs and the minimum covering fraction estimates for the IR-selected TDEs \cite{masterson24}. The ENT covering fractions are also remarkably consistent with the typical covering fraction of AGNs with $M_{BH} \sim 10^{8-8.5}$ (e.g., \cite{ma13}). The nuclei of the ENT hosts are clearly dusty and likely harbor an AGN-like ``dusty torus'', which is unsurprising given the existing evidence for AGN activity in the host galaxies of Gaia16aaw and Gaia18cdj.

\paragraph*{Emission Lines} \label{sec:lines}

All of the ENTs in our sample show emission lines. In our spectra of Gaia16aaw and Gaia18cdj, the most prominent line is the broad Mg II line. AT2021lwx exhibits a number of strong and relatively narrow emission lines, but also broader lines including Mg II \cite{subrayan23, wiseman23}. As this line is common among the ENTs, we fit it as a single component Gaussian to determine the line width and luminosity. The Mg II line and the fits are shown in figure \ref{fig:MgII}. It is noteworthy that all three ENTs have Mg II emission, as this line is conspicuously absent from all near-UV spectra of TDEs to date (e.g., \cite{cenko16, brown18, hung21}) but ubiquitous among AGNs (e.g., \cite{grandi79, mclure04, wang09}), although this may just be evidence of a pre-existing AGN-like gas reservoir. In Fig.\ \ref{fig:comparison_spec}, the ANT AT2019brs also shows weak Mg II emission but ASASSN-15lh, the other object with the requisite wavelength coverage, does not. Each of the nuclear transients in Fig.\ \ref{fig:comparison_spec} except ASASSN-15lh shows broad H$\alpha$ emission.

The fitted FHWMs for Gaia16aaw, Gaia18cdj, and AT2021lwx are $1.0 \times 10^{4}$ km s$^{-1}$, $5.8 \times 10^{3}$ km s$^{-1}$, and $1.2 \times 10^{4}$ km s$^{-1}$. These are high, but consistent with the broadest end of the line-width distribution for AGNs across a range of luminosities (e.g., \cite{dong09, shen11, le20}). Our fits give integrated fluxes of $3.2 \times 10^{-15}$ erg s$^{-1}$ cm $^{-2}$, $1.3 \times 10^{-15}$ erg s$^{-1}$ cm $^{-2}$, and $1.3 \times 10^{-15}$ erg s$^{-1}$ cm $^{-2}$ respectively for Gaia16aaw, Gaia18cdj, and AT2021lwx. Compared to the continuum, these fluxes correspond to EWs of 97 \AA, 11 \AA, and 13\AA. The combination of FWHM and EW for Gaia16aaw is typical of an AGN, but the EWs for Gaia18cdj and AT2021lwx are smaller than normal given their broad FWHMs \cite{dong09}.

The line luminosities of $1.8 \times 10^{43}$ erg s$^{-1}$, $6.1 \times 10^{42}$ erg s$^{-1}$, and $6.7 \times 10^{42}$ erg s$^{-1}$ are typical of luminous AGNs \cite{dong09, le20}. Among our sample of ENTs, Gaia16aaw is the most spectroscopically similar to typical AGNs, while Gaia18cdj and AT2021lwx exhibit broader lines and lower EWs than is typical for AGNs. This may be a result of the later phase at which the spectrum of Gaia16aaw was taken. We compare the Mg II line parameters for the ENTs to a sample of SDSS quasars \cite{shen11} in figure \ref{fig:emission_line_comp}.

Gaia18cdj and AT2021lwx also exhibit broad H$\alpha$ emission lines in their follow-up spectra. We similarly fit these broad features as a single component Gaussian. Both Gaia18cdj and AT2021lwx have an H$\alpha$ FWHM of $\sim 5000-6000$ km s$^{-1}$, typical of broad-line AGNs (e.g., \cite{liuh19}) and consistent with the lower end of the FWHM distribution for TDEs \cite{charalampopoulos22} and SNe IIn \cite{taddia13}. The H$\alpha$ luminosities are $\sim 1 \times 10^{43}$ erg s$^{-1}$ for Gaia18cdj and $\sim 5 \times 10^{43}$ erg s$^{-1}$ for AT2021lwx, consistent with typical AGNs \cite{liuh19}. These H$\alpha$ luminosities are 1 to 2 orders of magnitude more luminous than typical TDEs \cite{charalampopoulos22}, but the ENTs are similarly bolometrically overluminous, so this does not preclude a TDE origin. We show the H$\alpha$ line parameters for the ENTs as compared to the SDSS quasars \cite{shen11} and the TDEs from \cite{charalampopoulos22} in figure \ref{fig:emission_line_comp}.

To examine the narrow lines present in the spectra AT2021lwx, we compare a near-peak spectrum with a late-time spectrum in figure \ref{fig:21lwx_spec_comp}. The strong emission lines fade over time. After comparing the C III], C II], O III, H$\delta$, and H$\gamma$ lines between the two spectra, we find that the late-time emission lines are a factor of $8.0 \pm 1.1$ fainter on average. This is similar to, but slightly smaller than the factor of $\approx 11$ change in the $g$-band (rest-frame near-UV) continuum flux, confirming that these narrow lines are transient in origin.

\paragraph*{X-ray Emission} \label{sec:xray}

In addition to the pre-flare X-ray measurements from ROSAT presented previously, each of the ENTs in our sample has X-ray data from \textit{Swift} XRT. Additionally, AT2021lwx has follow-up X-ray observations from XMM-Newton \cite{jansen01} and Chandra \cite{weisskopf02}. Gaia16aaw is detected in \textit{Swift} XRT at a rest-frame $0.3-10$ keV luminosity of $(1.4 \pm 0.3) \times 10^{45}$ erg s$^{-1}$, slightly fainter but ultimately consistent with the pre-flare X-ray luminosity measured by ROSAT. Gaia18cdj is undetected in the XRT data, with a 3$\sigma$ upper-limit of $<3.6 \times 10^{44}$ erg s$^{-1}$. The luminosity estimates for Gaia16aaw and Gaia18cdj assume an absorbed power-law model with $\Gamma = 2$ and the Galactic $N_H$ column density to convert the XRT count rates to fluxes. As the XMM-Newton and Chandra spectra for AT2021lwx had low count rates, we fit them simultaneously with an absorbed power-law model. This fit yields a column density of $N_{H} = 1.3 \times 10^{21}$ cm$^{-2}$ and a hard photon index of $\Gamma = 0.6$, indicating the formation of a corona. We use this spectral shape to convert the count rates for AT2021lwx into fluxes. Throughout the flare, AT2021lwx shows X-ray emission at levels of a few times $10^{44}$ erg s$^{-1}$. The X-ray emission of AT2021lwx becomes harder as the source fades, consistent with AGN-like X-ray emission (e.g., \cite{auchettl18}). The source X-ray luminosities are provided in table \ref{tab:xrays}.

We can also use the ratio of optical to X-ray emission to place these ENTs in context with other accretion-powered transients. We interpolated the UV/optical bolometric light curve to the time of the X-ray observations and compared the three ENTs to a sample of TDEs and ANTs well-observed in both the X-ray and UV/optical. The comparison TDEs are ASASSN-14li \cite{brown17a}, ASASSN-15oi \cite{holoien18a}, ASASSN-18ul \cite{wevers19}, ASASSN-19dj \cite{hinkle21a}, and AT2019dsg (ZTF19aapreis; \cite{vanvelzen21}). The ANTs are ASASSN-17jz \cite{holoien21}, ASASSN-18jd \cite{neustadt20}, ASASSN-18el \cite{trakhtenbrot19b, hinkle23b}, AT2019pev (ZTF19abvgxrq; \cite{frederick21}), and ASASSN-20hx \cite{hinkle22a}. We interpolate these events in the same manner and remove upper limits.

The results are shown in Figure \ref{fig:xray_opt_ratio}. Each of the ENTs is overluminous in both X-ray and UV/optical emission as compared to the TDE and ANT samples. This is especially true when considering that the first X-ray constraints for the ENTs come at a rest-frame phase of $\sim 850$, $\sim 400$, and $\sim 300$ days after peak for Gaia16aaw, Gaia18cdj, and AT2021lwx, respectively, when the emission has likely faded.

Gaia16aaw in particular exhibits an X-ray to UV/optical luminosity ratio of $\approx 20$ during the epoch of its X-ray observation, notably higher than the other ENTs and any of the well-observed TDEs or ANTs in the comparison sample. However, the X-ray luminosity of Gaia16aaw as measured by \textit{Swift}/XRT is consistent given the uncertainties with the pre-flare X-ray luminosity measured by ROSAT, so the high X-ray to UV/optical ratio may be the result of consistent X-ray fluxes and a declining UV/optical light curve. Assuming the X-ray emission is constant in time, the ratio at the UV/optical peak would be 0.7, more consistent with the other transients in Fig.\ \ref{fig:xray_opt_ratio}. Gaia18cdj is fully consistent with the range of ratios seen for other transients. This remains true for any X-ray luminosity within roughly 2 orders of magnitude of the upper limit. AT2021lwx exhibits a ratio of $\approx 0.3$, similar to the majority of the TDE population and consistent with the distribution of the ANTs. All of the ENT ratios are broadly similar to the ratios seen for typical AGNs (e.g., \cite{just07, auge23}).

Combining the X-ray and UV/optical measurements, we can also estimate the peak Eddington ratios of the ENTs as $\approx 0.11$ for Gaia16aaw, $\approx 0.16$ for Gaia18cdj, and $\geq 0.2$ for AT2021lwx. These Eddington ratios are similar to TDEs (e.g., \cite{wevers17, mockler19}), strong AGNs (e.g., \cite{ricci17}), and many ANTs (e.g,, \cite{neustadt20, frederick21, holoien21}).

\subsection*{Rate Estimate} \label{sec:rates}

We can attempt to constrain the physical mechanism powering these ENTs through an estimate of their rates as compared to the rates of other classes of events. We calculate the rates following \cite{schmidt68} as

\begin{equation}
R = \sum_{i=1}^{N} R_i = \sum_{i=1}^{N} \frac{1}{t_{survey,i} / (1 + z_i)} \frac{1}{V_{max, i} f_{loss}}
\end{equation}

\noindent where $t_{survey,i}$ is the time span of the survey detecting the $i$th transient, corrected into the rest-frame by $(1 + z_i)$. $V_{max, i}$ is the maximum co-moving volume in which that transient would be detected by the survey. The completeness factor $f_{loss}$ is taken to be the fraction of the sky observed by the survey since these transients exist for much longer than a single observing season. Finally, N is the total number of observed transients. We list the assumed $t_{survey}$, limiting magnitude, and $f_{loss}$ for each source of data in table \ref{tab:rates_input}.

Given a typical decay rate of $\approx 0.1$ mag month$^{-1}$, we require that the source reaches a peak of 0.3 mag brighter than the limiting magnitude to count as a detection. This gives at least 3 months over which the source is above the limiting magnitude, sufficient for detection even in lower cadence surveys such as Gaia. We note that our calculated rates have only a weak (15\% per 0.1 mag) dependence on this assumption, much smaller than the Poisson errors. We estimated a rate uncertainty by dividing the estimated rate by the number of sources contributing and re-scaling using the 1$\sigma$ Poisson confidence intervals computed by \cite{gehrels86} for that number of events. This yields a rate of $1.0^{+1.0}_{-0.6} \times 10^{-3}$ Gpc$^{-3}$ yr$^{-1}$ based on the three observed ENTs Gaia16aaw, Gaia18cdj, and AT2021lwx. Gaia did not trigger on AT2021lwx, which means that our search of Gaia for ENTs is not complete and that $f_{loss}$ should be smaller. Since we are unable to quantify the completeness based on the internal survey operations, we will treat this rate estimate as a lower limit.

Regardless of the specific methodology, the estimated rate of these events is far below the local rates of luminous transients like TDEs ($\approx$100-500 Gpc$^{-3}$ yr$^{-1}$; \cite{vanvelzen14, holoien16a, godoy-rivera17, yao23}) and SLSNe-I ($\approx$10-100 Gpc$^{-3}$ yr$^{-1}$; \cite{quimby13, zhao21}) and are significantly lower than SNe Ia ($\approx$$2 \times 10^{4}$ Gpc$^{-3}$ yr$^{-1}$; e.g., \cite{desai24}). However, a more fair comparison to make is between the rates of these events at $z \sim 1$. The rate of SLSNe at this redshift grows to $\approx 50-170$ Gpc$^{-3}$ yr$^{-1}$ \cite{prajs17} and the TDE rate is expected to drop by a factor of roughly 6 \cite{kochanek16b} due to the declining SMBH number density at higher redshifts. Thus, even at $z \sim 1$, the ENTs are at least a factor of several thousand rarer than typical SLSNe or TDEs. We additionally re-estimate the rate for ASASSN-15lh-like events given the longer ASAS-SN baseline without any additional detections. This yields a rate roughly 40 times higher than the estimated ENT rate.

\subsection*{Examining Potential Physical Origins for the Flares}

\paragraph*{Gravitational Lensing} \label{sec:lensing}

First, we consider if the extreme luminosities inferred for the ENTs may in fact be due to gravitational lensing. While high redshift supernovae have been observed to be magnified by factors of 10 or more (e.g., \cite{kelly15}), these high magnifications require an intervening massive galaxy or galaxy cluster. The photometry of the ENT host-galaxies presented in fig.\ \ref{fig:color_images} shows no evidence for foreground lens galaxies or clusters. Indeed, the CIGALE fits for Gaia16aaw and Gaia18cdj have reduced $\chi^2$ values of 6.3 and 0.5 respectively, indicating that they are well modeled as single galaxies. Similarly, the spectra for Gaia16aaw and Gaia18cdj show no signs of foreground galaxies and none of the apparent lower-redshift absorption doublets seen in the spectrum of AT2021lwx are well-matched to Mg II or other typical absorption lines.

The lensing cross-section scales as $\sigma_v^4$, where $\sigma_v$ is the velocity dispersion of the lens galaxy (e.g., \cite{kochanek96}). Since we find no evidence for a massive foreground galaxy, any potential lens galaxy must be low mass, which greatly decreases the likelihood of lensing. Nevertheless, we need to consider the so-called magnification bias \cite{turner80}, which increases the probability of observing a lensed intrinsically-faint source. As Gaia16aaw is detected at a flux of $4 \times 10^{-13}$ erg s$^{-1}$ cm$^{-2}$ in pre-flare ROSAT coverage, we will assume that the number counts of the ENT host galaxies can be reasonably well-described by power-law distribution of $dN/dF \sim F^{-2.5}$, as for similarly bright AGNs \cite{georgakakis08, lehmer12}. Such a slope corresponds to a factor of $\approx 10$ in terms of magnification bias \cite{turner80}. This bias allows a galaxy with $\sim$60\% of the velocity dispersion, or 15\% of the mass \cite{faber76}, to produce the same lensing probability as a calculation without accounting for magnification bias. Given our strong constraints on the lack of massive foreground systems from the host galaxy fits and follow-up spectra and the fact that even the most luminous normal supernovae require a magnification of $>$10, we find a lensing explanation unlikely.

\paragraph*{Luminous Supernovae} \label{sec:supernovae}

Most supernovae are powered by the radioactive decay of $^{56}$Ni into $^{56}$Co and later the decay of $^{56}$Co into stable $^{56}$Fe. If we assume that the ENTs in our sample are powered by $^{56}$Ni decay, we can use the scaling between nickel mass and energy production \cite{nadyozhin94} to estimate the initial mass of $^{56}$Ni. Even for our least energetic event, Gaia16aaw, the nickel mass required is $280$ M$_{\odot}$. This quickly increases to more than 1000 M$_{\odot}$ for Gaia18cdj and AT2021lwx. Such an explanation is clearly unphysical.

There are classes of supernovae that are not solely powered by radioactive decay. In particular, the Type I and Type II SLSNe \cite{quimby11, gal-yam12, gal-yam19} have peak luminosities higher than can be explained by radioactive decay. In the case of SLSNe I, a plausible explanation is the spin-down of a magnetar \cite{ostriker71}. The rotational energy of a typical magnetar, with a mass of 1.4 M$_{\odot}$ and a spin period of 1 ms is $3 \times 10^{52}$ erg, which is insufficient to power any of the ENTs in our sample. Even if we assume a magnetar with the mass of the most massive known neutron star, PSR J0952-0607 at 2.35 $_{\odot}$ \cite{romani22}, and the maximum neutron star spin period of 0.9 ms \cite{hashimoto94, burgio03, kaaret07}, the maximum rotational energy is $\approx 7 \times 10^{52}$ erg. This is well below the energies of Gaia18cdj and AT2021lwx and would require an unrealistically high conversion efficiency to UV/optical emission for Gaia16aaw. Thus, magnetars cannot power the ENTs.

For SLSNe II, a large amount of kinetic energy in the supernova ejecta is converted to light when it shocks with the surrounding circumstellar medium (CSM). Following \cite{inserra18}, we can estimate the mass required to power the observed luminosity of these ENTs. The relevant physical parameters are the peak luminosity, rise time, and photospheric velocity. Given the range of observed luminosities and rise times and assuming a typical velocity of 6000 km s$^{-1}$ we require over 1000 M$_{\odot}$ of CSM to power the ENTs. The CSM mass can be lowered to $\sim 100$ M$_{\odot}$ for a velocity of $\sim 20000$ km s$^{-1}$, but this velocity is higher than typical SLSNe II \cite{inserra18} and the ENT spectra show no signs of substantial velocity offsets in their spectra. Thus, we also rule out a supernova origin for ENTs.

\paragraph*{AGN Flares} \label{sec:agn_flares}

With strong gravitational lensing and various classes of SNe ruled out, we arrive at accretion onto a black hole as the most likely explanation for these ENTs. This is not particularly surprising as the ENTs are located in their host nuclei. More importantly, accretion is the most efficient known method of converting large amounts of mass into energy in an astrophysical system.

Each of the ENT hosts shows some evidence for hosting an AGN or AGN-like behavior in the flare evolution itself. Therefore, in addition to a TDE origin for the flares, we must consider whether or not AGN activity can produce these transients. We first consider if these flares can be the extreme end of typical AGN stochastic variability. Extrapolating from Figure 3 of \cite{macleod12}, we estimate the fraction of AGNs undergoing variability at levels similar to the ENT flares on similarly long timescales.

For the case of Gaia16aaw, with a $\Delta m = -2$ mag flare, the fraction of AGNs with a similarly large brightening on long timescales is 0.01\%. For Gaia18cdj and AT2021lwx, with $\Delta m = -3$ mag, the fraction is much lower, at $9 \times 10^{-5}$ \%. While these estimates are for continuous rather than impulsive variability, they still provide a useful constraint on the likelihood of normal AGNs exhibiting such large flares. These fractions combined with the number density of AGNs at $z = 1$ with $L \approx 10^{45}$ erg s$^{-1}$ (roughly 300 Gpc$^{-3}$; \cite{ueda14, zou24}), and a $\sim 10$ yr total flare timescale imply rates of AGN flares in this amplitude range of $3 \times 10^{-5} \mbox{ -- } 3 \times 10^{-3}$ Gpc$^{-3}$ yr$^{-1}$. The upper end of this range suggests that Gaia16aaw is possibly consistent with a stochastic AGN flare. However, smooth AGN flares, like the ones seen for the ENTs, are rare. In a study of nuclear transients, \cite{auchettl18} found that $<4$\% of coherently declining nuclear accretion flares are likely to be powered by stochastic AGN variability rather than TDEs. Therefore, we scale the rate estimates by $0.04$ to estimate the corresponding rate of smooth AGN flares, which yields $1 \times 10^{-6} \mbox{ -- } 1 \times 10^{-4}$ Gpc$^{-3}$ yr$^{-1}$. This is at least an order of magnitude too low to explain the ENTs and rules out AGN variability in producing these ENTs. 

We note that the \cite{auchettl18} estimate of $<4$\% of smooth nuclear flares arising from stochastic AGN activity is computed from sparse X-ray light curves rather than the higher-cadence UV/optical data analyzed for the ENTs. While probing a different wavelength regime, the sparse X-ray light curves can more easily hide short-term variability that we have already ruled out for the ENTs. Furthermore, the calculation of \cite{auchettl18} was for highly-variable AGNs and the lack of significant pre-flare AGN variability in any of the ENT hosts makes it unlikely that smooth AGN flares can explain the ENT light curves.

There are other exotic mechanisms that have the potential to produce luminous flares in AGNs. One model is a transient powered by the interaction of an AGN disk wind with the surrounding broad line region (BLR) clouds \cite{moriya17}. Qualitatively this is similar to the scenario for Type IIn SNe or SLSNe-II, where shocks between outflowing mass and a dense medium produce a strong radiative transient. Under the assumption that the kinetic energy is fully converted into radiated luminosity and assuming a typical disk wind velocity of $0.1c$ \cite{guistini19}, our measured energies imply $\sim 6 - 28$ solar masses of ejected material. These are lower limits, as the required ejecta masses increase with the inverse of the covering fraction of the BLR clouds \cite{dunn07} or relaxing the assumption that the conversion of kinetic to radiative energy is 100\% efficient. Even a modest 50\% efficiency of energy conversion (higher than estimated for SNe IIn; \cite{moriya13, tsuna19}) and a typical BLR covering fraction of 0.4 \cite{dunn07} increases these mass estimates by a factor of 5.

It is possible to avoid such high ejecta masses if the disk wind velocity is increased, but we find no evidence for higher velocities in the ENT spectra and only the fastest winds have considerably larger velocities \cite{guistini19}. Furthermore, while an increase in wind velocity decreases the ejecta mass required, it also decreases the emission timescales. Another potential problem with this model is that \cite{moriya17} suggest that the most likely mechanism for launching a disk wind is the limit-cycle oscillation (e.g., \cite{honma91, ohusga06}). This requires an AGN accreting near Eddington, which is ruled out for Gaia18cdj and highly unlikely for AT2021lwx. Although the duty cycle for limit-cycle oscillations in AGNs is not well constrained, it can be of order 10\% \cite{janiuk11, moriya17}, further exacerbating the situation. Given these issues, we find this model to be an unlikely explanation for the ENTs.

Another potential mechanism is a smooth flare powered by a disk instability which allows rapid accretion onto the SMBH (e.g., \cite{lipunova24}). These flares are proposed to occur in systems where the disk is truncated and the inner portions of the disk have a high temperature and low density \cite{yuan14}. In this model, both the peak accretion rate and timescale of the flare depend on the truncation radius of the disk.  Before the flare, the host galaxy of Gaia16aaw is measured to have an X-ray luminosity of $4 \times 10^{45}$ erg s$^{-1}$, corresponding to an Eddington ratio of $\sim$10\%. This is in a regime where disk truncation flares will not occur \cite{lipunova24}. With the wide range of implied physical states for the pre-flare AGNs in the ENT host galaxies, it is unlikely that such a model would naturally produce our observed sample of ENTs, each with a similar peak luminosity and timescale, although it remains a plausible physical mechanism for Gaia18cdj and AT2021lwx. Further development of models and testable predictions on how such flares would look will elucidate if this is a viable option for driving ENT flares.

\paragraph*{Tidal Disruption Events} \label{sec:tdes}

Next, we examine the possibility that the ENTs are extreme examples of tidal disruption events. Typical TDEs have peak luminosities up to $\sim 5 \times 10^{44}$ (e.g., \cite{hinkle21b, vanvelzen21, hammerstein23}) although the recently-discovered class of featureless TDEs can be more luminous \cite{hammerstein23}. In contrast to the ENTs, known TDEs rarely have decay timescales more than $\approx$100 days. However, the smooth evolution of essentially all TDEs makes them a promising candidate to explain the smooth ENT flares.

The tidal disruption of a massive main-sequence (MS) star naturally results in a more luminous flare as the peak accretion rate at fixed SMBH mass scales as $\dot{M}_{peak} / \dot{M}_{edd} \sim M_{*}^{1.1}$ \cite{stone13, kochanek16b} for typical stellar mass-radius relations \cite{demircan91, tout96}. However, at fixed SMBH mass, the disruption of a more massive star is expected to result in a shorter flare, although this is a much weaker effect with $t_{fb} \sim M_{*}^{-0.1}$. Thus, it is plausible to power a luminous and long-lived flare through the tidal disruption of a massive star on a massive SMBH. The estimated SMBH masses for the ENTs are close to the Hills mass \cite{hills75}, but a TDE can still occur given the higher stellar masses required. Additionally, there is a large scatter on the SMBH-galaxy scaling relations that makes the SMBH estimates uncertain at the level of $\sim0.5$ dex. 

Converting the emitted energies to accreted mass provides a constraint on the minimum stellar mass that could possibly have powered these ENTs. The highest ENT energies correspond to $\approx 1.4$ M$_{\odot}$ of accreted mass for 10\% efficiency. As roughly half the original stellar mass becomes unbound in a TDE, this places a lower limit on the stellar mass of $\gtrsim 3$ M$_{\odot}$. Following \cite{kochanek16b}, the TDE of 3 M$_{\odot}$ MS star on a SMBH with mass similar to the ENTs would result in a flare with a characteristic timescale of $\approx 600$ days and a peak Eddington ratio of $\approx 0.1$, reasonably well-matched to the observed ENT flares. More massive MS stars, up to roughly 10 M$_{\odot}$, yield timescales and Eddington ratios compatible with the ENT observables.

The timescales and luminosities of the ENTs are also similar to predictions of the TDEs resulting from extremely massive stars in the early universe \cite{chowdhury24} and agree well with recent models of massive star TDEs \cite{bandopadhyay24}. Quantitatively, \cite{bandopadhyay24} predict that the rise time for a TDE on a $10^{8.4}$ M$_{\odot}$ SMBH is 1 yr, which agrees well with the $\approx 300 \mbox{ -- } 400$ day rise times for Gaia16aaw and Gaia18cdj. The rise time for AT2021lwx is slightly shorter, at $\approx$ 170 days, which remains consistent with the upper limit on its SMBH mass. Another key prediction of these models is high peak mass accretion rates for massive star disruptions. Around a $10^6$ M$_{\odot}$ SMBH, \cite{bandopadhyay24} predict a peak mass accretion rate of 15 M$_{\odot}$ yr$^{-1}$ for a $3 \mbox{ -- } 5$ M$_{\odot}$ star, depending on its evolutionary stage. After correcting this to the higher SMBH masses of the ENTs, following the peak timescale scaling, this becomes $\approx 1$ M$_{\odot}$ yr$^{-1}$, in good agreement with the $\approx 0.3 \mbox{ -- } 1$ M$_{\odot}$ yr$^{-1}$ accretion rates derived from the peak luminosities and assuming a 10\% accretion efficiency.

Since the timescale and energetics of a high-mass TDE appear consistent with the ENTs, the next consideration is whether the expected rates of such events are compatible with our estimated ENT rate. Let us assume that the ENTs are the result of $\sim$$3 \mbox{--} 10$ M$_{\odot}$ stars, which satisfies the constraints from the emitted energy, timescale, and peak Eddington ratio. From the estimated local rate of TDEs \cite{vanvelzen14, holoien16a, yao23} and the expected redshift evolution \cite{kochanek16b}, we find a minimum TDE rate at $z = 1$ of $\approx 15$ Gpc$^{-3}$ yr$^{-1}$. For standard initial mass functions \cite{salpeter55, kroupa01, chabrier03}, the high-mass slope is $\alpha = -2.35$. Thus, there are 12 times fewer $3 - 10$ M$_{\odot}$ stars than $0.5 \mbox{--} 2$ M$_{\odot}$ stars, a range consistent with the local TDE population \cite{mockler19}. While this calculation assumes a typical stellar IMF, there is evidence that the stellar population of the Milky Way Galactic center prefers a top-heavy IMF (e.g., \cite{maness07, bartko10, lu13}), with a shallower high-mass slope. Similarly, the population of TDEs appears consistent either with top-heavy IMFs in the nuclei of some TDE host galaxies or the preferential disruption of moderately massive stars \cite{mockler22}. Either of these effects will increase the expected rate of massive star TDEs relative to the above scaling.

The stellar lifetimes are also important to consider, as stars with short lifetimes may not live long enough to be scattered onto an orbit that results in a TDE. The lifetimes of  $3 - 10$ M$_{\odot}$ stars are approximately 90 times lower than those of $0.5 - 2$ M$_{\odot}$ stars. If we scale the rates by the birth abundance and the lifetimes, the intrinsic rate of $3 - 10$ M$_{\odot}$ TDEs would be roughly 1000 times lower than $0.5 - 2$ M$_{\odot}$ star TDEs. Scaling from the expected rate of solar mass TDEs at $z = 1$, the $3 - 10$ M$_{\odot}$ TDE rate at this redshift is $\approx 1.5 \times 10^{-2}$ Gpc$^{-3}$ yr$^{-1}$, sufficient to explain the ENTs we find. 

\bibliography{SciA_ENTs.bib}
\bibliographystyle{sciencemag}

\clearpage

\section*{Acknowledgements} 

We thank Christopher Storfer, John Tonry, and Alexa Anderson for their helpful discussions. We thank Phillip Wiseman for sharing the published spectra of AT2021lwx. 

This work presents results from the European Space Agency (ESA) space mission Gaia. Gaia data are being processed by the Gaia Data Processing and Analysis Consortium (DPAC). Funding for the DPAC is provided by national institutions, in particular the institutions participating in the Gaia MultiLateral Agreement (MLA). The Gaia mission website is \\ https://www.cosmos.esa.int/gaia. The Gaia archive website is https://archives.esac.esa.int/gaia. This work is also based in part on observations obtained with the Samuel Oschin Telescope 48-inch and the 60-inch Telescope at the Palomar Observatory as part of the Zwicky Transient Facility project. ZTF is supported by the National Science Foundation under Grant No.\ AST-2034437 and a collaboration including Caltech, IPAC, the Weizmann Institute for Science, the Oskar Klein Center at Stockholm University, the University of Maryland, Deutsches Elektronen-Synchrotron and Humboldt University, the TANGO Consortium of Taiwan, the University of Wisconsin at Milwaukee, Trinity College Dublin, Lawrence Livermore National Laboratories, and IN2P3, France. Operations are conducted by COO, IPAC, and UW. The ZTF forced-photometry service was funded under the Heising-Simons Foundation grant \#12540303 (PI: Graham).

This publication makes use of data products from the Wide-field Infrared Survey Explorer, which is a joint project of the University of California, Los Angeles, and the Jet Propulsion Laboratory/California Institute of Technology, funded by the National Aeronautics and Space Administration. The CSS survey is funded by the National Aeronautics and Space Administration under Grant No. NNG05GF22G issued through the Science Mission Directorate Near-Earth Objects Observations Program.  The CRTS survey is supported by the U.S.~National Science Foundation under grants AST-0909182.

This project used public archival data from the Dark Energy Survey (DES). Funding for the DES Projects has been provided by the U.S. Department of Energy, the U.S. National Science Foundation, the Ministry of Science and Education of Spain, the Science and Technology Facilities Council of the United Kingdom, the Higher Education Funding Council for England, the National Center for Supercomputing Applications at the University of Illinois at Urbana–Champaign, the Kavli Institute of Cosmological Physics at the University of Chicago, the Center for Cosmology and Astro-Particle Physics at the Ohio State University, the Mitchell Institute for Fundamental Physics and Astronomy at Texas A\&M University, Financiadora de Estudos e Projetos, Fundação Carlos Chagas Filho de Amparo à Pesquisa do Estado do Rio de Janeiro, Conselho Nacional de Desenvolvimento Científico e Tecnológico and the Ministério da Ciência, Tecnologia e Inovação, the Deutsche Forschungsgemeinschaft and the Collaborating Institutions in the Dark Energy Survey.

The Collaborating Institutions are Argonne National Laboratory, the University of California at Santa Cruz, the University of Cambridge, Centro de Investigaciones Enérgeticas, Medioambientales y Tecnológicas–Madrid, the University of Chicago, University College London, the DES-Brazil Consortium, the University of Edinburgh, the Eidgenössische Technische Hochschule (ETH) Zürich, Fermi National Accelerator Laboratory, the University of Illinois at Urbana-Champaign, the Institut de Ciències de l’Espai (IEEC/CSIC), the Institut de Física d’Altes Energies, Lawrence Berkeley National Laboratory, the Ludwig-Maximilians Universität München and the associated Excellence Cluster Universe, the University of Michigan, the National Optical Astronomy Observatory, the University of Nottingham, The Ohio State University, the OzDES Membership Consortium, the University of Pennsylvania, the University of Portsmouth, SLAC National Accelerator Laboratory, Stanford University, the University of Sussex, and Texas A\&M University.

Based in part on observations at Cerro Tololo Inter-American Observatory, National Optical Astronomy Observatory, which is operated by the Association of Universities for Research in Astronomy (AURA) under a cooperative agreement with the National Science Foundation.

This work uses data from Pan-STARRS. The Pan-STARRS1 Surveys (PS1) and the PS1 public science archive have been made possible through contributions by the Institute for Astronomy, the University of Hawaii, the Pan-STARRS Project Office, the Max-Planck Society and its participating institutes, the Max Planck Institute for Astronomy, Heidelberg and the Max Planck Institute for Extraterrestrial Physics, Garching, The Johns Hopkins University, Durham University, the University of Edinburgh, the Queen's University Belfast, the Harvard-Smithsonian Center for Astrophysics, the Las Cumbres Observatory Global Telescope Network Incorporated, the National Central University of Taiwan, the Space Telescope Science Institute, the National Aeronautics and Space Administration under Grant No. NNX08AR22G issued through the Planetary Science Division of the NASA Science Mission Directorate, the National Science Foundation Grant No. AST-1238877, the University of Maryland, Eotvos Lorand University (ELTE), the Los Alamos National Laboratory, and the Gordon and Betty Moore Foundation.

This work uses data from the University of Hawaii's ATLAS project, funded through NASA grants NN12AR55G, 80NSSC18K0284, and 80NSSC18K1575, with contributions from the Queen's University Belfast, STScI, the South African Astronomical Observatory, and the Millennium Institute of Astrophysics, Chile.

Based in part on observations obtained at the Southern Astrophysical Research (SOAR) telescope, which is a joint project of the Ministério da Ciência, Tecnologia e Inovações do Brasil (MCTI/LNA), the US National Science Foundation’s NOIRLab, the University of North Carolina at Chapel Hill (UNC), and Michigan State University (MSU). This paper includes data gathered with the 6.5-meter Magellan Telescopes located at Las Campanas Observatory, Chile. 

Some of the data presented herein were obtained at the W. M. Keck Observatory, which is operated as a scientific partnership among the California Institute of Technology, the University of California and the National Aeronautics and Space Administration. The Observatory was made possible by the generous financial support of the W. M. Keck Foundation.

We acknowledge the use of public data from the Swift data archive. This research has made use of data obtained from the Chandra Data Archive and the Chandra Source Catalog, and software provided by the Chandra X-ray Center (CXC) in the application packages CIAO and Sherpa. Based in part on observations obtained with XMM-Newton, an ESA science mission with instruments and contributions directly funded by ESA Member States and NASA.

This work is based in part on observations made by ASAS-SN, ATLAS, Pan-STARRS, and Keck. The authors wish to recognize and acknowledge the very significant cultural role and reverence that the summits of Haleakal\=a and Maunakea have always had within the indigenous Hawaiian community.  We are most fortunate to have the opportunity to conduct observations from these mountains.

This research was supported in part by grant NSF PHY-2309135 to the Kavli Institute for Theoretical Physics (KITP). 

\textbf{Funding:} JTH and BJS are supported by NASA grant 80NSSC23K1431. KA is supported by the Australian Research Council Discovery Early Career Researcher Award (DECRA) through project number DE230101069. CSK is supported by NSF grants AST-1907570 and AST-2307385. CA acknowledges support by NASA grants JWSTGO-02114, JWST-GO-02122, JWST-GO-03726, JWSTGO-04436, and JWST-GO-04522. AD and BJS are supported by National Science Foundation grant AST-1911074. WBH is supported by the National Science Foundation Graduate Research Fellowship Program under Grant No. 2236415. Any opinions, findings, and conclusions or recommendations expressed in this material are those of the author(s) and do not necessarily reflect the views of the National Science Foundation. 

\textbf{Author contributions:} JTH led the analysis and drafting the manuscript. BJS, CSK, KA, and CA helped prepare the manuscript. KA reduced and analyzed the X-ray data. CSK and JMMN conducted modeling of the IR lags. AP, JS, and TW-SH\ obtained follow-up spectroscopy of the sources. JTH, BJS, CSK, and JS contributed to the physical interpretation. JTH, MEH, MAT, CA, TdJ, DDD, AD, WBH, and AVP contributed to the SCAT vetting of sources. All authors provided feedback on the manuscript. 

\textbf{Competing interests:} The authors declare no competing interests. 

\textbf{Data Availability:} All survey photometry is publicly available. The follow-up spectra from this work are available at the Weizmann Interactive Supernova Data Repository (WISeREP, https://www.wiserep.org/). All other data needed to evaluate the conclusions in the paper are present in the paper and/or the Supplementary Materials.

\bigskip

\begin{center}
\Large List of Supplementary Materials: 
\end{center}

\indent Supplementary Text \\
\indent Figures S1 through S8 \\
\indent Tables S1 through S3 \\
\indent References 179 through 201 \\

\clearpage

\begin{figure*}
\centering
 \includegraphics[width=\textwidth]{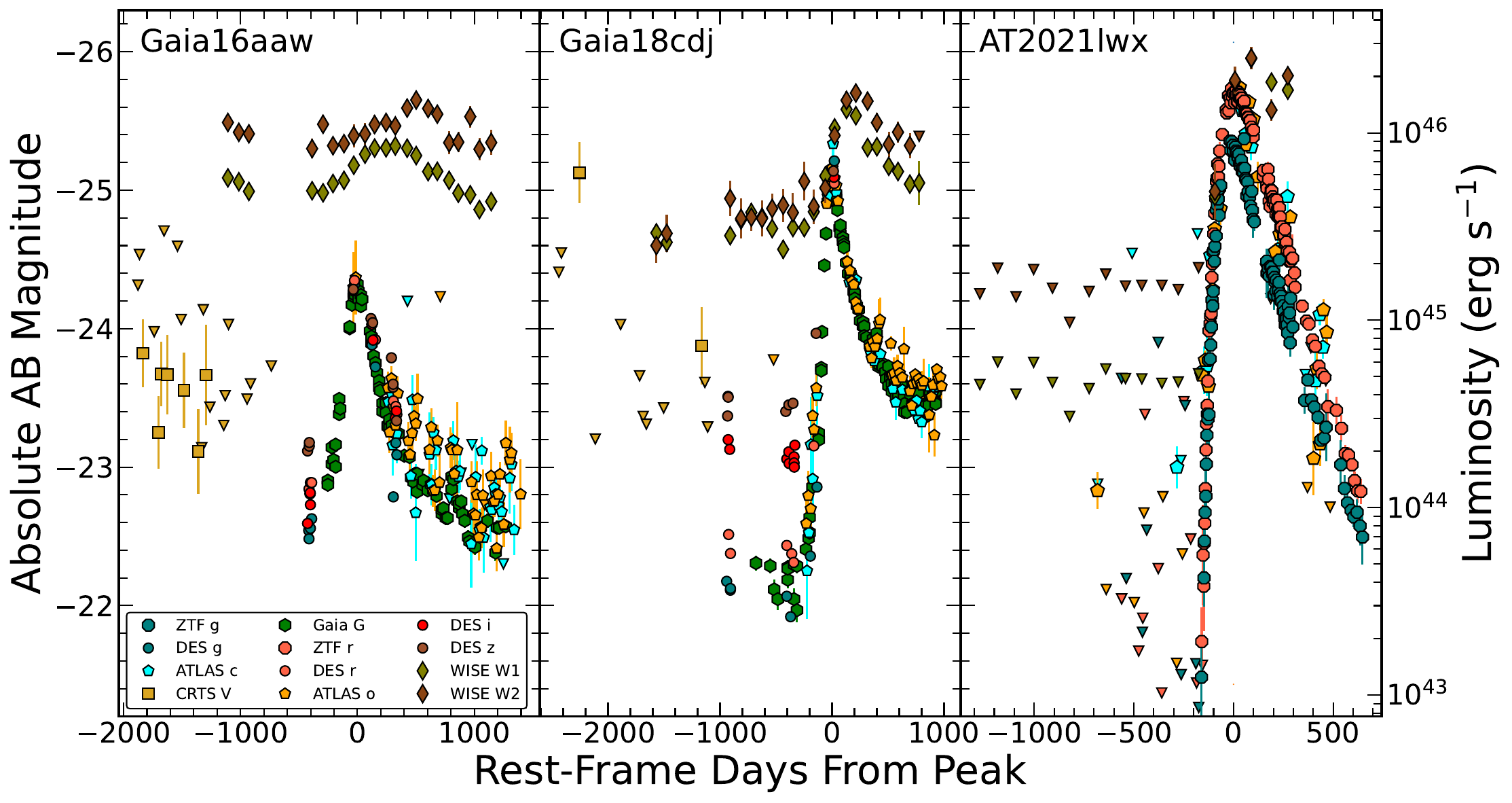}
 \caption{\textbf{Optical and IR light curves of Gaia16aaw (left), Gaia18cdj (middle), and AT2021lwx (right).} These light curves are corrected for Galactic foreground extinction, but have not had any host contribution removed. Shown are DES (circles, $griz$), ATLAS (pentagons, $co$), CRTS (squares, $V$), Gaia (hexagons, $G$), ZTF (octagons, $gr$), and WISE (diamonds, $W1W2$). Downward-facing triangles indicate 3$\sigma$ upper limits for filters of the same color. All data are in the AB magnitude system.}
 \label{fig:opt_ir_lcs}
\end{figure*}

\clearpage

\begin{figure*}
\centering
 \includegraphics[width=0.94\textwidth]{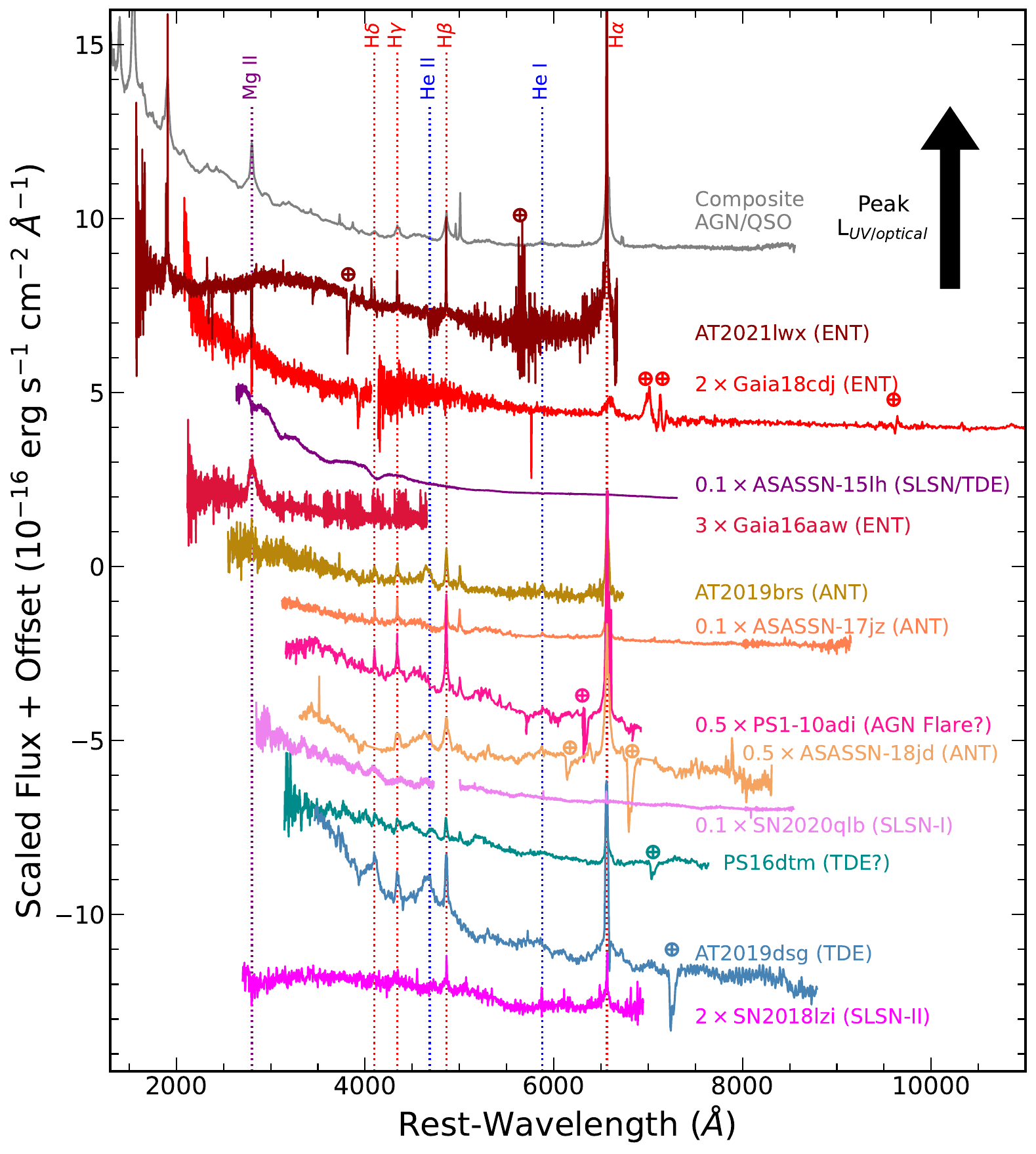}
 \caption{\textbf{Spectra of the ENTs Gaia16aaw, Gaia18cdj, and AT2021lwx (in red shades) as compared to other luminous transients.} More luminous events are towards the top of the figure. The comparison objects are a composite AGN/QSO spectrum from SDSS \cite{vandenberk01}, ASASSN-15lh (purple; \cite{dong16,leloudas16}), the ANTs AT2019brs \cite{frederick21}, ASASSN-17jz \cite{holoien21}, and ASASSN-18jd \cite{neustadt20} (orange shades), the TDE candidates PS16dtm \cite{blanchard17} and AT2019dsg \cite{vanvelzen21} (blue shades), the SLSNe SN2020qlb \cite{west23} and SN2018lzi \cite{pessi24} (violet shades), and PS1-10adi \cite{kankare17} (pink). Atmospheric telluric features are marked with an $\oplus$. Vertical dashed lines mark H, He, and Mg features in red, blue, and purple respectively. The spectra are scaled and offset as needed to enhance readability.}
 \label{fig:comparison_spec}
\end{figure*}

\clearpage

\begin{figure*}
\centering
 \includegraphics[width=0.95\textwidth]{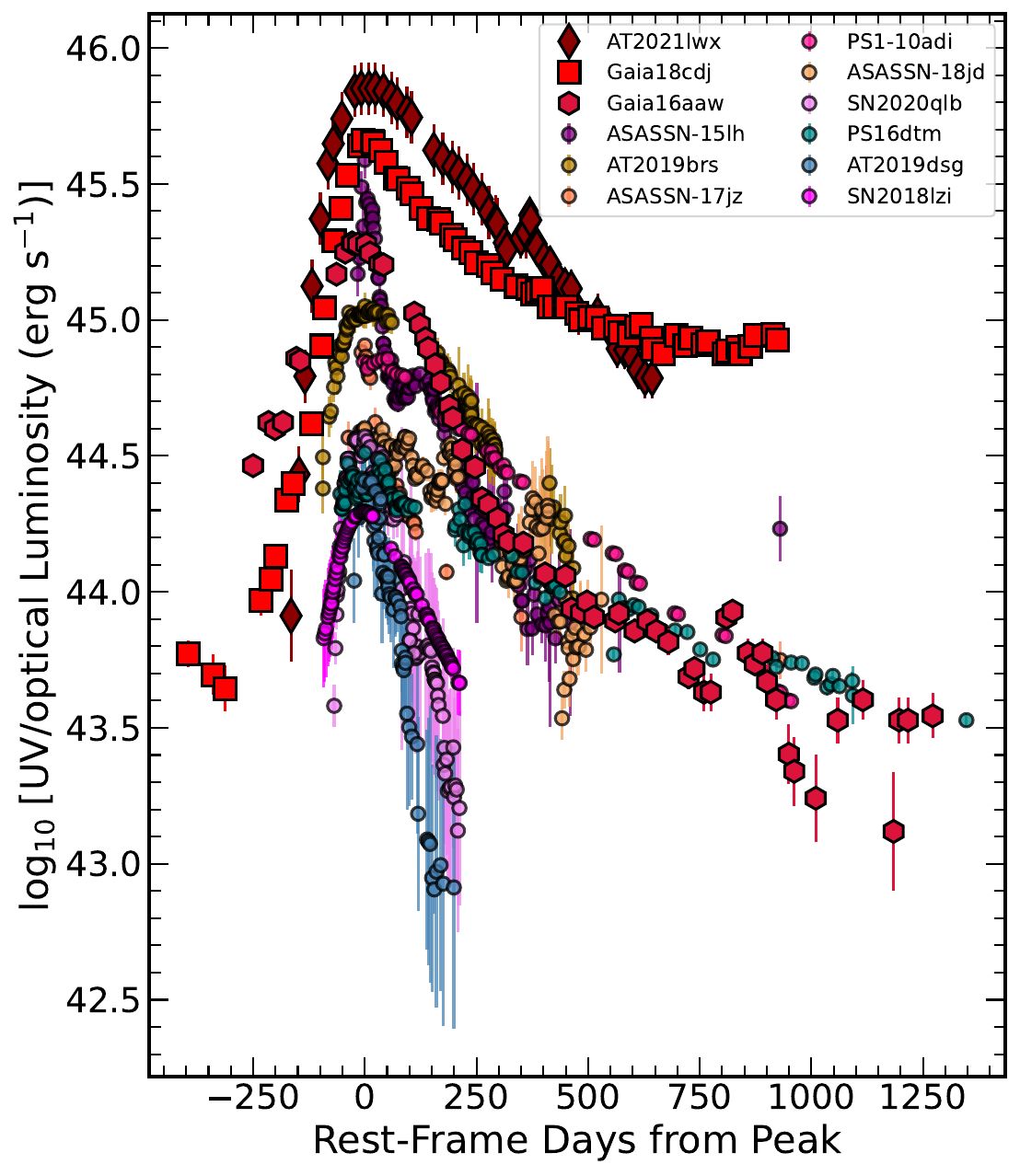}
 \caption{\textbf{UV/optical bolometric light curves of the ENTs Gaia16aaw, Gaia18cdj, and AT2021lwx (in red shades) as compared to other luminous transients.} The comparison objects are ASASSN-15lh (purple; \cite{dong16,leloudas16}), the ANTs AT2019brs \cite{frederick21}, ASASSN-17jz \cite{holoien21}, and ASASSN-18jd \cite{neustadt20} (orange shades), the TDE candidates PS16dtm \cite{blanchard17} and AT2019dsg \cite{vanvelzen21} (blue shades), the SLSNe SN2020qlb \cite{west23} and SN2018lzi \cite{pessi24} (violet shades), and PS1-10adi \cite{kankare17} (pink). All data have had the host contribution removed and are corrected for Galactic foreground extinction.}
 \label{fig:comparison_phot}
\end{figure*}

\clearpage

\begin{figure*}
\centering
 \includegraphics[width=\textwidth]{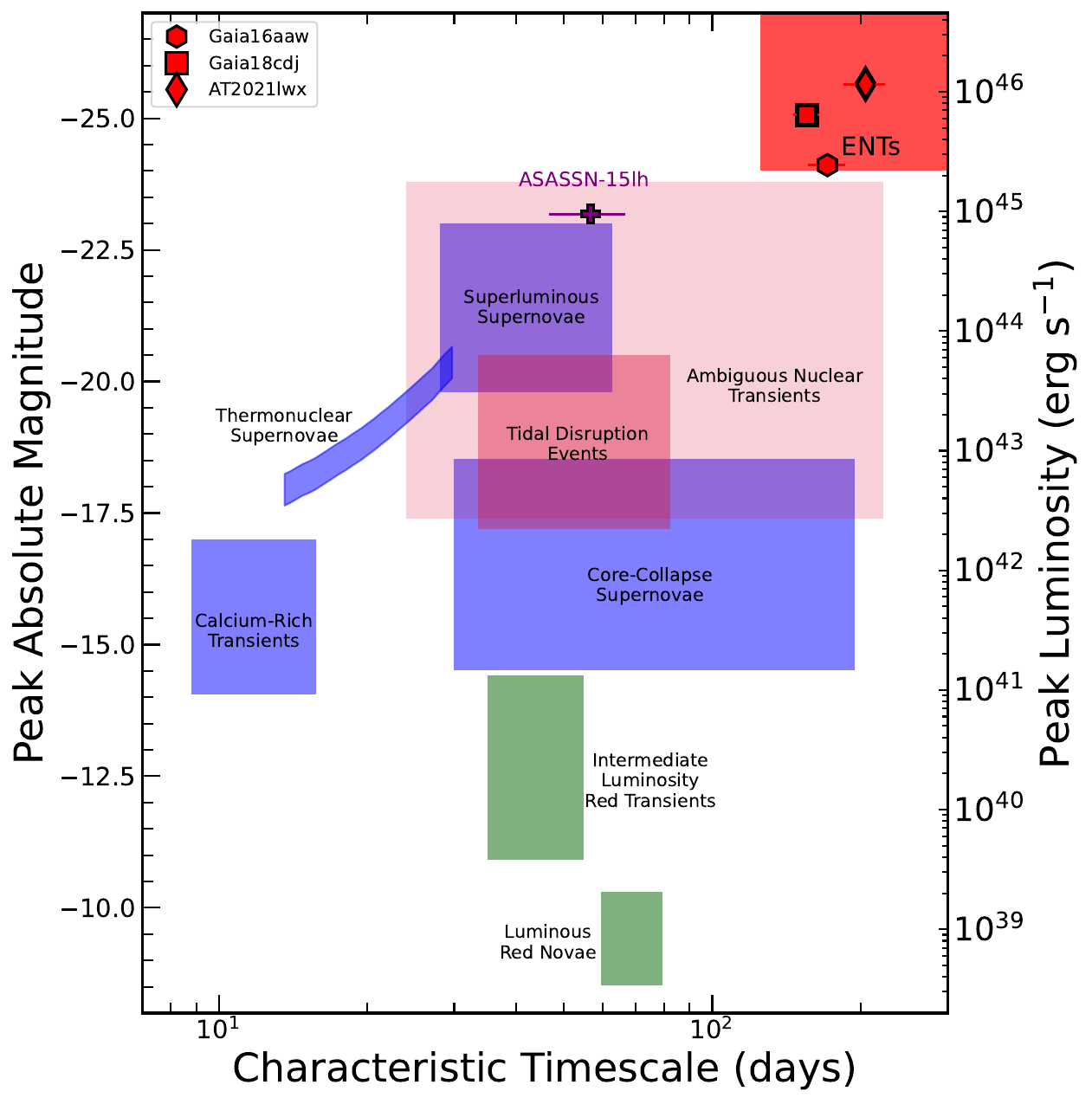}
 \caption{\textbf{Optical absolute magnitude as compared to the characteristic timescale for a range of transient classes.} Blue regions show various types of supernovae, green regions show classes of stellar mergers and/or mass transfer, and red-shaded regions show events powered by accretion onto SMBHs. The overluminous nuclear transient ASASSN-15lh is shown as a purple plus and our sample of ENTs are presented as red symbols.}
 \label{fig:transients}
\end{figure*}

\clearpage

\begin{figure*}
\centering
 \includegraphics[width=\textwidth]{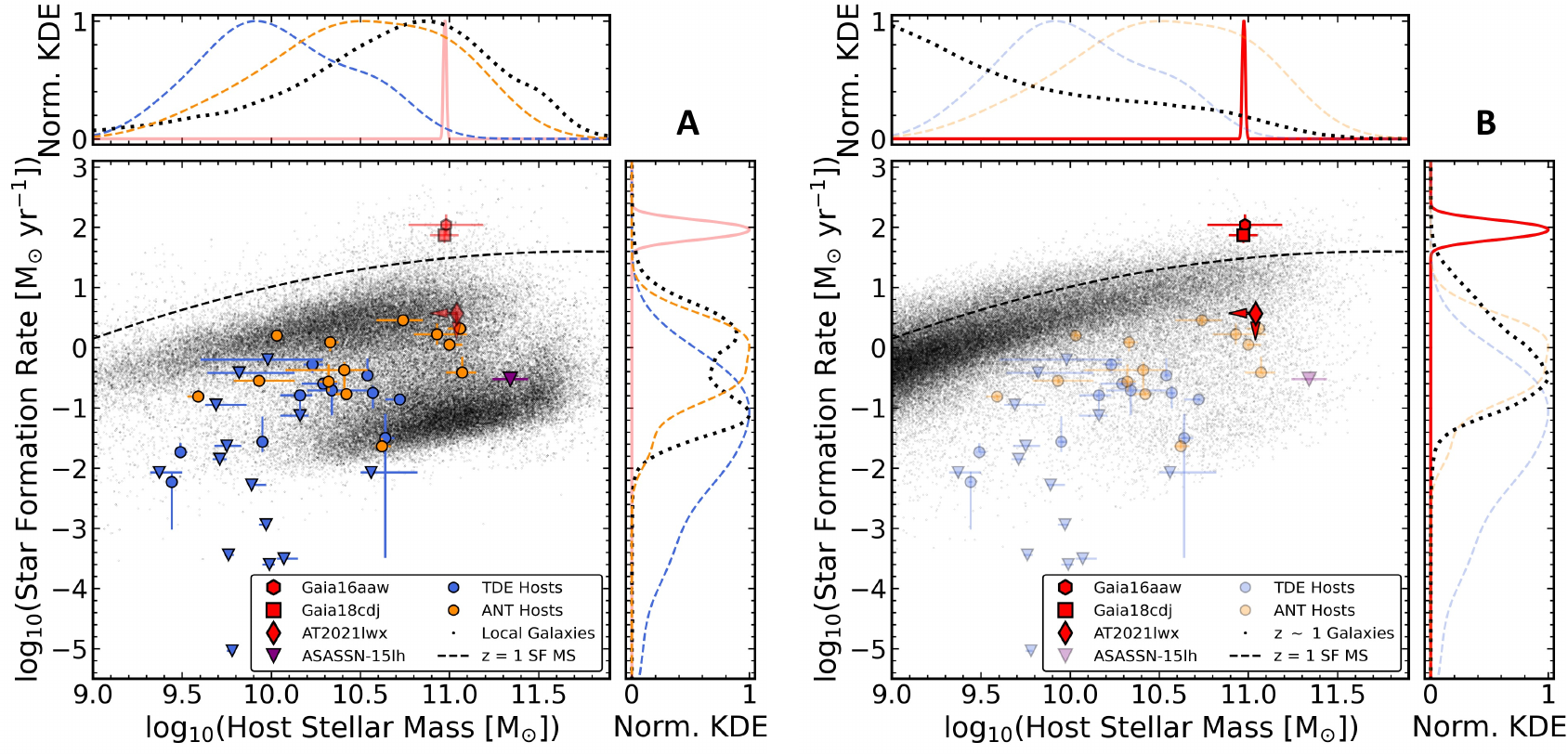}
 \caption{\textbf{Host-galaxy star formation rate as compared to stellar mass.} The red points are the ENTs Gaia16aaw, Gaia18cdj, and AT2021lwx. The blue points are a comparison sample of TDEs and the gold points are a comparison sample of ANTs. The background gray points are samples of local galaxies from SDSS (A) and z $\sim 1$ galaxies from COSMOS (B). In the two outer panels, we show normalized KDEs of the distributions for the various samples, excluding the limit for AT2021lwx in the ENT KDEs. The black dashed line indicates the star-forming main sequence at $z = 1$.}
 \label{fig:host_props}
\end{figure*}

\clearpage

\begin{figure*}
\centering
 \includegraphics[width=\textwidth]{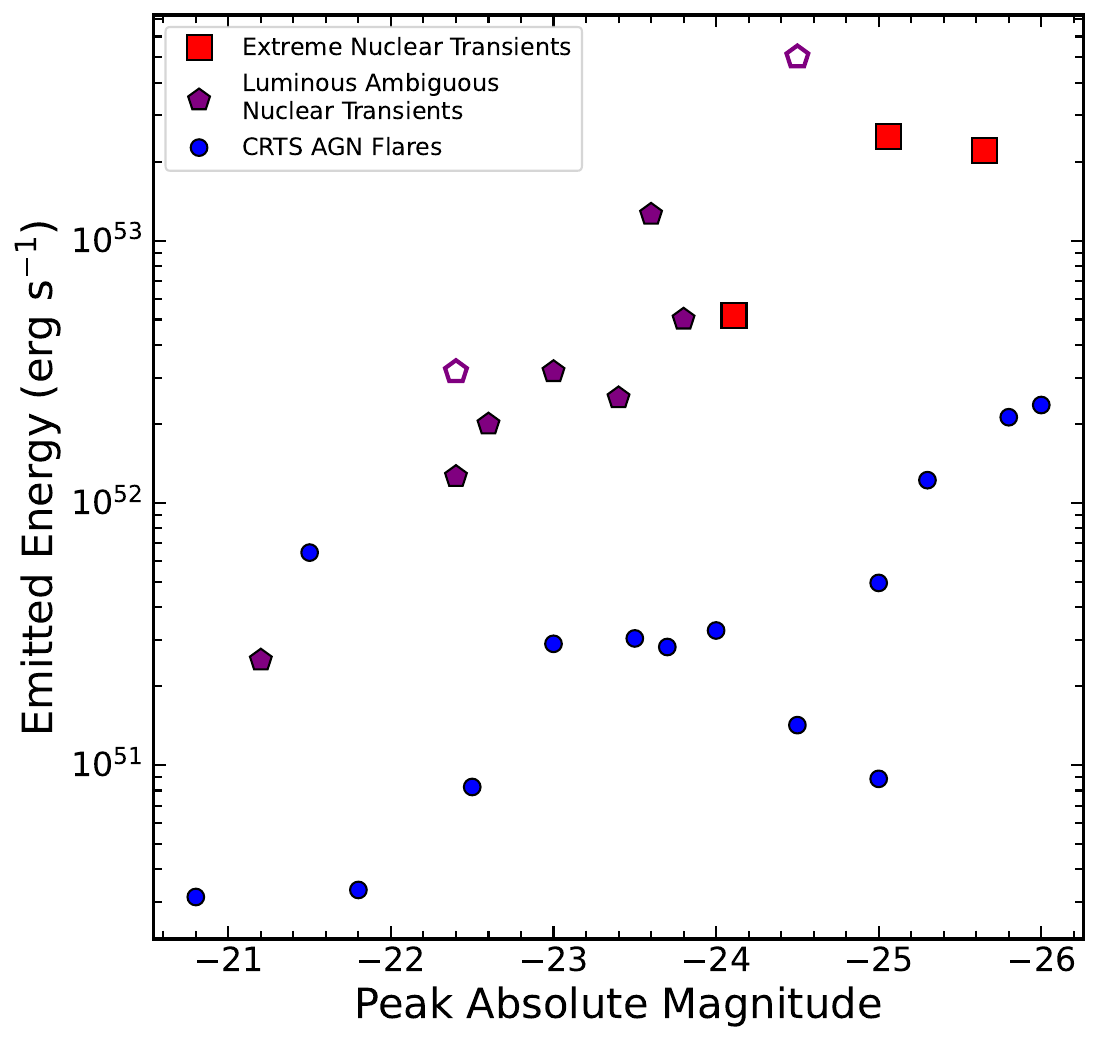}
 \caption{\textbf{Emitted energy as compared to the peak observed optical absolute magnitude for several groups of luminous nuclear flares.} The ENTs from this work are shown as red squares, the ANTs from the sample of \cite{wiseman25} are shown as purple pentagons, and the smooth, monotonic AGN flares identified in CRTS data by \cite{graham17} are shown as blue circles. The open purple symbols represent ANTs with non-monotonicity in recent ZTF photometry. A continuum appears present from the ANTs to the ENTs, but the CRTS AGN flares are roughly an order of magnitude lower in energy.}
 \label{fig:luminous_flares_comp}
\end{figure*}

\clearpage

\begin{figure*}
\centering
 \includegraphics[width=\textwidth]{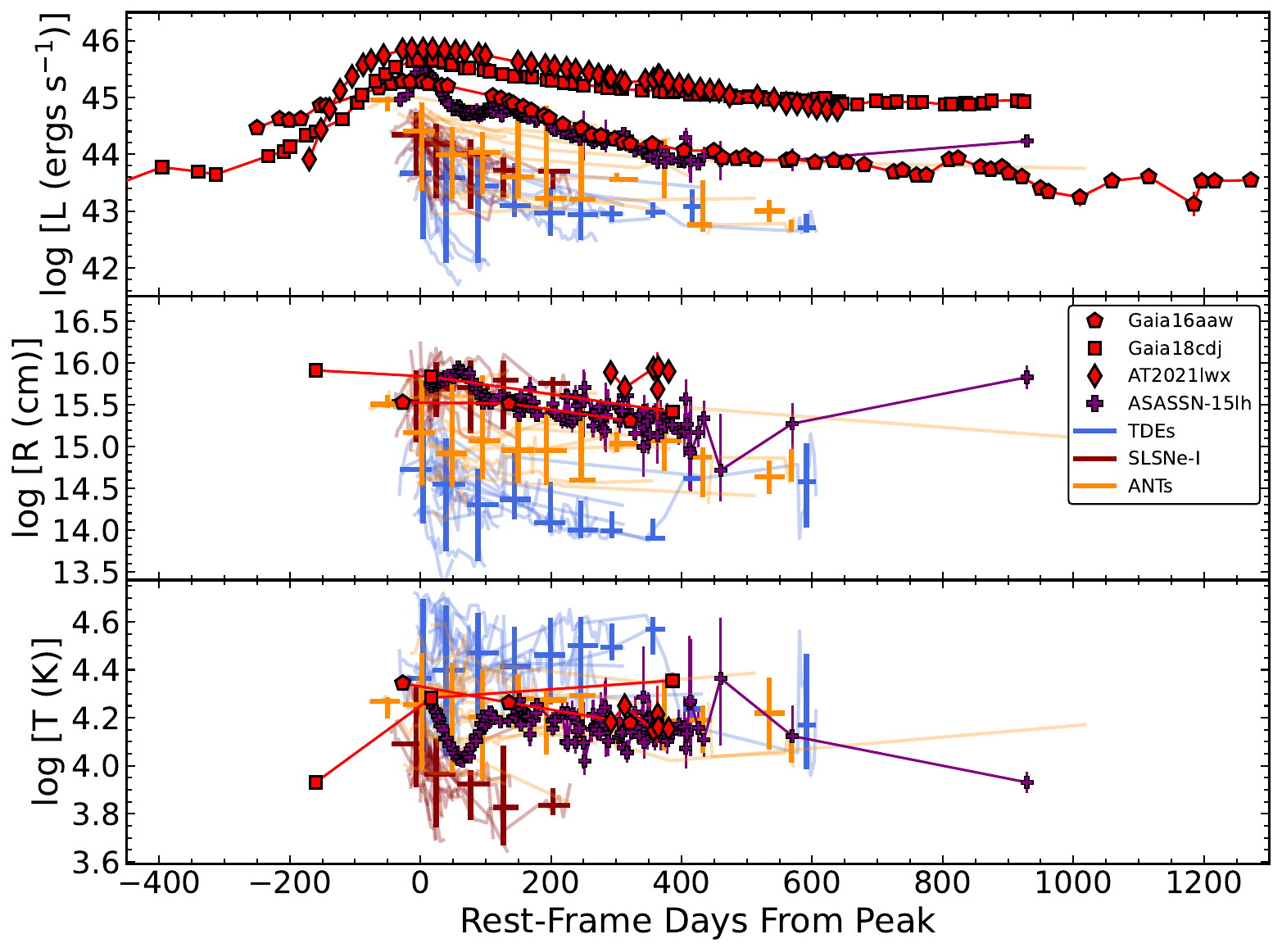}
 \caption{\textbf{Evolution of the blackbody luminosity (top panel), radius (middle panel), and temperature (bottom panel) for the ENTs (red points) and comparison samples of TDEs (blue), SLSNe-I (dark red), and ANTs (gold).} Time is in rest-frame days relative to the peak luminosity. In addition to the individual source evolutions, we have binned the comparison samples in 50-day bins, with the vertical extent indicating the 90\% confidence interval and the horizontal extent indicating the range over which data were combined.}
 \label{fig:BB_comp}
\end{figure*}

\clearpage

\begin{figure*}
\centering
 \includegraphics[width=\textwidth]{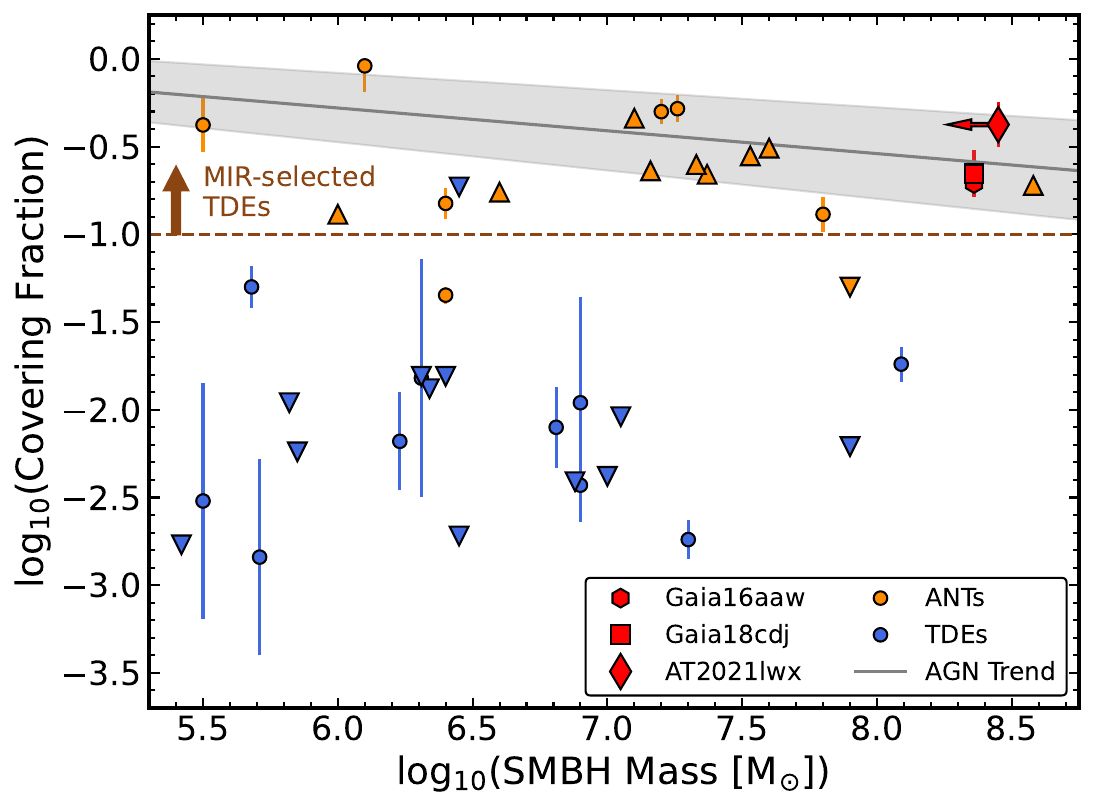}
 \caption{\textbf{Dust covering fraction as compared to SMBH mass for the ENTs (red symbols) and comparison samples of ANTs (gold symbols) and TDEs (blue symbols).} For the comparison data, squares mark detections, upward-facing triangles denote lower limits, and the downward-facing triangles represent upper limits. The solid gray line is the best-fit trend between AGN dust covering fraction and SMBH mass from \cite{ma13}, with the shaded error representing the 90\% confidence interval on their linear fit. The dashed brown line indicates the estimated minimum covering fraction for the MIR-selected TDEs from \cite{masterson24}.}
 \label{fig:dust_covering}
\end{figure*}

\clearpage

\begin{figure*}
\centering
 \includegraphics[width=\textwidth]{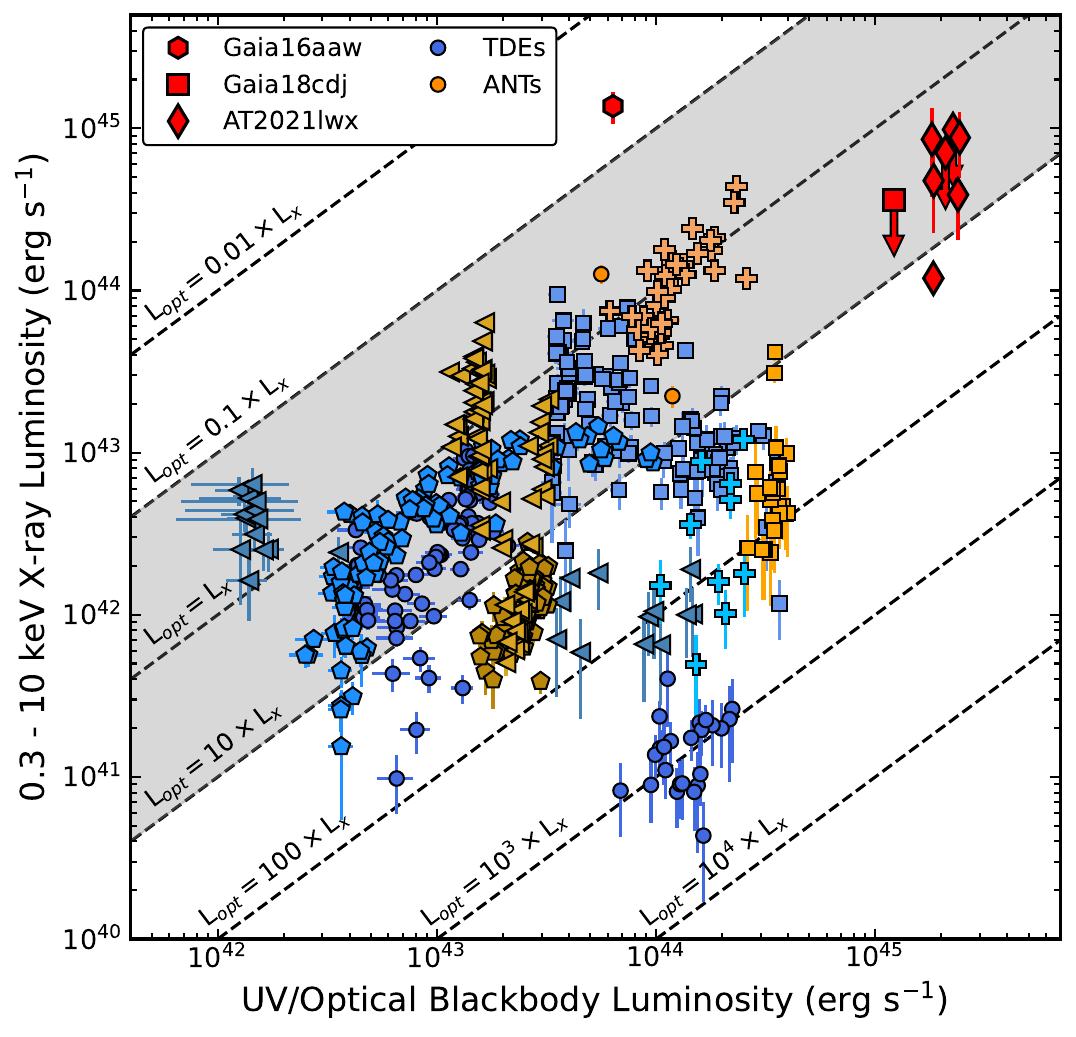}
 \caption{\textbf{The rest-frame 0.3-10 keV X-ray luminosity as compared to the contemporaneous UV/optical luminosity for the ENTs (red symbols).} The comparison samples of TDEs and ANTs are shown in blue and orange shades, respectively. Within the comparison sample, each individual object is shown with a different symbol and color. The dashed diagonal lines represent constant X-ray to UV/optical luminosity ratios. Downward-facing arrows indicate 3$\sigma$ upper limits. The gray shaded band represents the ratio range of typical AGNs \cite{just07, auge23}.}
 \label{fig:xray_opt_ratio}
\end{figure*}

\clearpage

\begin{center}
\Large Supplementary Materials
\end{center}

\renewcommand{\thefigure}{S\arabic{figure}}
\setcounter{figure}{0}

\renewcommand{\thetable}{S\arabic{table}}

\noindent \textbf{This PDF file includes:} \\
\indent Supplementary Text \\
\indent Figures S1 through S8 \\
\indent Tables S1 through S3 \\
\indent References 179 through 201 \\

\clearpage

\subsection*{Observational Data}

\paragraph*{Sample Selection} \label{sec:supp_sample}

When implementing our search criteria we supplemented the Gaia photometry with archival photometry from the All-Sky Automated Survey for Supernovae (ASAS-SN; \cite{shappee14, kochanek17, hart23}), Asteroid Terrestrial-impact Last Alert System  (ATLAS; \cite{tonry18}), the Catalina Real-Time Transient Survey (CRTS; \cite{drake09}), and the Zwicky Transient Facility (ZTF; \cite{bellm19, masci19}). We also obtained optical spectra with the UH 2.2-m telescope and the Spectral Classification of Astronomical Transients (SCAT; \cite{tucker22}) to vet several other smooth nuclear transients which were ultimately rejected because their redshifts indicated lower peak luminosities not sufficient to meet criterion (4). We finally used the Fermi Large Area Telescope Catalog \cite{fermi15} to search for gamma-ray detections and Vizier \cite{ochsenbein00} to search several radio catalogs across the sky.

\begin{figure*}
\centering
 \includegraphics[height=0.49\textwidth]{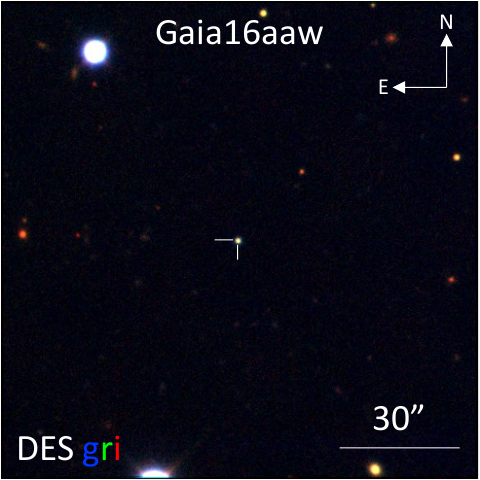}
 \includegraphics[height=0.49\textwidth]{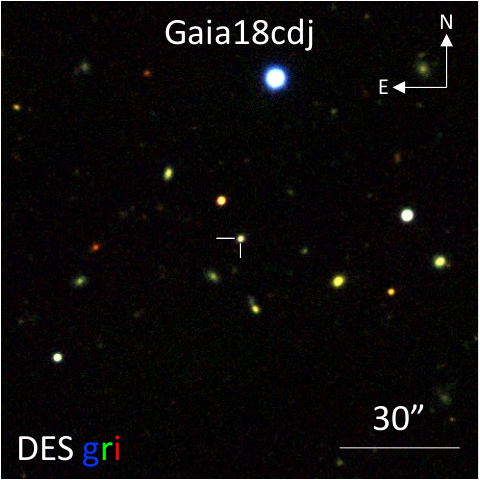}
 \includegraphics[height=0.49\textwidth]{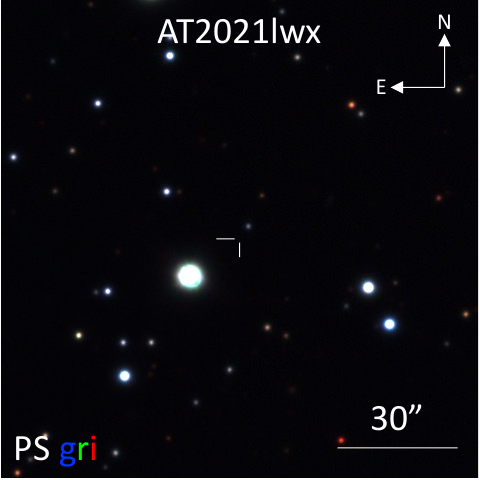}
 \caption{\textbf{Pre-flare $gri$ color images Gaia16aaw (top left; DES), Gaia18cdj (top right; DES), and AT2011lwx (bottom; Pan-STARRS).} The white reticle in each image marks the location of the transient.}
 \label{fig:color_images}
\end{figure*}

\begin{figure*}
\centering
 \includegraphics[width=\textwidth]{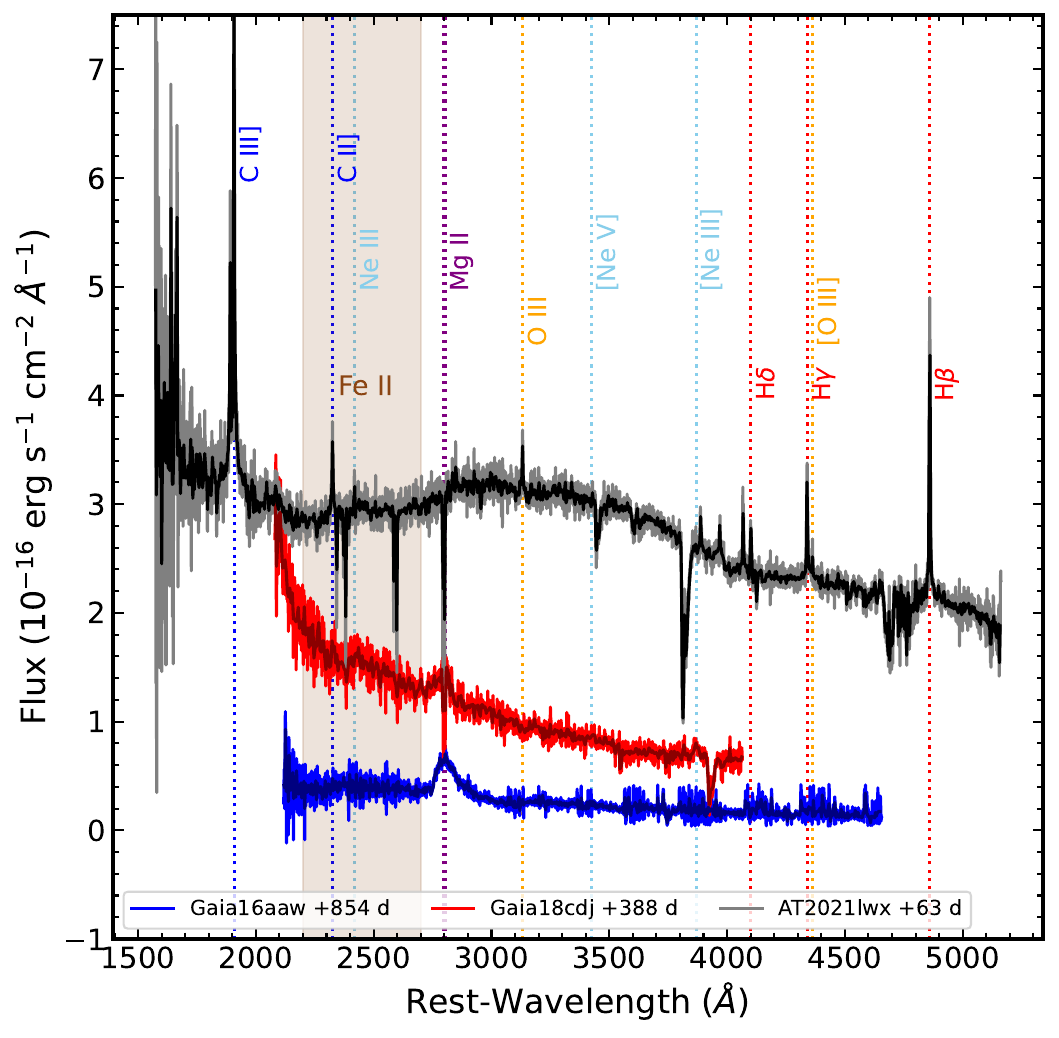}
 \caption{\textbf{Spectra of Gaia16aaw (blue), Gaia18cdj (red), and AT2021lwx (gray).} At the redshift of these sources, the observer-frame optical covers the rest-frame UV and blue portion of the optical. Marked are emission lines of C (blue), Ne (sky blue), Mg (purple), O (orange), H (red), and the broad feature due to Fe lines (brown). The legend indicates the time of observation in rest-frame days from peak.}
 \label{fig:spectra}
\end{figure*}

\begin{figure*}
\centering
 \includegraphics[width=\textwidth]{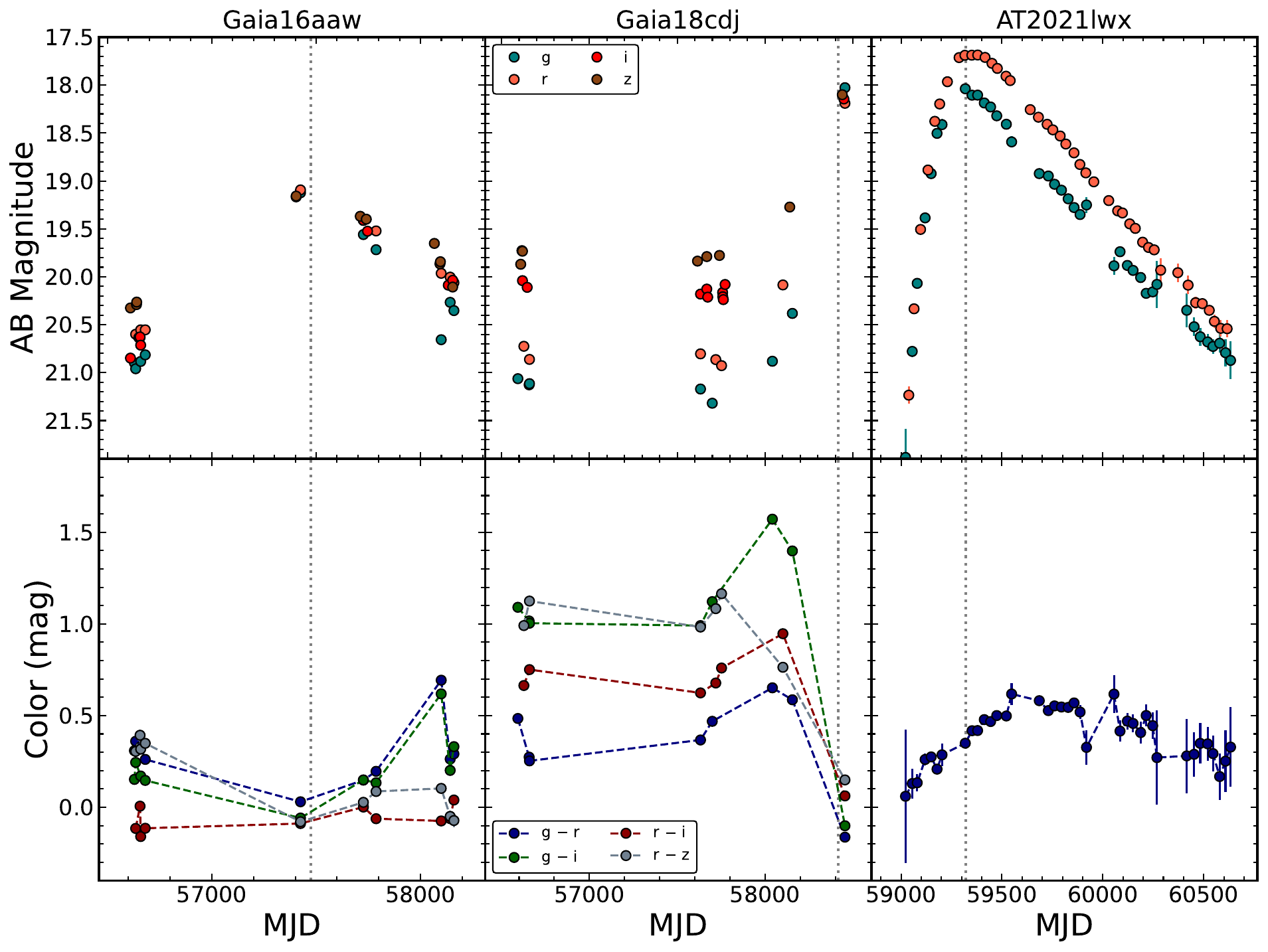}
 \caption{\textbf{Observed-frame optical light curves (top panel) and colors (bottom panel) for Gaia16aaw (left), Gaia18cdj (middle), and AT2021lwx (right).} For Gaia16aaw and Gaia18cdj, the data is from DES and for AT2021lwx the data is from ZTF. These light curves are corrected for Galactic foreground extinction, but have not had any host flux or extinction removed. The dashed vertical gray line marks the time of peak for each source.}
 \label{fig:color_plots}
\end{figure*}

\begin{figure*}
\centering
 \includegraphics[width=0.49\textwidth]{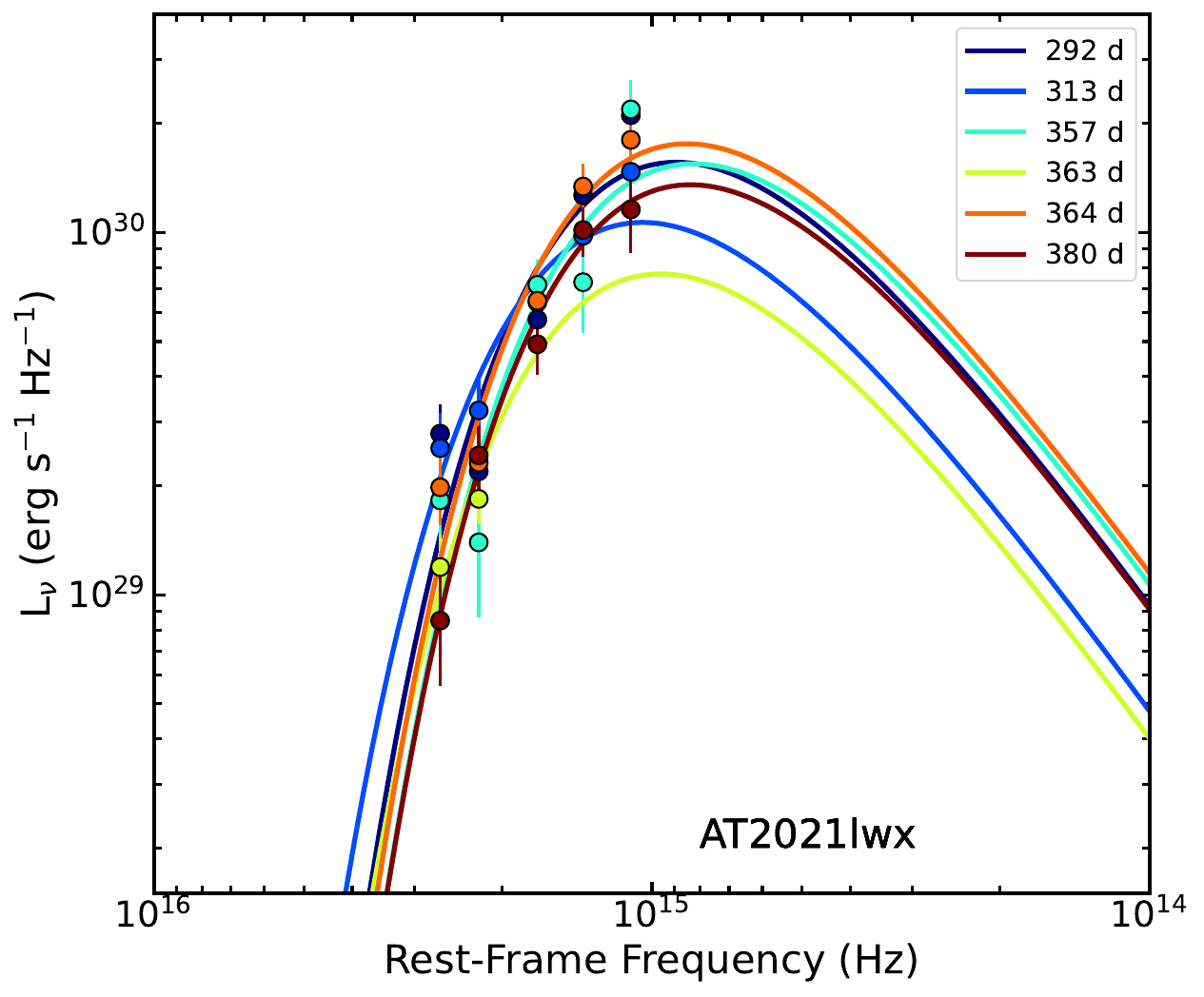}
 \includegraphics[width=0.49\textwidth]{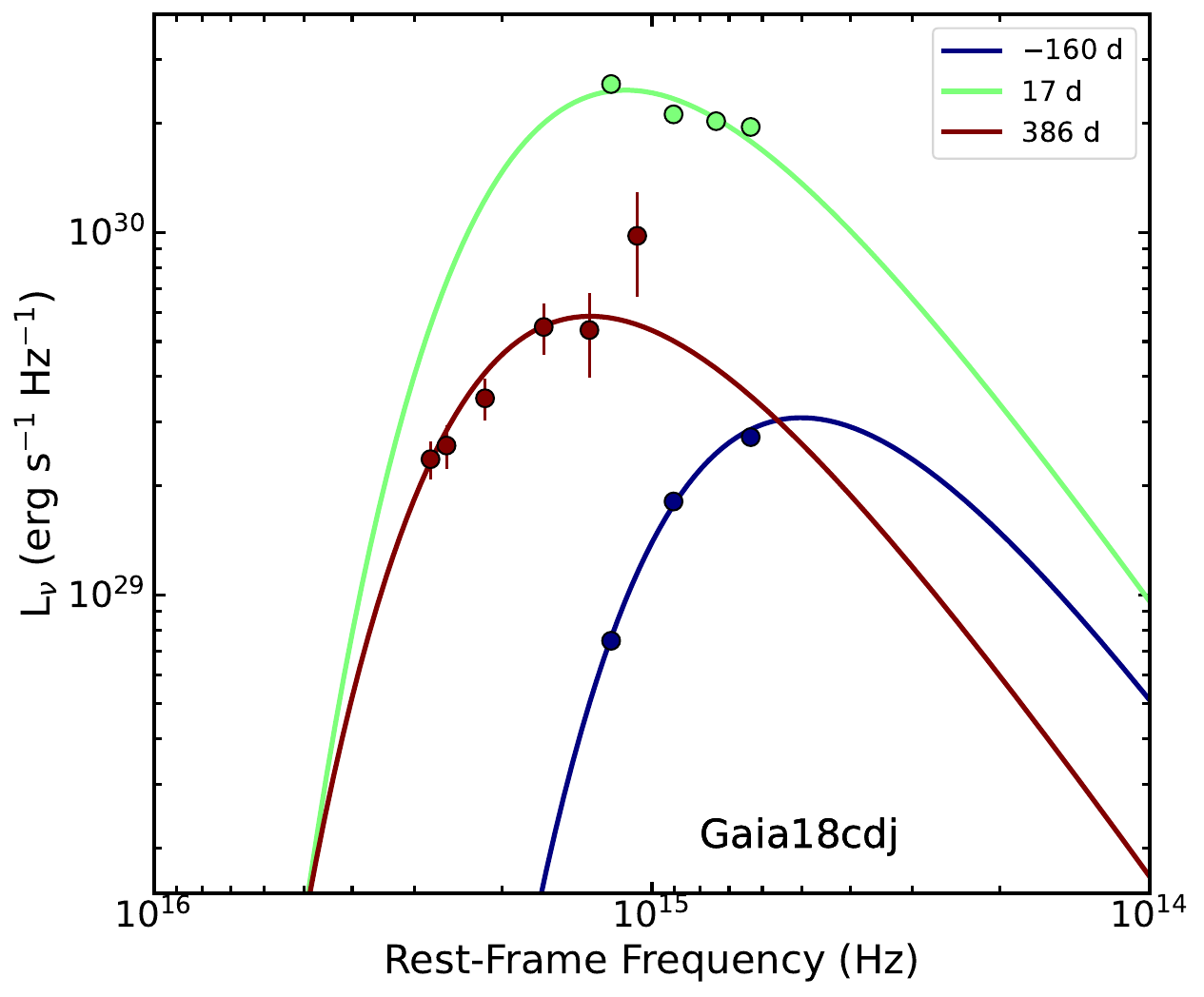}
 \includegraphics[width=0.49\textwidth]{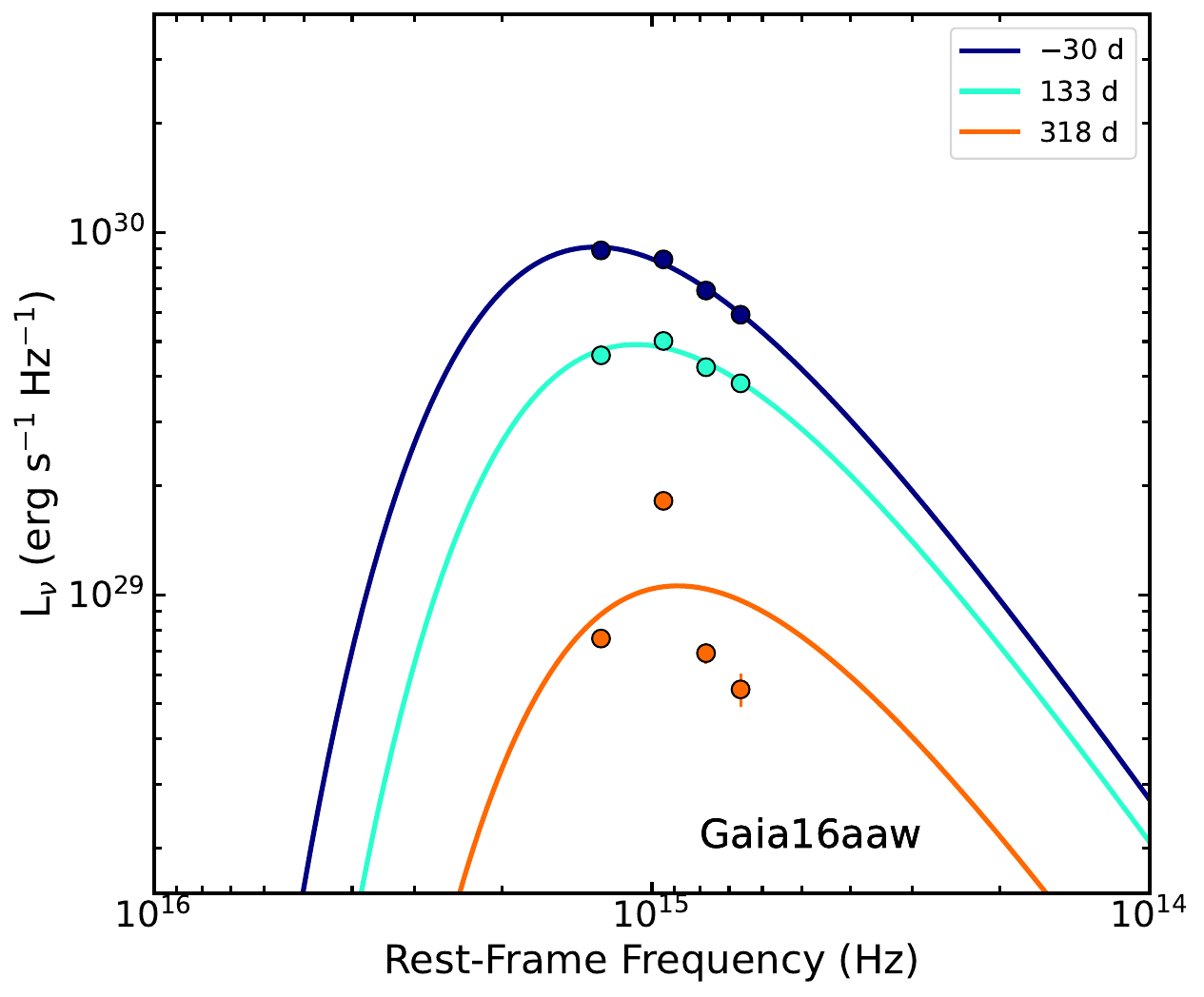}
 \caption{\textbf{Rest-frame UV/optical photometry and corresponding best-fitting blackbody models for AT2021lwx (top left), Gaia18cdj (top right), and Gaia16aaw (bottom).} For AT2021lwx, the data fit is from Swift UVOT, for Gaia18cdj the data comes from Swift and DES, and the data for Gaia16aaw is from DES. The times given in the legend are rest-frame phases from peak emission.}
 \label{fig:BB_fits}
\end{figure*}

The three transients we study in detail are Gaia16aaw, Gaia18cdj, and ZTF20abrbeie. Gaia16aaw, $(\alpha,\delta) =$ (04:11:57.000, $-$42:05:30.80), was discovered on 2016 January 23.5 by the Gaia Alerts team \cite{hodgkin13}. The discovery was announced publicly on the Transient Name Server (TNS) and given the identification AT2016dbs \cite{16aaw_tns}. Gaia18cdj, $(\alpha,\delta) =$ (02:09:48.140, $-$42:04:37.02), was discovered on 2018 August 12.1 by the Gaia Alerts team \cite{hodgkin13} and announced on TNS with the identification AT2018fbb \cite{18cdj_tns}. ZTF20abrbeie, $(\alpha,\delta) =$ (21:13:48.408, $+$27:25:50.48), was discovered on 2021 April 13.5 by ZTF. It was announced to TNS with the identification AT2021lwx \cite{21lwx_tns}. Color images created using \textsc{aplpy} \cite{robitaille12} from $gri$ images taken by either the Dark Energy Survey (DES; for Gaia16aaw and Gaia18cdj) \cite{DES05} or the Panoramic Survey Telescope and Rapid Response System (Pan-STARRS; for AT2021lwx) \cite{chambers16} are shown in figure \ref{fig:color_images}.

As ZTF20abrbeie has had previous papers published on its evolution under the name AT2021lwx \cite{subrayan23, wiseman23}, we elect to use that name in this manuscript to avoid confusion. For the two Gaia sources, we elect to use their survey names in this manuscript.

\paragraph*{Archival and Transient Photometry} \label{sec:supp_phot}

We first searched for available archival photometry to constrain the evolution of these objects prior to the ENT flare. For Gaia16aaw and Gaia18cdj, we obtained $V$-band photometry from the CRTS. As the typical signal-to-noise ratio (S/N) per CRTS epoch was low, we binned these data in monthly bins to search for pre-flare variability. These data are shown in Figure \ref{fig:opt_ir_lcs} as tan squares. The sources are weakly detected and show no significant evidence for strong variability prior to the flares as any pre-flare variability in CRTS data for Gaia16aaw and Gaia18cdj is consistent with noise.

To constrain both the pre-flare characteristics of the objects and their behavior during the flare, we also obtained photometry from the Wide-field Infrared Survey Explorer (WISE; \cite{wright10}). This includes data taken during both the AllWISE \cite{wright10} and NEOWISE \cite{mainzer11, NEOWISE_phot} portions of the WISE mission. We have focused on the $W1$ (3.4 $\mu m$) and $W2$ (4.6 $\mu m$) bands as they span the full lifetime of the WISE mission. Gaia16aaw and Gaia18cdj are well-detected in the NEOWISE single exposure catalog and we therefore construct their $W1$ and $W2$ light curves by binning the individual exposures within a given epoch.

For AT2021lwx, there is no persistent source detected in the NEOWISE single exposure catalog, consistent with the apparent lack of a host galaxy in Pan-STARRS imaging \cite{subrayan23, wiseman23}. We therefore performed aperture photometry on the NEOWISE images to obtain our WISE light curves. To avoid contamination by nearby, bright stars, we used a 4" radius aperture with an aperture correction estimated from the computed WISE curve of growth. The local background was estimated using a sigma-clipped median within an annulus around the target. Despite the small source aperture, the $W1$ band retained a low level of residual flux from the nearby, bright star. To mitigate this, we subtracted the mean pre-flare flux from the light curves and added the pre-flare scatter in quadrature with the estimated flux uncertainties for each epoch. The WISE light curves are shown in Figure \ref{fig:opt_ir_lcs} as olive and brown diamonds for $W1$ and $W2$ respectively. Given the lack of a pre-flare detection, we only show NEOWISE data for AT2021lwx.

DES \cite{DES05} imaging covered the locations of Gaia16aaw and Gaia18cdj. These images span from late 2013 to late 2018 and cover these locations both prior to and during the ENT. We created DES light curves in the $griz$ bands by performing aperture photometry with a 3" radius aperture, subtracting the local background estimated through a sigma-clipped median in an annulus centered on the source, and calibrating to nearby stars with catalog photometry from ATLAS Refcat2 \cite{tonry18b}. We elected to use a relatively large aperture to ensure that the entire galaxy was contained within the aperture. The DES data are shown in Figure \ref{fig:opt_ir_lcs} as circles, with the color corresponding to the filter.

For each ENT we also obtained ATLAS light curves from their forced point-spread function photometry service. This yielded light curves in the `cyan' ($c$, 4200--6500 \AA) and `orange' ($o$, 5600--8200 \AA) filters \cite{tonry18}. To ensure more robust detections, we stacked the ATLAS data in monthly bins throughout the flares. The ATLAS photometry is plotted in Figure \ref{fig:opt_ir_lcs} as cyan and orange pentagons.

Each of the sources was observed by the Neil Gehrels Swift Gamma-ray Burst Mission (\textit{Swift}; \cite{gehrels04}) at least once during its evolution. These observations utilized both the UltraViolet and Optical Telescope (UVOT; \cite{roming05}) and X-ray Telescope (XRT; \cite{burrows05}). We combined all UVOT images taken in a given filter per epoch using the HEASoft {\tt uvotimsum} package and used the {\tt uvotsource} package to extract source and background fluxes from these coadded images. For Gaia16aaw (PI: Hinkle) and Gaia18cdj (PI: Hinkle) we used the default UVOT aperture with a radius of 5" and background regions with radii of 50" and 40", respectively. For AT2021lwx (PI: Wang), there is a star of comparable brightness within the default 5" aperture and so we used a smaller 3" aperture to capture only the source flux. As AT2021lwx is relatively close to the Galactic plane and is therefore in a moderately crowded field, we used a background ellipse with an effective area of $\approx$27" chosen to avoid contamination from nearby stars.

Finally, we acquired light curves from the discovering survey for each source. For Gaia16aaw and Gaia18cdj, this was $G$-band photometry from Gaia, obtained through the Gaia Alerts service. We estimated the uncertainties on this photometry following \cite{riello21} and binned visits within 5 days for subsequent analyses. The Gaia light curves are shown in Figure \ref{fig:opt_ir_lcs} as green hexagons. For AT2021lwx, we procured ZTF photometry in the $g$ and $r$ bands from the ZTF forced photometry service \cite{masci19}. We stacked the pre-flare data in monthly bins to obtain deep limits. During the rise and through peak, we stacked the data 5-day bins, switching to monthly bins after the last seasonal break to better sample the decline. These data are shown in Figure \ref{fig:opt_ir_lcs} as teal and red octagons.

\subsection*{Host-Galaxy Properties}

\paragraph*{Stellar Mass and Star Formation Rate} \label{sec:supp_mass_and_sfr}

We further examine the SFRs of these ENT host galaxies using narrow emission lines and standard scaling relations. Gaia16aaw shows an emission feature near [O II], although there is contamination from night sky lines. Assuming that this line is indeed [O II], we derive a flux of $6.1 \times 10^{-17}$ erg s$^{-1}$ cm$^{-2}$. At the distance of Gaia16aaw and correcting for the $A_V = 0.92$ mag extinction from the CIGALE fits, this corresponds to a luminosity of $1.3 \times 10^{42}$ erg s$^{-1}$. From typical scaling relations \cite{kennicutt98, kewley04}, such a luminosity implies a SFR of $\sim 10 \mbox{--} 20$ M$_{\odot}$ yr$^{-1}$, roughly 2$\sigma$ below the estimate from the CIGALE fits. 

Gaia18cdj does not show an [O II] emission line, so we calculated an upper-limit on the [O II] luminosity using Equation (1) with a line width of 300 km s$^{-1}$. Correcting for the extinction from the CIGALE fits, this yields a luminosity upper-limit of $<4 \times 10^{41}$ erg s$^{-1}$, implying an SFR of $<3 \mbox{--} 6$ M$_{\odot}$ yr$^{-1}$ \cite{kennicutt98, kewley04}. While the SFRs estimated from spectroscopic scaling relations are low relative to the estimates from CIGALE, we note (1) that these lines are in regions of the spectra with S/N below 5 and (2) that the SFRs estimated from the rest-frame UV absolute magnitudes of $M \approx -22.5$ mag \cite{kennicutt98} are consistent with the CIGALE estimates.

\paragraph*{Dust Covering Fraction} \label{sec:supp_dust}

Anisotropies in the dust distribution can cause the ratio of IR to UV/optical luminosity to underestimate low covering factors and overestimate high covering factors. While the emitting geometry of these ENTs is unknown, we apply the corrections of \cite{stalevski16} to examine their effects. Using Table 1 of \cite{stalevski16} and assuming an aligned disk and torus with $\tau_{9.7 \mu m} = 3$, their minimum computed optical depth, we find covering fractions of $f_c \geq 0.42$, $f_c = 0.44$, and $f_c = 0.58$ for Gaia16aaw, Gaia18cdj, and AT2021lwx respectively. While these corrections increase the covering fraction estimates for our ENTs, their position relative to the TDE and ANT samples as well as the typical AGN trend in Fig.\ \ref{fig:dust_covering} is not dramatically different.

\begin{figure*}
\centering
 \includegraphics[height=0.38\textwidth]{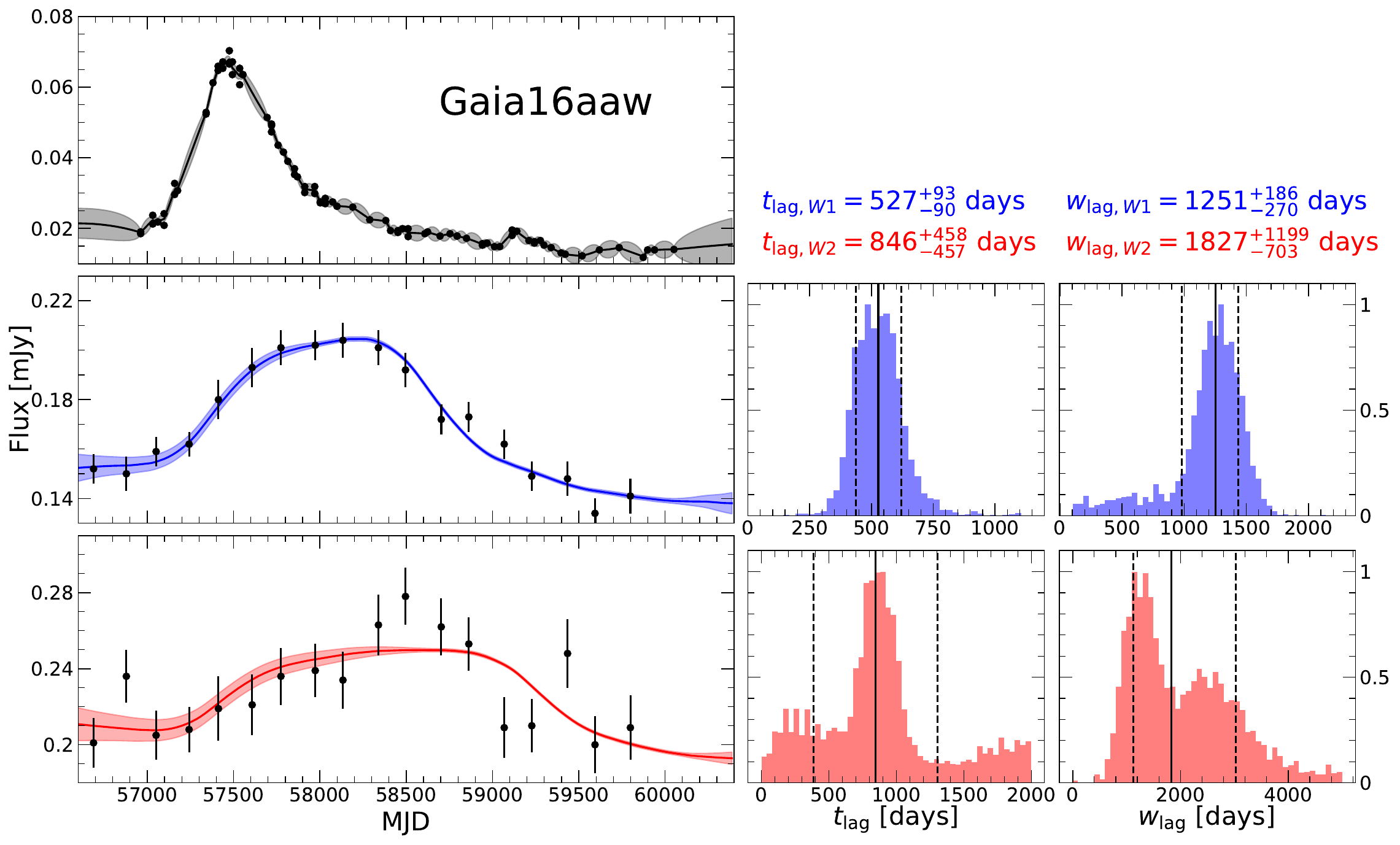}
 \includegraphics[height=0.38\textwidth]{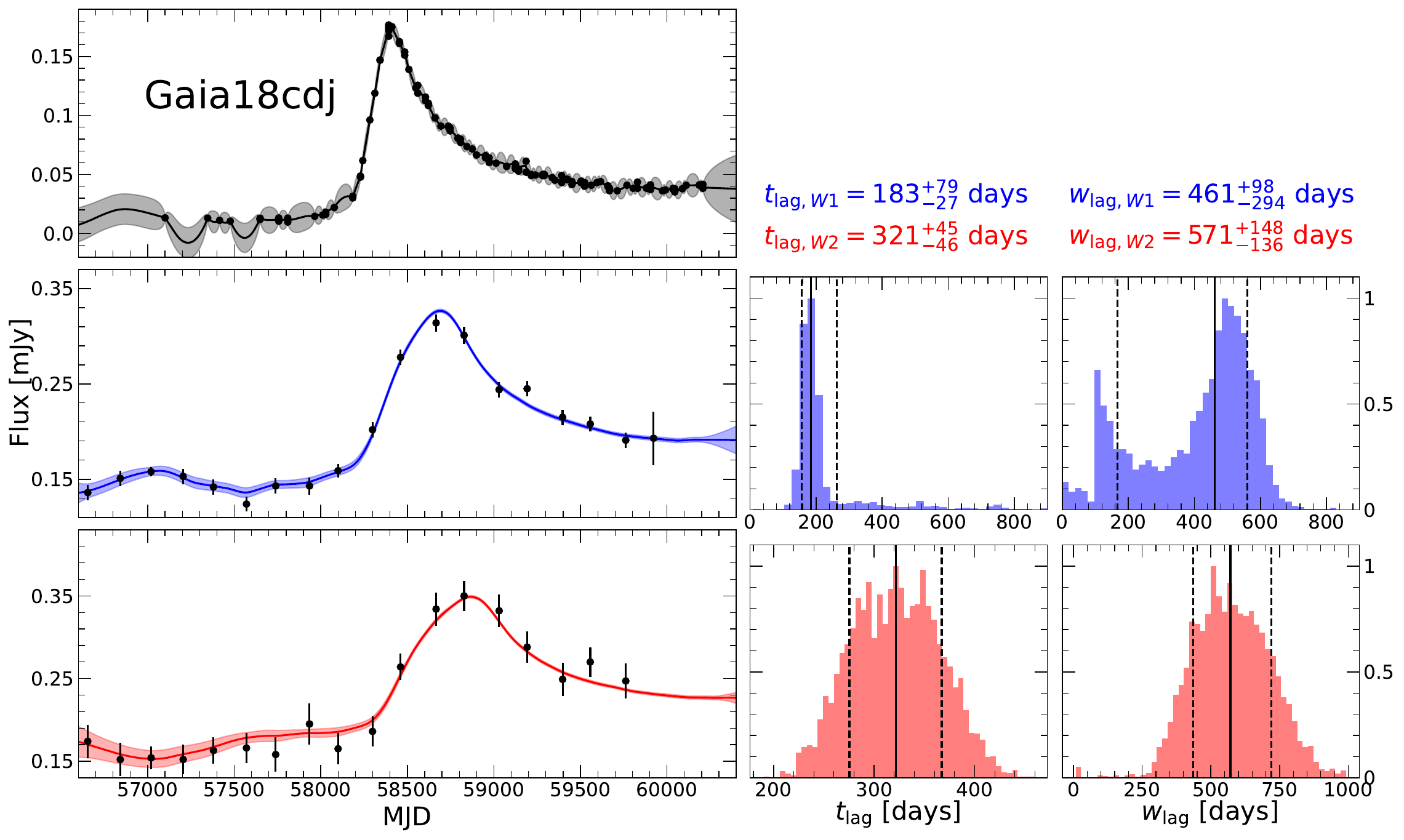}
 \includegraphics[height=0.38\textwidth]{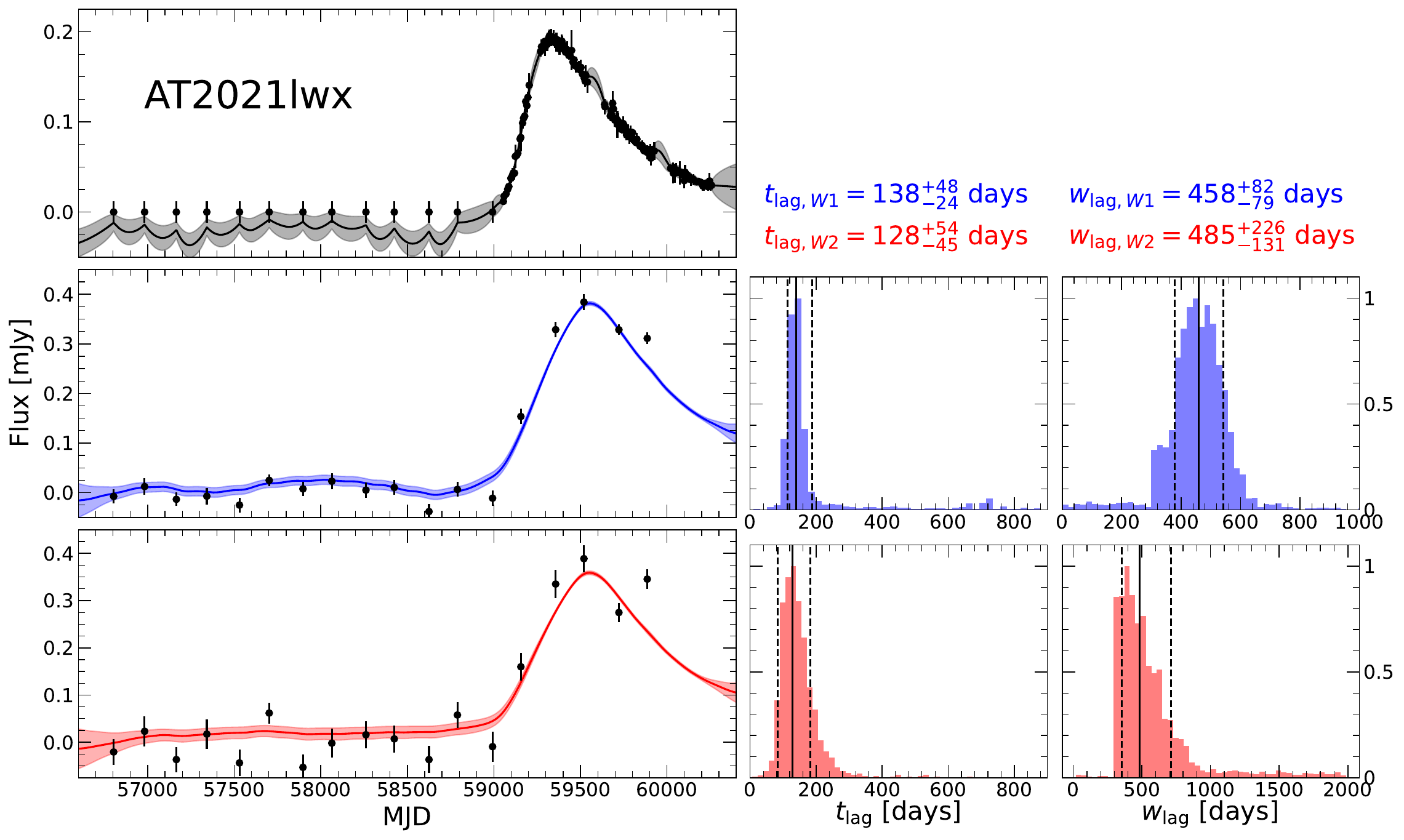}
 \caption{\textbf{JAVELIN \cite{zu11, zu13} fits to the observer-frame optical (black), NEOWISE $W1$ (blue), and NEOWISE $W2$ (red) light curves for Gaia16aaw (top panel), Gaia18cdj (middle panel), and AT2021lwx (bottom panel).} We use the Gaia $G$-band light curve for Gaia16aaw and Gaia18cdj and ZTF $r$-band light curve for AT2021lwx. The right panels within each figure give histograms for the lags (t$_{lag}$) and top hat smoothing length (W$_{lag}$).}
 \label{fig:javelin}
\end{figure*}

\begin{figure*}
\centering
 \includegraphics[width=\textwidth]{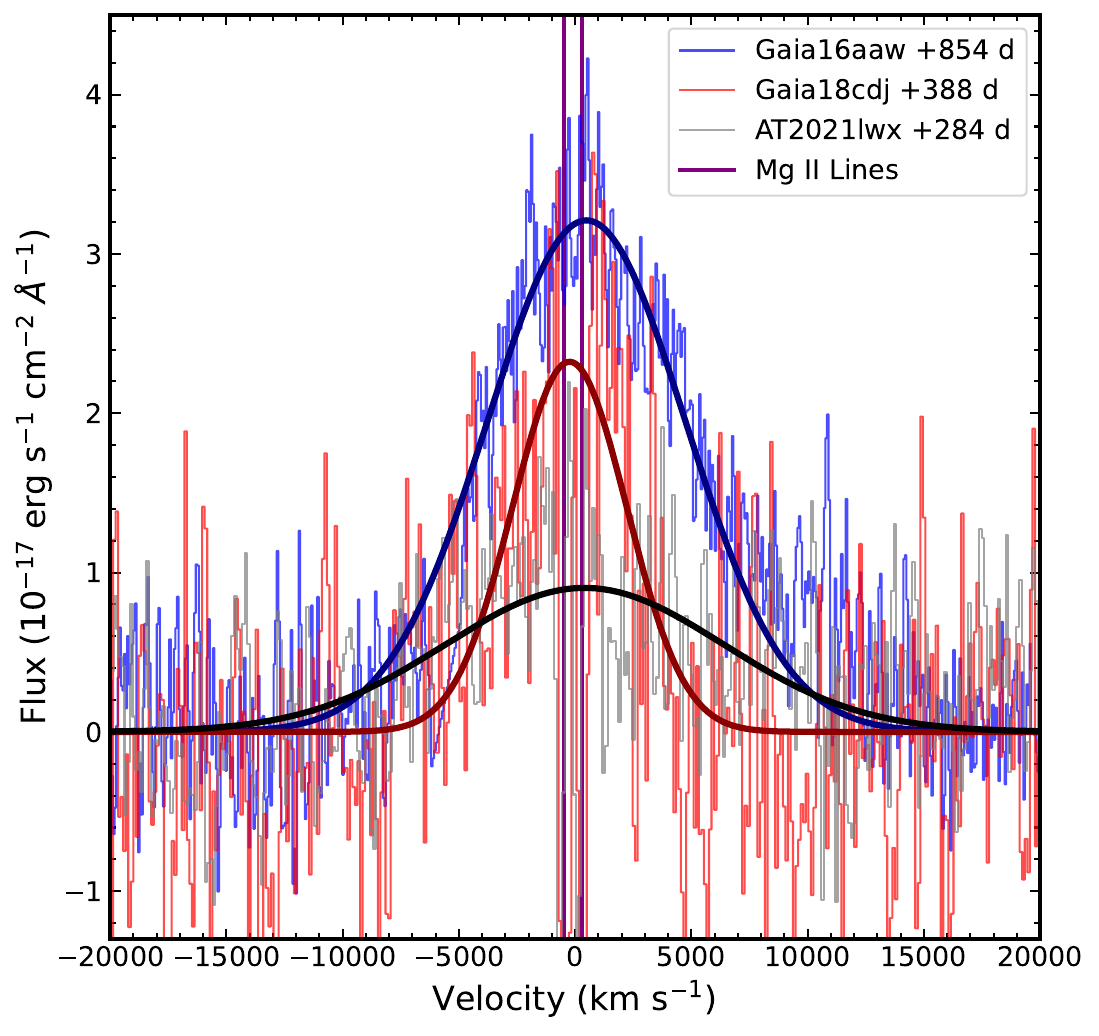}
 \caption{\textbf{The broad Mg II feature for Gaia16aaw (blue), Gaia18cdj (red), and AT2021lwx (gray).} Single-component Gaussian fits are shown as the solid line, with the color corresponding to the object.}
 \label{fig:MgII}
\end{figure*}

\begin{figure*}
\centering
 \includegraphics[width=0.48\textwidth]{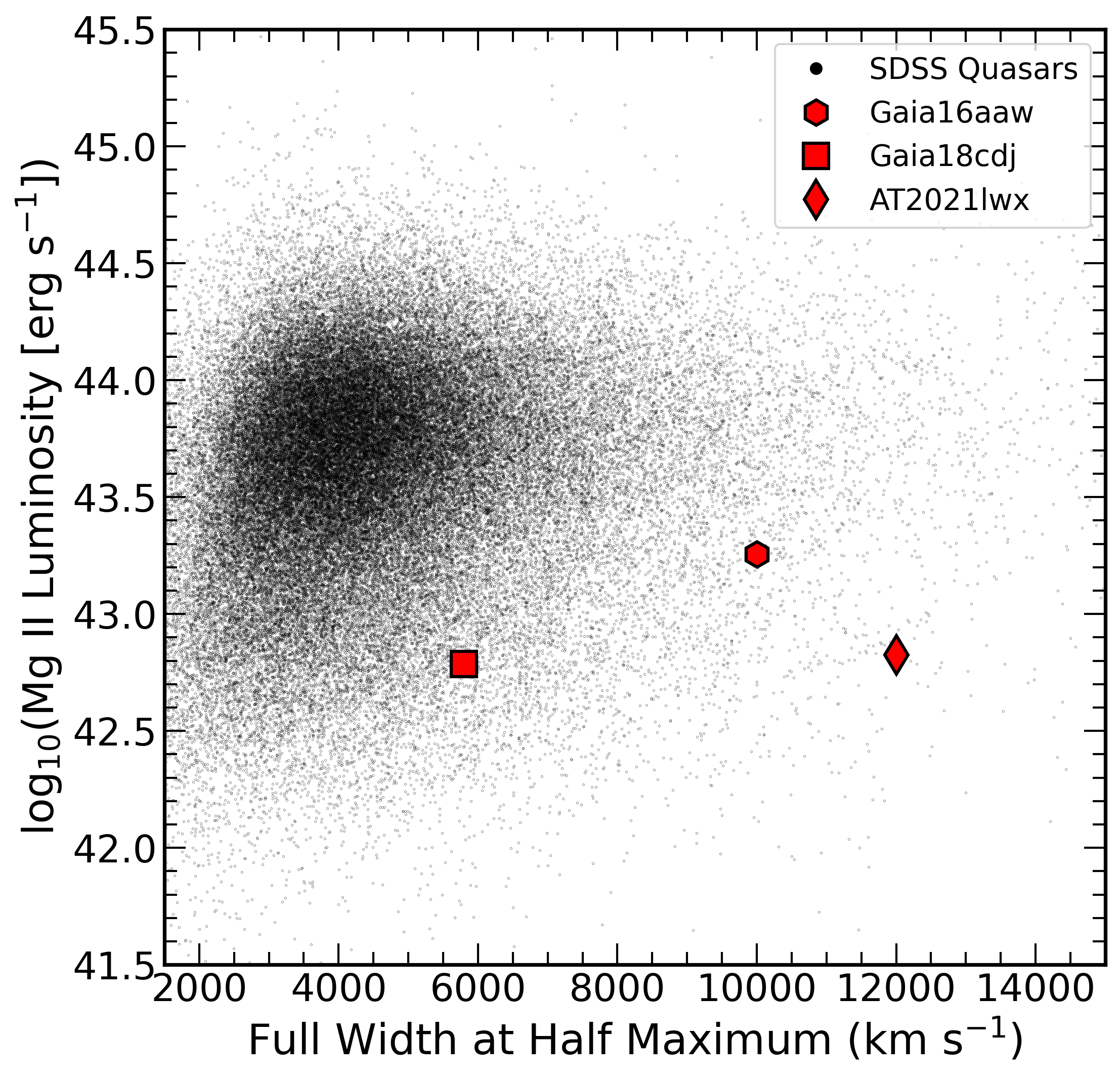}
 \includegraphics[width=0.48\textwidth]{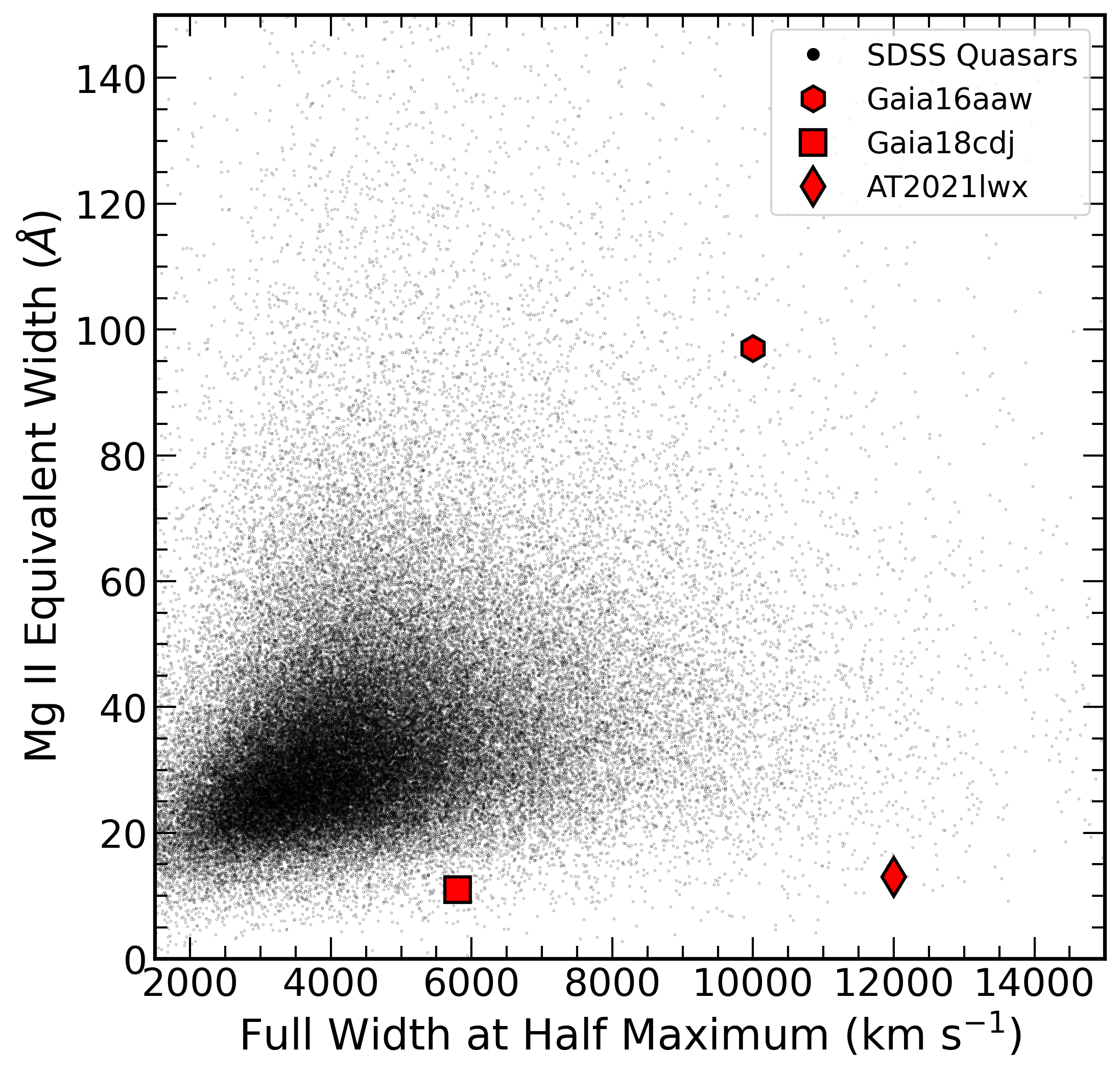}
 \includegraphics[width=0.48\textwidth]{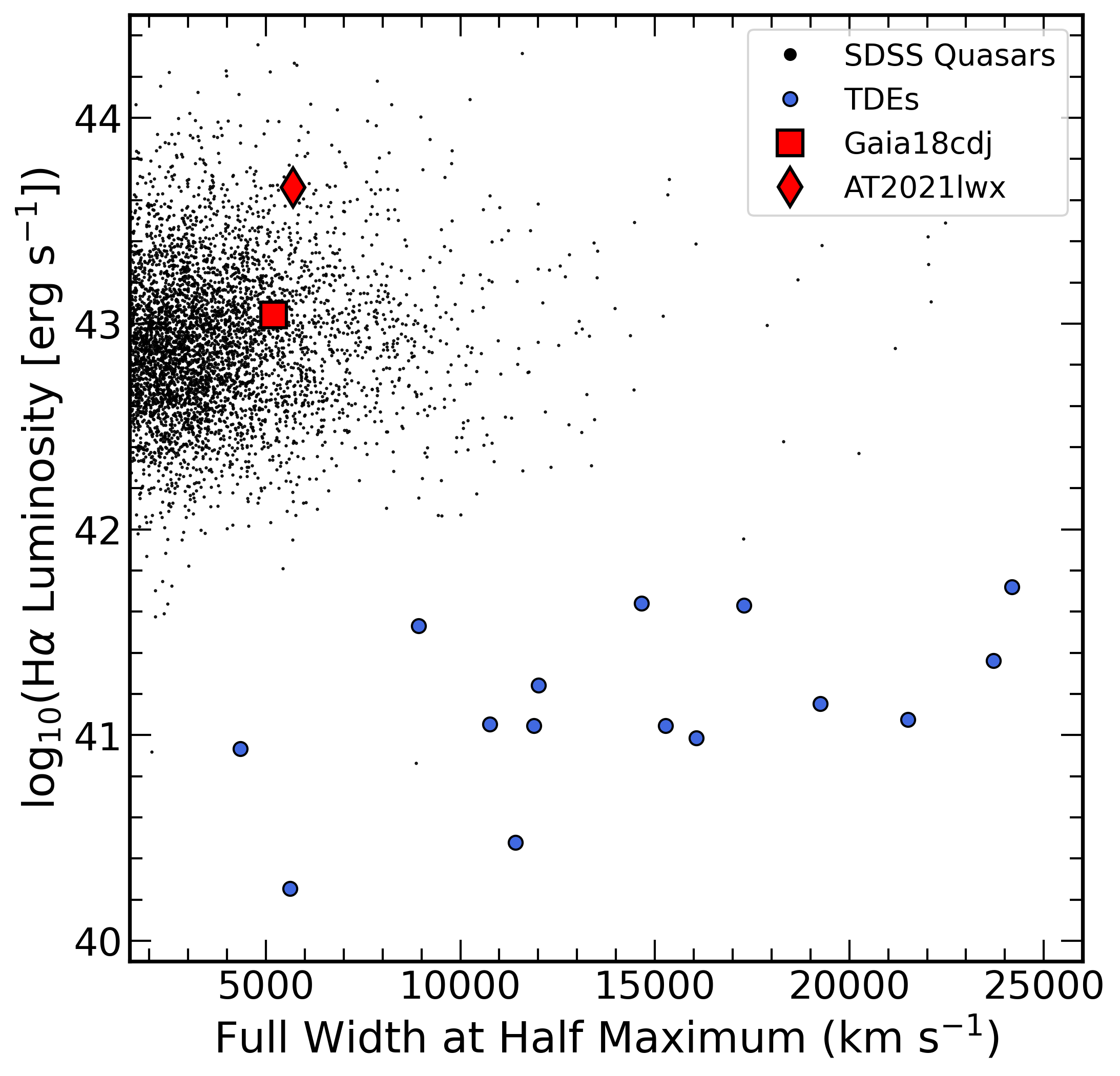}
 \caption{\textbf{ENT emission line parameters compared to AGNs and TDEs.} Mg II line luminosity as compared to FWHM (top left) and MG II equivalent width as compared to FWHM (top right) for the ENTs and a sample of SDSS quasars from \cite{shen11}. The bottom panel compares the H$\alpha$ line luminosity to the FWHM again for the ENTs and SDSS quasars, along with TDEs from \cite{charalampopoulos22}.}
 \label{fig:emission_line_comp}
\end{figure*}

\begin{figure*}
\centering
 \includegraphics[width=\textwidth]{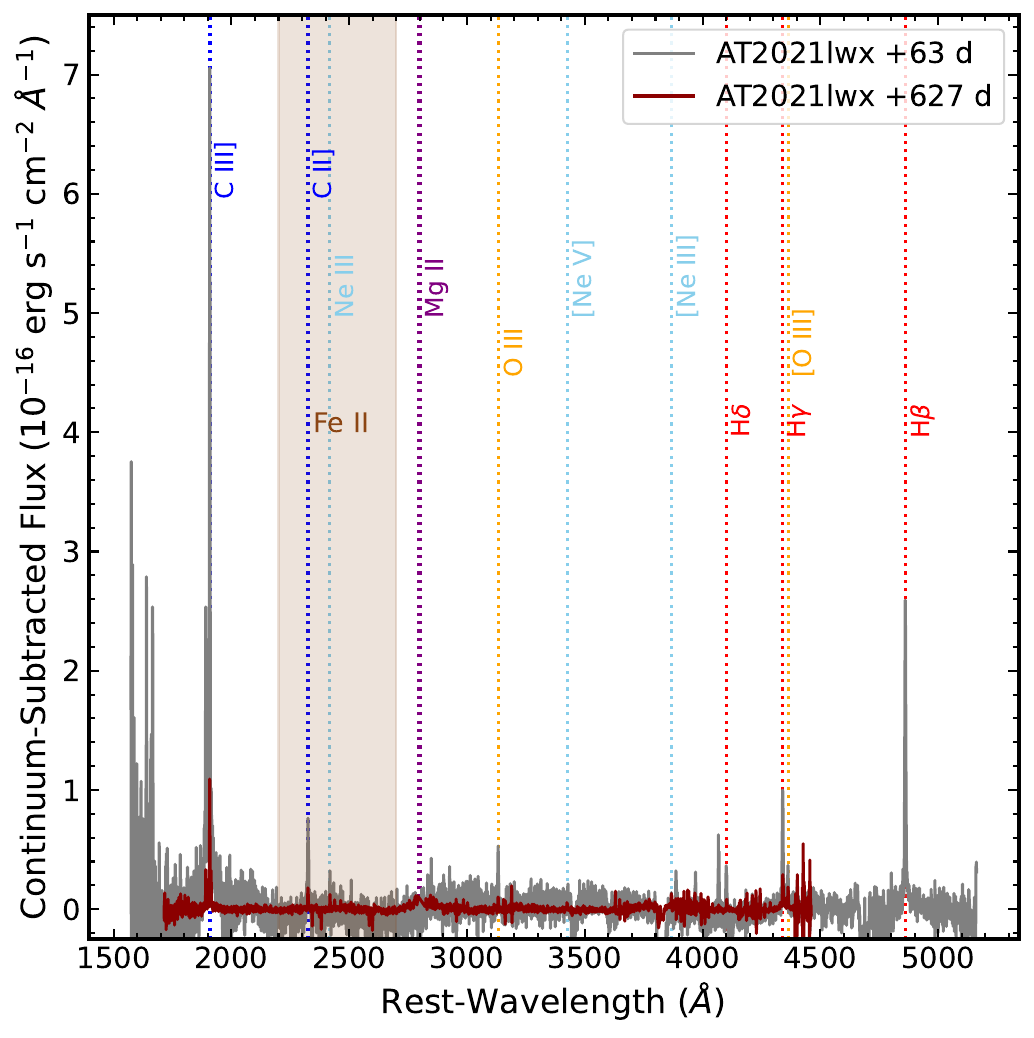}
 \caption{\textbf{Continuum-subtracted spectra taken of AT2021lwx at rest-frame phases of $+63$ d and $+627$ d.} Strong narrow lines of C, O, Ne, and H are seen both near peak and at late times. The emission line flux decrease is consistent with the decrease in the transient continuum.}
 \label{fig:21lwx_spec_comp}
\end{figure*}

We also used \textsc{JAVELIN} \cite{zu11, zu13} to compute temporal lags between the optical and IR flares, employing a top hat smoothing function. For AT2021lwx, as the source is undetected pre-flare, we added artificial optical measurements at the time of the NEOWISE observations with zero flux and uncertainties equal to the first optical flux uncertainty. This was necessary to avoid significantly negative fluxes in the model optical light curve at early times. These fits are shown in figure \ref{fig:javelin}. We find rest-frame lags of $\approx$265 days for Gaia16aaw, $\approx$150 days for Gaia18cdj, and $\approx$70 days for AT2021lwx, where these are the weighted average of the lags between the $W1$ and $W2$ bands and the optical band. The $W2$ lags are generally longer than the $W1$ lags as we would expect since the $W2$ emission will include contributions from cooler, more distant dust. The width of the smoothing top hats were roughly twice the mean lag, which is the lag distribution of a spherical shell. This suggests that the dust subtends a large solid angle around the transient.

We can alternatively estimate the dust covering fractions as $f_c \approx (L_{IR} \Delta t)/(\tau_{dust} E_{flare})$ where $L_{IR}$ is the peak IR luminosity of the dust echo, $\Delta t$ is the temporal lag, $\tau$ is the dust optical depth, and $E_{flare}$ is the emitted energy of the transient. Assuming $\tau_{dust} = 0.25$, we find dust covering fractions of $\sim$0.7 for Gaia16aaw, $\sim$0.2 for Gaia18cdj, and $\sim$0.3 for AT2021lwx, each in good agreement with the estimates from the luminosity ratios.

While it is not exact and depends on the dust composition in detail, the fraction of the observed optical light that consists of photons scattered by the dust should be comparable to the absorbed fraction. Hence, another argument that the overall fraction of the absorbed light must be modest is that if the mean optical depth of the dust producing the IR echoes is high, scattered photons from the peak UV/optical emission would overproduce the tail of the light curve. As it is, some fraction of the tail is likely from photons scattered off the dust. This should be a general consideration for models of mid-IR dust echoes from nuclear transients (e.g., \cite{jiang21a}), although the detailed modeling of the dust radiative transfer is beyond the scope of this paper.

\clearpage

\clearpage

\clearpage

\clearpage

\clearpage

\clearpage

\clearpage

\clearpage

\clearpage

\begin{table}
\centering
\caption{UV/optical Blackbody Fits}
 \begin{tabular}{cccccccccccc}
  \hline
Object & Source & MJD & T$_{16}$ & T$_{50}$ & T$_{84}$ & L$_{16}$ & L$_{50}$ & L$_{84}$ & R$_{16}$ & R$_{50}$ & R$_{84}$ \\
& &  & (K) & (K) & (K) & (erg/s) & (erg/s) & (erg/s) & (cm) & (cm) & (cm) \\
\hline
Gaia16aaw & DES & 57414.6$^{a}$ & 4.33 & 4.34 & 4.36 & 45.26 & 45.28 & 45.30 &  15.51 & 15.53 &  15.54 \\
Gaia16aaw & DES & 57745.2$^{a}$ & 4.25 & 4.26 & 4.27 &  44.92 &  44.93 & 44.94 & 15.50 & 15.51 &  15.53 \\
Gaia16aaw & DES & 58121.7$^{a}$ & 4.17 & 4.18 & 4.19 & 44.18 & 44.19  &  44.20 & 15.29 &  15.31 &  15.33 \\
Gaia18cdj & DES  & 58105.7$^{a}$ & 3.92 & 3.93 & 3.94 & 44.39 & 44.40 & 44.41  & 15.89  & 15.91   & 15.93 \\
Gaia18cdj &  DES  & 58448.9$^{a}$  & 4.27  & 4.28  & 4.29  & 45.64 & 45.66 & 45.67 & 15.82 &   15.84 &  15.85 \\
Gaia18cdj  & \textit{Swift} &  59164.3  & 4.33  & 4.36 & 4.39  &  45.06 & 45.10 &  45.14 & 15.33 &  15.42  &  15.49 \\
AT2021lwx & \textit{Swift} & 59923.6 & 4.13 & 4.18 & 4.23  & 45.27 & 45.35 & 45.44  & 15.76  & 15.89 & 16.02 \\
AT2021lwx & \textit{Swift} & 59966.5 & 4.20 & 4.25 & 4.30 &  45.17 & 45.26 & 45.34 & 15.56 & 15.70 & 15.83 \\
AT2021lwx & \textit{Swift} & 60054.1 & 4.11 & 4.15 & 4.19 &  45.23 & 45.32 & 45.41 &  15.82 & 15.94 &   16.06 \\
AT2021lwx & \textit{Swift} & 60067.1 & 4.10 & 4.21 &  4.33 &  44.83 &  45.08 & 45.50 &  15.32 & 15.69 &  16.13 \\
AT2021lwx & \textit{Swift} & 60069.1 & 4.13 & 4.16 & 4.19 & 45.32 & 45.39 & 45.45 & 15.86 & 15.94  & 16.03 \\
AT2021lwx & \textit{Swift} & 60100.5 & 4.12 &  4.15 & 4.18 & 45.20 & 45.27 & 45.33 & 15.82 & 15.90 &   15.98 \\
\hline
 \end{tabular}\\
\begin{flushleft}UV/optical luminosity, effective radius, and temperature estimated from blackbody fits to the host-subtracted and extinction-corrected DES and/or \textit{Swift} data. The values are given in common logarithms when scaled by the units in the second row of the header.
\bigskip
\\ $^{a}$The MJDs given for the DES epochs are the average of the MJDs for the individual bands used in that fit.
\end{flushleft}
\label{tab:BB_fits} 
\end{table}

\clearpage

\begin{table}
\centering
\caption{X-ray Luminosities}
 \begin{tabular}{ccccc}
  \hline
Object &  Source & MJD & log(L$_x$ / (erg s$^{-1}$)) & L$_x$ Uncertainty \\
\hline
Gaia16aaw & ROSAT & 48100.0 & 45.60 & 0.15 \\
Gaia16aaw & ROSAT & 50500.0 & 45.26 & 0.17 \\
Gaia16aaw & \textit{Swift} & 59175.0 & 45.14 & 0.10 \\
Gaia18cdj & ROSAT & 48100.0 & 45.42 & --- \\
Gaia18cdj & \textit{Swift} & 59164.3 & 44.56 & --- \\
AT2021lwx & ROSAT & 48200.0 & 46.02 & --- \\
AT2021lwx & \textit{Swift} & 59923.6 & 44.99 & --- \\
AT2021lwx & \textit{Swift} & 59966.5 & 44.93 & 0.24 \\
AT2021lwx & \textit{Swift} & 60054.1 & 44.85 & --- \\
AT2021lwx & \textit{Swift} & 60067.1 & 44.59 & 0.20 \\
AT2021lwx & \textit{Swift} & 60069.1 & 44.94 & 0.19 \\
AT2021lwx & \textit{Swift} & 60100.5 & 44.68 & 0.23 \\
AT2021lwx & XMM-Newton & 60101.7 & 44.08 & 0.04 \\
AT2021lwx & Chandra & 60300.0 & 44.44 & 0.08 \\
\hline
 \end{tabular}\\
\begin{flushleft} Rest-frame 0.3 - 10 keV X-ray luminosities for the ENTs in our sample. A value of ``---'' in the uncertainty column indicates a 3$\sigma$ upper limit.\end{flushleft}
\label{tab:xrays} 
\end{table}

\clearpage

\begin{table}
\centering
\caption{Inputs for Rate Calculations}
 \begin{tabular}{cccc}
  \hline
Survey & $t_{survey}$ (yr) & Limiting Magnitude & $f_{loss}$ \\
\hline
Gaia & 9.5 & 20.7 & 1.0 \\
ZTF & 5.3 & 20.5 & 0.73 \\
\hline
 \end{tabular}\\
\begin{flushleft}Time spans, limiting magnitudes, and loss factors corresponding to the survey coverage used to estimate the rates of the ENTs.\end{flushleft}
\label{tab:rates_input} 
\end{table}

\end{document}